\documentclass{emulateapj}
\usepackage{placeins}
\usepackage{graphicx}
\usepackage{amsmath}
\usepackage{natbib}

\shorttitle{Global MHD Simulations} \shortauthors{Zhu \& Stone}

\newcommand\msunyr{\rm M_{\odot}\,yr^{-1}}
\newcommand\be{\begin{equation}}
\newcommand\en{\end{equation}}

\newcommand\etal{{\rm et al}.\ }

\begin{document}

\title{Global Evolution of an Accretion Disk with Net Vertical Field: Coronal Accretion, Flux Transport, and Disk Winds}

\author{Zhaohuan Zhu\altaffilmark{1}, AND  James M. Stone\altaffilmark{2}}

\altaffiltext{1}{Department of Physics and Astronomy, University of Nevada, Las Vegas, 
4505 South Maryland Parkway, Las Vegas, NV 89154, USA}
\altaffiltext{2}{Department of Astrophysical Sciences, 4 Ivy Lane, Peyton Hall,
Princeton University, Princeton, NJ 08544, USA}
 
\email{zhzhu@physics.unlv.edu, jstone@astro.princeton.edu }

\begin{abstract}
We report results from global ideal MHD simulations that study thin accretion disks (with thermal scale height $H/R$=0.1 and 0.05) threaded by net vertical magnetic fields.  Our computations span three orders of magnitude in radius, extend all the way to the pole, and are evolved for more than one thousand innermost orbits.  We find that: (1) inward accretion occurs mostly in the upper magnetically dominated regions of the disk at $z\sim R$, similar to predictions from some previous analytical work and the "coronal accretion" flows found in  GRMHD simulations.  (2) A quasi-static global field geometry is established in which flux transport by inflows at the surface is balanced by turbulent diffusion.  The resulting field is strongly pinched inwards at the surface.  A steady-state advection-diffusion model, with turbulent magnetic Prandtl number of order unity, reproduces this geometry well. (3) Weak unsteady disk winds are launched beyond the disk corona with the Alfv\'{e}n radius $R_{A}/R_{0}\sim3$.  Although the surface inflow is filamentary and the wind is episodic, we show the time averaged properties are well described by steady wind theory.  Even with strong fields, $\beta_{0}=10^3$ at the midplane initially, only 5\% of the angular momentum transport is driven by the wind, and the wind mass flux from the inner decade of radius is only $\sim$ 0.4\% of the mass accretion rate.  (4) Within the disk, most of the accretion is driven by the $R\phi$ stress from the MRI and global magnetic fields. Our simulations have many applications to astrophysical accretion systems.
\end{abstract}

\keywords{accretion, accretion disks - dynamo - diffusion - magnetohydrodynamics (MHD) - 
instabilities - turbulence - protoplanetary disks -  }

\section{Introduction}
Accretion and outflow have been observed in a wide range of astrophysical systems,
ranging from protoplanetary disks \citep{Hartmann1998} to 
 supermassive black holes \citep{Begelman1984}.   
Two mechanisms are likely to drive accretion in most systems: MHD turbulence or a magnetized disk wind. 
Turbulence can lead to net $R-\phi$ stress within the disk that transports angular momentum radially,
while a disk wind can lead to a $z-\phi$ stress at the disk surface which carries angular momentum away vertically. 
It is thought
the main mechanism driving turbulence is the magnetorotational instability (MRI, \citealt{BalbusHawley1991,BalbusHawley1998}),
while in a Newtonian potential the main mechanism which produces a wind is
the magnetocentrifugal effect associated with vertical fields \citep{BlandfordPayne1982}. 
Since many astrophysical disks are poorly ionized (e.g. protoplanetary disks),
the effects of non-ideal MHD on MRI (e.g. the review by \citealt{Turner2014a}) and disk winds \citep{Konigl1989,WardleKoenigl1993,Konigl2010,Salmeron2011,BaiStone2013,Bai2017,Simon2013a, Simon2013b} is also important in these systems.
 
Outflow/wind/jet launching mechanisms require large scale poloidal magnetic fields. In the magnetocentrifugal wind
model, the magnetic field is anchored in the rotating disk. If the poloidal component of the field
makes an angle of less than 60$^o$ with the disk surface, the centrifugal force can break the
potential barrier and accelerates matter outwards, leading to outflow. The stress at the wind base
launches the outflow and, at the same time torques the disk, 
leading to accretion below the wind region. Such a picture has been confirmed by numerical
simulations with prescribed poloidal magnetic fields (e.g. the review by \citealt{Pudritz2007}). 

Net vertical magnetic fields also play an important role in MRI turbulence.
It is known that the turbulent stress increases as the net field increases \citep{Hawley1995}, thus net field promotes both outflow and disk accretion.

To address the relative importance of turbulence and a disk wind in driving accretion, we need to rely on numerical simulations.
However, simultaneously resolving small scale turbulence and capturing the large scale disk wind is challenging. In order to resolve the MRI, most previous simulations with net vertical flux only study a 
small patch of the disk using the shearing box approximation \citep{SuzukiInutsuka2009,BaiStone2013,Fromang2013}. 
However, the resulting wind in such simulations can flow in either radial direction \citep{BaiStone2013,Lesur2014}, and the outflow rate
drops dramatically when a taller box has been used \citep{Fromang2013}. Global simulations of turbulence within an accretion disk with net vertical fields 
have been carried out
recently \citep{SuzukiInutsuka2014,Gressel2015}
 and strong outflows and high disk accretion rates
have been observed. However,
these simulations have limited radial range, only cover a wedge around the disk midplane $z\in[-0.5R,0.5R]$, and have not been evolved long enough for a true steady-state to emerge. On the other hand, global (usually two-dimensional) simulations of disk winds that do not capture the MRI in the disk interior have been reported 
 \citep{StoneNorman1994,OuyedPudritz1997a,Krasnopolsky1999,Krasnopolsky2003,Kato2002,PorthFendt2010,RamseyClarke2011}.  
In such work, either only the wind region has been studied or explicit resistivity has been assumed in the disk region as a "sub-grid" of the MRI (e.g. \citealt{CasseKeppens2002,CasseKeppens2004,FendtCemeljic2002,Zanni2007,Tzeferacos2009,Tzeferacos2013}).  In this work, we carry out global three-dimensional simulations which capture both the MRI and disk wind self-consistently by
using both mesh-refinement and special polar boundary conditions. 

An important question that can only be addressed in global simulations is the net rate of transport of vertical magnetic field.
Net vertical fields are both advected by the large-scale flow, and diffuse
due to turbulence. In the phenomenological model by \cite{Vanballegooijen1989}
disk accretion and field diffusion is modeled using turbulent viscosity and resistivity, however the result depends sensitively on the relative amplitude of both.
As clearly summarized in \cite{GuanGammie2009},
the evolution of the poloidal fields is governed by
\begin{align}
\partial_{t} A_{\phi}=&-v_{R} B_{z}-\frac{\eta}{R}\partial_{z} B_{R}+\eta\partial_{R} B_{z}\nonumber\\
\partial_{t} A_{\phi}=&v_{R}\frac{1}{R}\partial_{R} (RA_{\phi})+\eta\partial^2_{z} A_{\phi}+\eta\partial_{R}\left[ \frac{1}{R} \partial_{R} (RA_{\phi})\right]\,.\label{eq:vectorap}
\end{align}
if only radial motion is considered. In Equation \ref{eq:vectorap}, $A_{\phi}$ is the azimuthal component of the vector potential.  On the right hand side, the first term is advection by the accretion flow. Since $v_{R}\sim \nu/R$
in the viscous model, this term is roughly $\nu B_{z}/R $. The second term is the
vertical diffusion of radial field and is roughly $\eta B_{R}/z$. If we assume the large scale field
enters the disk at an angle of $\sim45^{o}$, as in the disk wind model, we have $B_{R}/z\sim B_{z}/H$.
The third term is radial
diffusion, roughly $\eta B_{z}/R$.
Thus, the first and second terms dominate, and they balance each other when $\nu/\eta\sim R/H$
or the turbulent magnetic Prandtl number (Pr) $\sim R/H$. However, local shearing box simulations suggest that 
Pr$\sim $1 \citep{GuanGammie2009,LesurLongaretti2009,FromangStone2009}, implying that 
large-scale fields will diffuse outwards faster than the inward advection
\citep{Lubow1994}. To maintain large scale magnetic fields, either turbulent diffusion is significantly reduced
\citep{SpruitUzdensky2005,RothsteinLovelace2008} or inward accretion is increased. 
An important insight is provided by \cite{Beckwith2009} (see also simulations by \citealt{StoneNorman1994}) which observed
that magnetic flux is mainly transported in the corona of the disk and magnetic fields are pinched within the corona, thus 2-D
treatment of the magnetic flux transport is desired. 
\cite{Lovelace2009, GuiletOgilvie2012, GuiletOgilvie2013}  have carried out such study of the magnetic
flux transport in 2-D and found that the fast flux transport at the disk surface may indeed sustain a steady magnetic field configuration, potentially solving the too-fast field diffusion
problem. Since the 
radial distribution of global magnetic fields also determines the collimation of disk winds \citep{OgilvieLivio2001, Anderson2005,Pudritz2006}, the processes which
determine and maintain large scale magnetic fields in disks are essential for sustaining outflow or even accretion.

In \S 2,  the theoretical framework on describing turbulence and disk wind is presented.
Our numerical method is introduced in \S 3. The results are presented in \S 4.
After a short discussion in \S 5, we conclude in \S 6. 

\begin{figure*}[t]
\begin{center}
\includegraphics[trim=0cm 0cm 0cm 0cm, clip=true, width=1\textwidth]{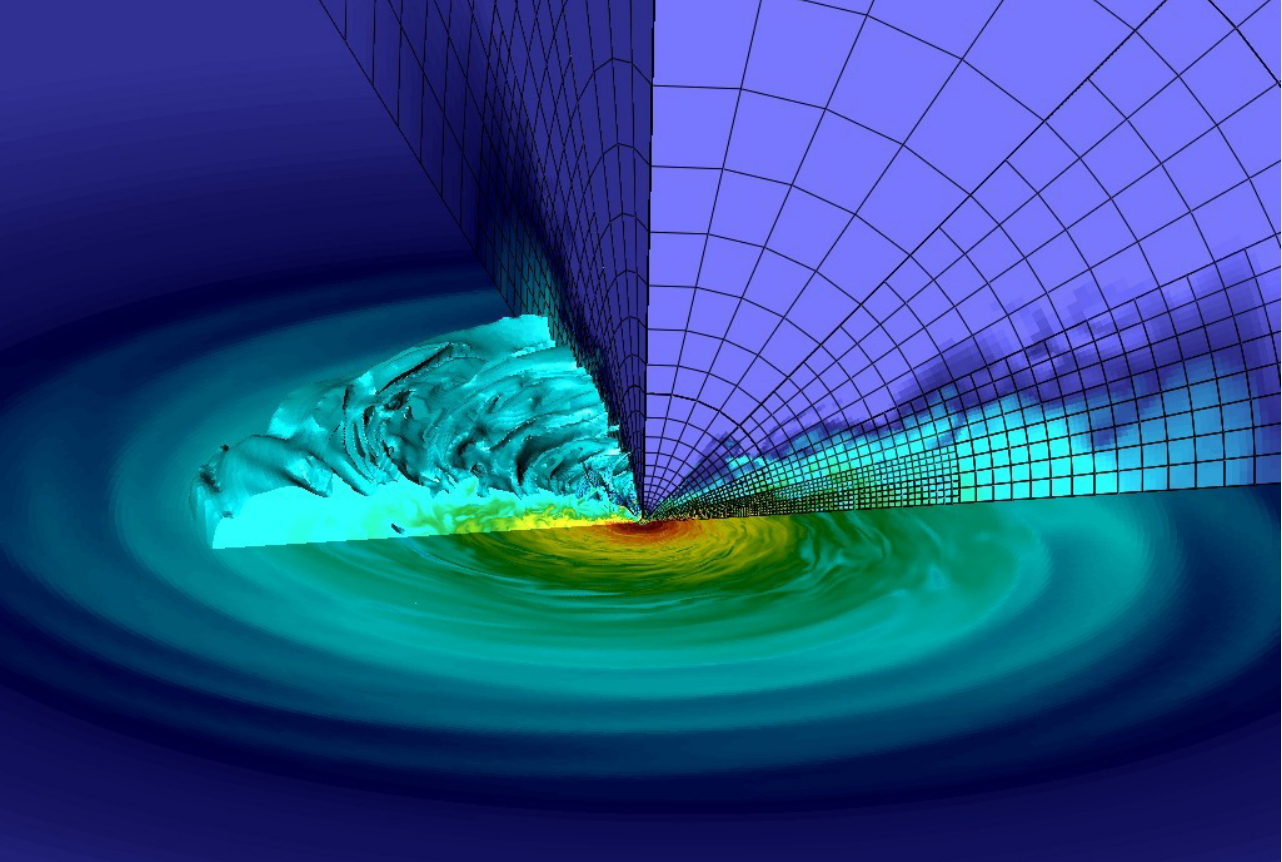}
\end{center}
\vspace{-0.1cm}
\caption{Density iso-surface (left side) and poloidal slice of density showing the nested grid (right side) from a snapshot at time 42 $T_{0}$ in our fiducial model with initial $\beta_0 = 1000$.
Static mesh refinement is used to capture MRI turbulence at the disk midplane, and to extend the calculation to the pole.   }
\label{global}
\end{figure*}

\section{Theoretical Framework}

Which is more important for disk accretion: turbulence or outflow?
If we average the angular momentum equation in the azimuthal direction and integrate it in the vertical direction, we can derive
\begin{align}
\frac{\partial \int \langle \rho v_{\phi}\rangle dz}{\partial t}&=-\frac{1}{R^2}\frac{\partial}{\partial R}\left( R^2 \int\left(\langle \rho v_{R}\delta v_{\phi}\rangle-\langle B_{R}B_{\phi}\rangle\right)dz\right) \nonumber\\
&-\frac{1}{2\pi R^2}\frac{\partial R v_{k} \dot{M}_{acc}}{\partial R}-\left(\langle \rho v_{z} v_{\phi}\rangle-\langle B_{z}B_{\phi}\rangle\right)\bigg |_{z_{min}}^{z_{max}}\label{eq:angcyl}
\end{align}
or 
\begin{align}
\frac{\partial \int \langle \rho \delta v_{\phi}\rangle dz}{\partial t}&=-\frac{1}{R^2}\frac{\partial}{\partial R}\left( R^2 \int\left(\langle \rho v_{R}\delta v_{\phi}\rangle-\langle B_{R}B_{\phi}\rangle\right)dz\right) \nonumber\\
&-\frac{\dot{M}_{acc}}{2\pi R^2}\frac{\partial R v_{k}}{\partial R}-\left(\langle \rho v_{z} \delta v_{\phi}\rangle-\langle B_{z}B_{\phi}\rangle\right)\bigg |_{z_{min}}^{z_{max}}\label{eq:angcyl2}
\end{align}
for the perturbed quantities ($\delta v_{\phi}=v_{\phi}-v_{k}$).
The symbol $\langle \rangle$ denotes that the quantity has been averaged over the $\phi$ direction \footnote{$v_{k}$ has been assumed to be
constant along $z$. Without this assumption, there will be an additional term related to $\dot{M}_{loss}\partial v_{k}/\partial z$.}, and
$\dot{M}_{acc}= 2\pi R\int \rho v_{R}dz$. We refer to
$\langle \rho v_{R}\delta v_{\phi}\rangle$ and $-\langle B_{R}B_{\phi}\rangle$ as the radial Reynolds and Maxwell stress respectively, to distinguish
them from the vertical stresses $T_{\phi z}$. If we normalize the stresses with pressure, we can identify different contributions to the total stress $\alpha$ as 
\begin{align}
\alpha_{R\phi, Rey}=\langle \rho v_{R}\delta v_{\phi}\rangle/\langle p\rangle\quad {\rm and}\quad \alpha_{R\phi, Max}=-\langle B_{R}B_{\phi}\rangle/\langle p\rangle \nonumber \\
\alpha_{\phi z, Rey}=\langle \rho v_{z}\delta v_{\phi}\rangle/\langle p\rangle\quad {\rm and}\quad \alpha_{\phi z, Max}=-\langle B_{z}B_{\phi}\rangle/\langle p\rangle \,.\nonumber
\end{align}
Stresses and the $\alpha$ parameters in spherical-polar coordinate can be defined in similar ways.
Since the vertically integrated $R-\phi$ stress determines the disk accretion rate as in Equation \ref{eq:angcyl2}, we can define
the vertically integrated $\alpha$ parameter as
\begin{equation}
\alpha_{int}=\frac{\int T_{R\phi}dz}{\Sigma c_{s}^2}\,.
\end{equation}
where $T_{R\phi}$ is the sum of both radial Reynolds and Maxwell stress.
If we choose $v_{k}$ so that $\langle\rho\delta v_{\phi}\rangle$=0 and assume that the magnetic stress dominates at the disk surface,
Equation \ref{eq:angcyl2} can be written as
\begin{equation}
\dot{M}_{acc}=-\frac{2\pi}{\partial R v_{k}/\partial R}\left(\frac{\partial}{\partial R}\left( R^2 \alpha_{R\phi,int}\Sigma c_{s}^2\right) -R^2\langle B_{z}B_{\phi}\rangle\bigg |_{z_{min}}^{z_{max}}\right)\label{eq:mdot}
\end{equation}

Equation \ref{eq:mdot} suggests that disk accretion is due to both 
$T_{R\phi}$ within the disk and $T_{\phi z}$ exerted
at the disk surface.
 Normally, the $R-\phi$ stress is from disk turbulence, 
and the $z-\phi$ stress at the surface is from a magnetocentrifugal disk wind. If both the $R-\phi$ and $z-\phi$ stresses 
have similar values, the second term on the right will be larger than the first term by a factor of $R/z$ which can be quite
large for a thin disk . 
On the other hand, with vigorous turbulence the internal stress may be larger than the surface stress, thus it is unclear if disk accretion is driven by turbulence or wind
when both processes are present. 

\section{Method}

We solve the magnetohydrodynamic (MHD) equations  in the ideal MHD limit using Athena++ (Stone \etal 2017, in preparation).
Athena++ is a newly developed grid based code using a higher-order Godunov scheme for MHD and 
the constrained transport (CT) to conserve the divergence-free property for magnetic fields. 
Compared with its predecessor Athena \citep{GardinerStone2005,GardinerStone2008,Stone2008},  Athena++ is highly optimized for 
speed and uses a flexible grid structure that enables mesh refinement, allowing global
numerical simulations spanning a large radial range. 
Furthermore, the geometric source terms in curvilinear coordinates 
(e.g. in cylindrical and spherical-polar coordinates) are carefully implemented so that angular momentum
is conserved to machine precision, a crucial feature to enable the angular momentum budget analysis as presented in \S 4.1.

Since a disk wind flows radially, we adopt a spherical-polar coordinate system ($r$, $\theta$, $\phi$)  
for our simulations, which should minimize
the effects of the domain boundary on the wind properties \citep{Ustyugova1999}.
Although we adopt spherical-polar coordinates for the simulations, we transform some quantities to cylindrical coordinates to study physical processes in disks. 
In this paper, we use ($R$, $\phi$, $z$) to denote positions in cylindrical coordinates and
($r$, $\theta$, $\phi$) to denote positions in spherical polar coordinates.  In both coordinate systems 
$\phi$ represents the azimuthal direction (the direction of disk rotation).

Our simulation domain extends from $r_{min}=0.1$ to $r_{max}=100$ with $\theta$ from 0 to $\pi$. 
$\phi$ extends from 0 to $2\pi$ except for the thin disk case for which $\theta$ is from 0 to $\pi/4$. 
We implement a special polar boundary in the $\theta$
direction allowing our simulation domain to extend all the way to the pole, which is different from previous similar simulations
where a hole was carved out close to the pole. This boundary condition prevents the loss of magnetic fields at the previously carved-out pole,
which is crucial to study transport of magnetic fields in disks. 
The details on
the implementation are given in the Appendix. 

The initial density profile at the disk midplane is
\begin{equation}
\rho_{0}(R,z=0)=\rho_{0}(R_{0},z=0)\left(\frac{R}{R_{0}}\right)^p\,,
\end{equation}
while the temperature is assumed to be constant on cylinders
\begin{equation}
T(R,z)=T(R_{0})\left(\frac{R}{R_{0}}\right)^q\,.
\end{equation}

Hydrostatic equilibrium in the $R-z$ plane requires that (e.g. Nelson \etal 2013)
\begin{equation}
\rho_{0}(R,z)=\rho_{0}(R,z=0) {\rm exp}\left[\frac{GM}{c_{s}^2}\left(\frac{1}{\sqrt{R^2+z^2}}-\frac{1}{R}\right)\right]\,,\label{eq:rho0}
\end{equation}
and
\begin{equation}
v_{\phi}(R,z)=v_{K}\left[(p+q)\left(\frac{c_{s}}{v_{\phi,K}}\right)^2+1+q-\frac{qR}{\sqrt{R^2+z^2}}\right]^{1/2}\,,\label{eq:vphi}
\end{equation}
where $c_{s}=\sqrt{p/\rho}$ is the isothermal sound speed, $v_{K}=\sqrt{GM_{*}/R}$, and $H=c_{s}/\Omega_{K}$.  A local isothermal equation of state is assumed during the simulation \footnote{To achieve this, we have used the adiabatic equation of state with $\gamma$=1.4, but instantaneous cooling is applied at each timestep. }.  Using Equations \ref{eq:rho0} and \ref{eq:vphi}, the density and velocity will become infinite at the pole. To avoid this, we use $R=\max(R,r_{min})$ in the above equations.

Considering the density at the disk's atmosphere is orders of magnitudes smaller than the density at the disk midplane, we adopt a density floor which varies with position 
\begin{equation}
\rho_{fl}=\left\{ \begin{array}{l} 
               \rho_{fl,0}\left(\frac{R}{R_{0}}\right)^p\left(\frac{1}{z^2}\right) \quad {\rm if}\quad R\geq r_{min}\\
                \rho_{fl,0}\left(\frac{r_{min}}{R_{0}}\right)^p\left(\frac{1}{z^2}\right) \quad {\rm if}\quad R< r_{min} \;{\rm and }\; r>3r_{min} \\
               \rho_{fl,0}\left(\frac{R}{R_{0}}\right)^p\left(\frac{1}{z^2}\right)\left(5-2\frac{r-r_{min}}{r_{min}}\right)\left(4\frac{r_{min}-R}{r_{min}}+1\right)\\ 
                  \quad \quad \quad\quad\quad\quad \quad {\rm if}\quad R< r_{min} \;{\rm and }\; r<3r_{min}\\
  \end{array}\right.
\end{equation}
This density floor becomes small at the disk atmosphere and becomes large close to the midplane and the inner radial boundary. 
When $\rho_{fl}$ gets too small ( smaller than $10^{-9}$ for $\beta_{0}=1000$ cases or $3\times10^{-9}$  for the $\beta_{0}=10^4$ case), 
we choose $10^{-9}$ or $3\times10^{-9}$ as the density floor.

Initial magnetic field is assumed to be vertical. To maintain $\nabla\cdot\bf{B}=0$, we use the
vector potential $\bf{A}$ to initialize the magnetic fields. We set
\begin{equation}
A_{\phi}=\left\{ \begin{array}{l} 
               \frac{1}{2}\times R \times B_{0}\left(\frac{r_{min}}{R_{0}}\right)^{m}  \quad {\rm if}\quad R\leq r_{min} \\
               \\
               \frac{B_{0}}{R_{0}^m}\frac{R^{m+1}}{ m+2}+\frac{B_{0}r_{min}^{m+2}}{R_{0}^m R}(\frac{1}{2}-\frac{1}{m+2})\quad {\rm if}\quad R> r_{min}
  \end{array}\right.
\end{equation}
where $m=(p+q)/2$.
Thus, the vertical magnetic fields are
\begin{equation}
B_{z}=\left\{ \begin{array}{l} 
           B_{0}\left(\frac{r_{min}}{R_{0}}\right)^{m} \quad {\rm if}\quad R\leq r_{min} \\
            \\
           B_{0}\left(\frac{R}{R_{0}}\right)^m \quad {\rm if}\quad R> r_{min}\,.
             \end{array}\right.
\end{equation}
With this setup, the plasma $\beta$ at the disk midplane beyond $r_{min}$ is a constant. 

We choose $p=-2.25$ and $q=-1/2$ in our simulation so that the disk surface density $\Sigma\propto R^{-1}$.  Then
$\rho_{0}(R_{0},z=0)$ is 1.  The time unit in the simulation is $T_{0}=2\pi/\Omega(R=1)$.
The grid is uniformly spaced in ln($r$), $\theta$, $\phi$ with 136$\times$64$\times$128 grid cells
in the domain of [ln(0.1), ln(100)]$\times$[$0$, $\pi$ ]$\times$[0, 2$\pi$] at the root level. Thus, $\Delta r/r$ equals 0.052 at the root level.
We use an open boundary condition in the radial direction (but mass inflow into the domain is not allowed), 
the polar boundary condition in the $\theta$ direction (Appendix), and
periodic boundary conditions in the $\phi$ direction. 

We have carried out three simulations, one with $(H/R)_{R=R_{0}}=0.1$ and initial plasma $\beta$=1000 at the disk midplane (the fiducial case), one with $(H/R)_{R=R_{0}}=0.1$ and $\beta=10^4$ (the weaker field case), and one with $(H/R)_{R=R_{0}}=0.05$ and $\beta=10^3$ (the thin disk case). The density floor parameter $\rho_{fl, 0}$  is $10^{-6}$ for the fiducial case and  $3\times10^{-7}$ for the other cases.
For the simulations with
$(H/R)_{R=R_{0}}=0.1$, three levels of refinement have been adopted towards the disk midplane, and $\Delta r/r$ at the midplane is 0.0065. Since $H/R$=0.1 at $R=1$, the disk scale height is resolved
by 15 grids at the finest level. For the thin disk case, 4 levels of mesh refinement have been used so that the disk scale height is still resolved by 15 grids. To save computational cost, $\phi$  in the thin disk case only extends from 0 to $\pi/4$.
The above choice of $H/R$ covers the normal thickness range of protoplanetary disks. The choice of the field strength is based on 
both numerical and analytical considerations. MRI is sufficiently resolved numerically with a stronger field ($\beta_{0}=1000$), 
while analytical work \citep{GuiletOgilvie2014, Okuzumi2014} suggests  a weaker net field threading protoplanetary disks ($\beta_{0}\sim 10^{4}-10^{7}$). 
We will mainly present results for our fiducial case 
($(H/R)_{R=R_{0}}=0.1$ and $\beta_{0}=1000$ ) since the MRI is better
resolved at the midplane and the simulation runs longer, but
other cases will be discussed  in \S 4.4.

In ideal MHD, the
wavelength for the fastest growing linear MRI mode satisfies 
\begin{equation}
\lambda=2\pi\sqrt{16/15}|v_{A}|/\Omega=9.17\beta^{-1/2} H
\end{equation}
(Hawley \etal 1995). With
$\beta_{0}$=1000, $\lambda$ is $\sim$ 0.29 H so that the most unstable wavelength
is resolved by 4.5 grid cells at $R=1$.

\begin{figure}[ht!]
\centering
\includegraphics[trim=0cm 6.5cm 0cm 0cm, width=0.5\textwidth]{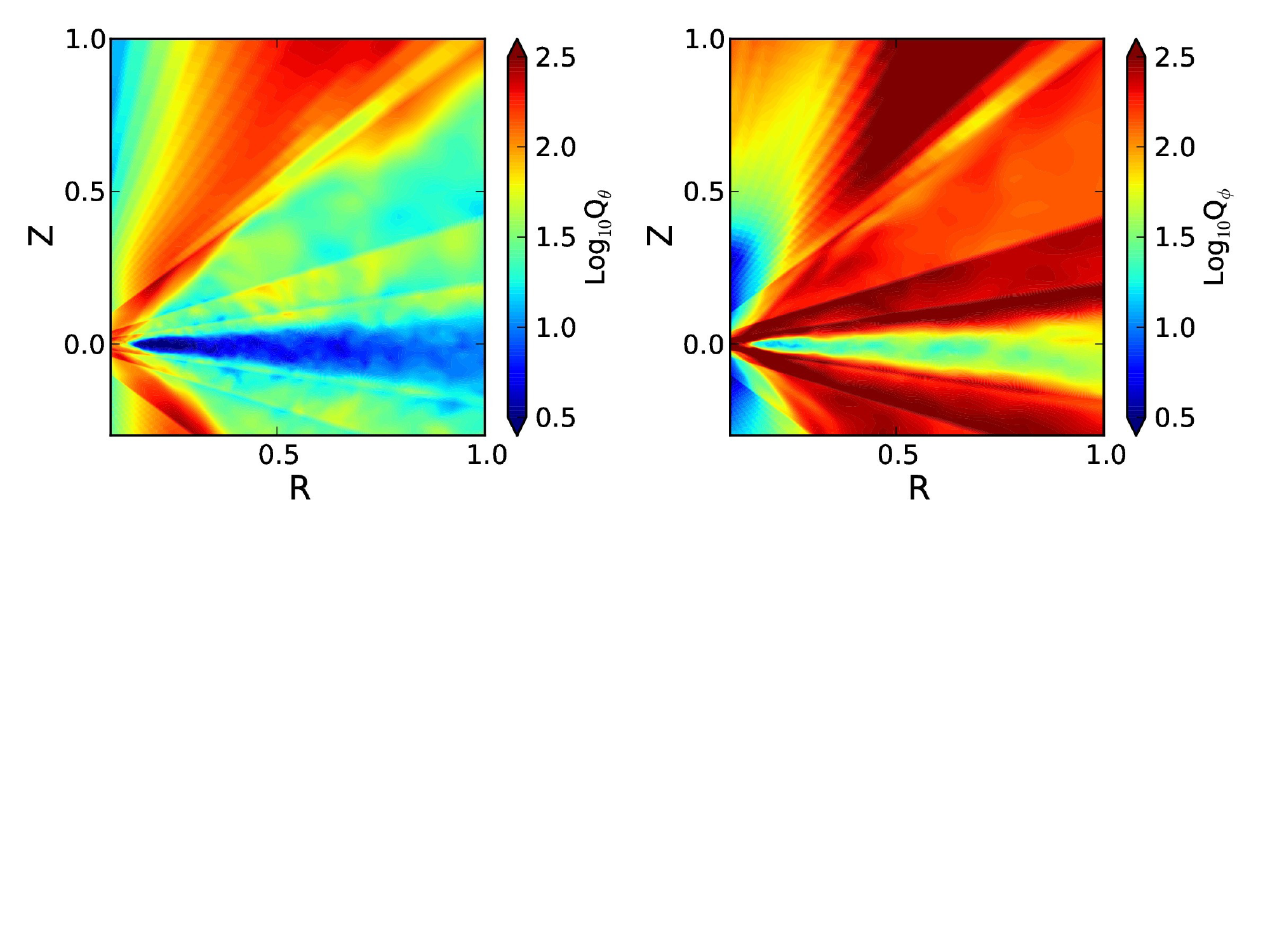} 
\vspace{-0.8 cm}
\caption{ The quality factor for $B_{\theta}$ and $B_{\phi}$ in our fiducial model at t=42 $T_{0}$. } \label{fig:twodq}
\end{figure}

To demonstrate that our fiducial case has reached the necessary resolution to capture MRI,
we plot the azimuthally averaged quality factor ($\langle Q_{\theta}\rangle$ and 
$\langle Q_{\phi}\rangle$) for $B_{\theta}$ and $B_{\phi}$ in Figure \ref{fig:twodq}. 
The quality factor is defined as the number of grid cells that resolve the fastest MRI growing mode \citep{Noble2010},
\begin{eqnarray}
Q_{\theta}=\lambda_{\theta}/(r \Delta \theta)\\
Q_{\phi}=\lambda_{\phi}/(r {\rm sin\theta} \Delta \phi)
\end{eqnarray}
where $\lambda_{\theta}=2\pi\sqrt{16/15}|v_{A,\theta}|/\Omega_{K}$
and $\lambda_{\phi}=2\pi\sqrt{16/15}|v_{A,\phi}|/\Omega_{K}$. The quantities $v_{A,\theta}$
and $v_{A,\phi}$ are the Alfven velocity calculated using the $\theta$ and $\phi$ components
of the magnetic field, and $\Omega_{K}$ is the Keplerian angular velocity at the disk midplane. 
To get converged results for MRI turbulence, \cite{Sorathia2012}
have shown that if $Q_{\phi}\approx 10$, $Q_{z}$ needs to be $\gtrsim 10-15$, and if $Q_{\phi}\gtrsim 25$, while
$Q_{z}$ can be smaller ($\sim$ 6 in their Figure 8). 
As shown in Figure \ref{fig:twodq}, the quality factor decreases towards the disk midplane despite the fact that it has been 
boosted by a factor of 8 towards the midplane due to mesh refinement.  
$Q_{\phi}$ is larger than 10 in the whole disk  ($>$50 at $R=1$), and $Q_{\theta}$ is around 10 at R=1.
Thus, our simulation should have adequate resolution to capture MRI turbulence except at the midplane of the innermost region ($R<0.3$).

\begin{figure*}[ht!]
\centering
\includegraphics[trim=0cm 1cm 0cm 0cm, width=1.0 \textwidth]{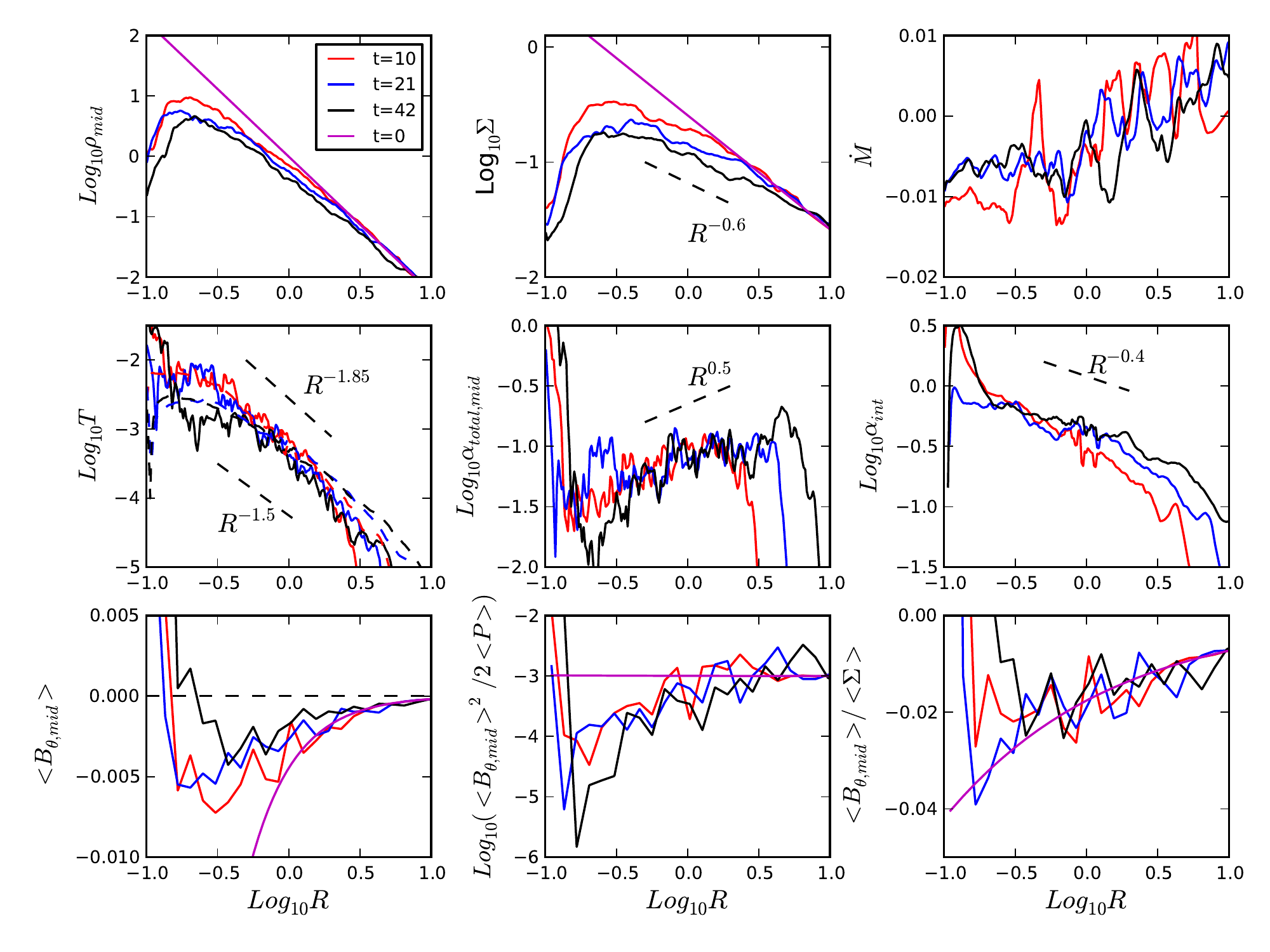} 
\vspace{0 cm}
\caption{The disk midplane density, surface density, mass accretion rate (upper  panels),
stresses (the dashed curves are vertically integrated $R\phi$ stresses and the solid curves are $r\phi$ stresses at the midplane), midplane $\alpha$, and vertically integrated $\alpha$ (middle panels),
 and net vertical magnetic field , $\langle B_{\theta,mid}\rangle^2/2\langle P\rangle$,  $\langle B_{\theta,mid}\rangle/\langle\Sigma\rangle$ (lower panels)  at different times. $\alpha_{total}$ and $\alpha_{int}$ are all calculated with the $r\phi$ or $R\phi$ stress. In the panels at the bottom row, quantities are also 
 averaged along the radial direction over 1 disk scale height.  } \label{fig:dB2}
\end{figure*}

\section{Results}

After running for t=45.6 $T_{0}$, which is equivalent to 1442 orbits at the inner boundary,
our fiducial disk reaches a steady state within R$\sim$3, i.e. the inner factor of $\sim30$ in radius, as evident in  the $\dot{M}$ panel of Figure \ref{fig:dB2} which shows that
 the disk accretion rate is almost a constant within R$\sim$3.  
A steady state throughout the whole disk cannot be established with our simulation setup since we do not supply new material at the outer boundary. The outer disk is still evolving.
As shown in the $\alpha_{total,mid}$ panel of Figure \ref{fig:dB2}, the MRI in the outer disk saturates at later times. 
 
 Once a quasi-steady state is reached at t=42 $T_{0}$ , the surface density profile follows $\sim 0.12 R^{-0.6}$, and 
 the midplane $\alpha_{r\phi}$ profile follows $\sim 0.1 R^{0.5}$. With these profiles and the given temperature profile, the 
 midplane stress then follows $\propto R^{-1.85}$. 
 The vertically integrated $\alpha_{r\phi}$ follows
 $\sim  0.5 R^{-0.4}$, leading to the profile of $R^{-1.5}$ for the vertically integrated stress. 
 With these values and Equation \ref{eq:mdot}, we can calculate $\dot{M}$=-0.0038,
 which is a constant with radii and  consistent with the direct $\dot{M}$ measurement
 in the upper right panel of Figure \ref{fig:dB2}. We can see that the vertically integrated $\alpha$ is much larger than the midplane $\alpha$ by a factor of $\sim$ 10. 
 This is because the stress is almost uniform vertically, or even increases towards the disk surface, while the density drops sharply towards the disk surface. 
 If the stress is perfectly uniform vertically to some disk heights, the vertically integrated $R\phi$ stress should follow $R^{-0.85}$ and the  midplane $\alpha$
 and the vertically integrated $\alpha$ should have the same slope. But the stress panel in Figure \ref{fig:dB2} suggests that the vertically integrated $R\phi$ stress
 is much steeper with $R^{-1.5}$, which is because the stress is larger
 at the inner disk's atmosphere where the fields are pinched as shown in \S 5.1. Since $\alpha_{int}\sim$ 0.5 at R=1,
42 $T_{0}$ is already longer than the viscous timescale ($R^2\Omega/(\alpha_{int}c_{s}^2)$) at R=1, 
explaining why the inner disk has a constant accretion rate.
 
 While the disk is losing mass through accretion, it is also losing magnetic field. The net vertical magnetic field
 decreases with time, shown in the $B_{\theta,mid}$ panel of Figure \ref{fig:dB2}. 
The plasma $\beta$
 calculated with the net vertical magnetic field increases from $10^3$ to $10^5$ at the inner disk (the lower center panel). 
 On the other hand, the ratio between the net vertical field and the disk surface density does not vary
 much over time (the $B/\Sigma$ panel). Although it is tempting to interpret this as flux freezing during
 the accretion process, we will show later that the global field transport is much more complicated. 
 
 \subsection{Accretion Structure}
\begin{figure*}[ht!]
\centering
\includegraphics[trim=0cm -1.cm 0cm 0cm, width=0.9\textwidth]{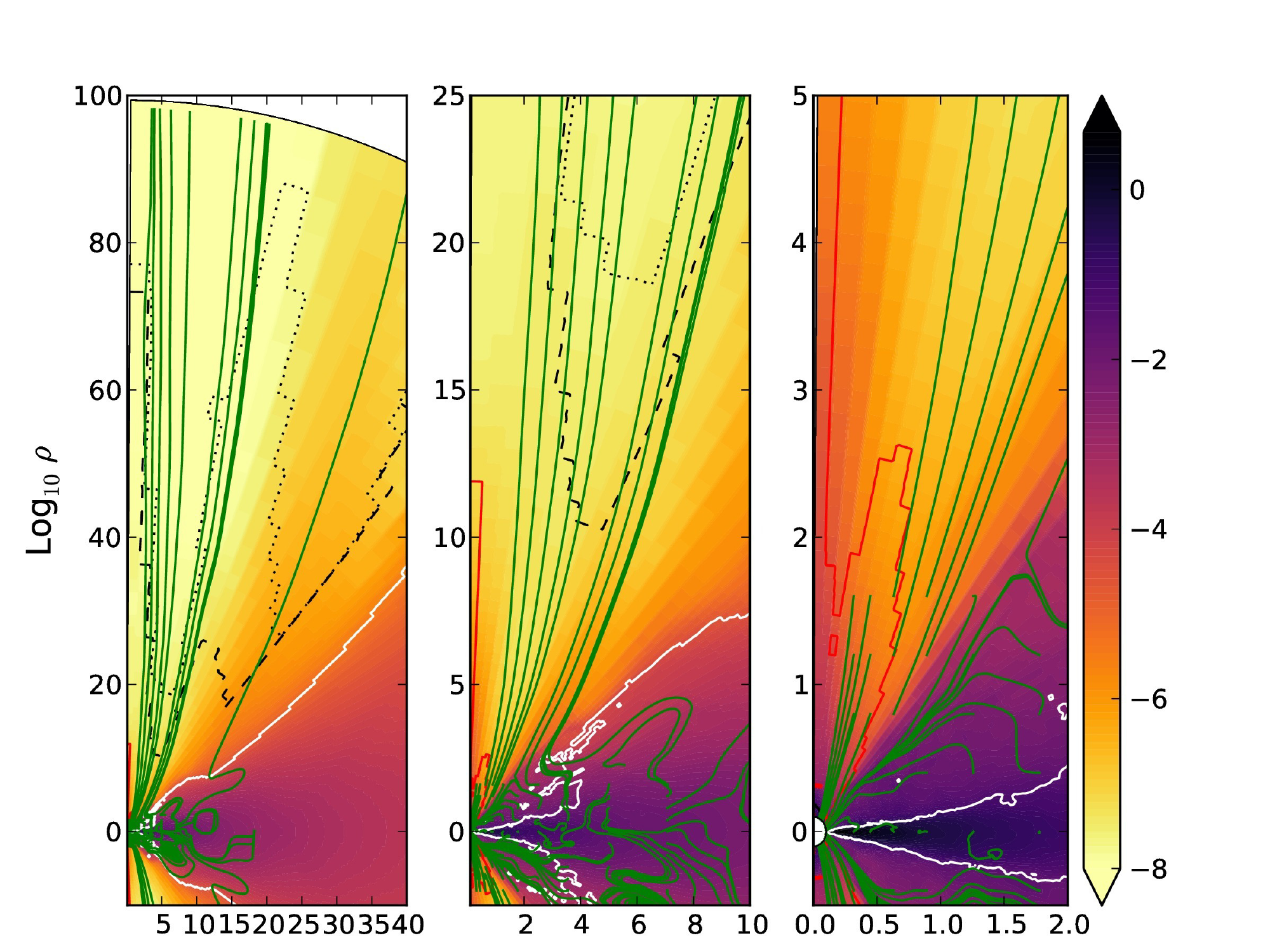} 
\vspace{-0.8 cm}
\caption{The azimuthally averaged density at different scales at t=42 $T_{0}$. The green curves are the velocity field lines calculated
with azimuthally averaged velocities. The dashed curves label the Alfven surface while the dotted curves label 
the fast magnetosonic surface. Clearly our simulation domain is beyond all these critical points. 
The white curves label where $\overline{\beta}=1$.
The red curves label where the azimuthally averaged density is only larger than the density floor by 10$\%$, indicating the 
majority grids in the azimuthal direction have reached the density floor there. Only the disk atmosphere at the very inner region 
has reached the density floor. 
} \label{fig:twodrhovel}
\end{figure*}

Although the whole disk accretes inwards, the accretion flow has a complicated structure within the disk
as shown in Figure \ref{global} and \ref{fig:twodrhovel}. 
Figure  \ref{global} suggests that the disk surface is highly filamentary, especially in the corona region, similar to \cite{Suriano2017}. 
Thus, to study the statistical properties of the disk, we average the quantities over the azimuthal direction, as shown in Figure \ref{fig:twodrhovel}.
The green curves in the figure are the velocity field lines and the white curves are where the plasma $\langle\beta\rangle$=1. 
The most striking feature is that the disk accretes 
through the disk atmosphere where the disk is magnetically
dominated ($\beta<1$) as shown in the rightmost panel. 
Such surface accretion has been seen as early as \cite{StoneNorman1994} and
recently been studied in GRMHD simulations by
\cite{Beckwith2009} where it was termed "coronal accretion".
Analytical works by \citep{GuiletOgilvie2012, GuiletOgilvie2013} have also seen such surface accretion
when the turbulent viscosity is considered in a constant H/R disk.

\begin{figure*}[ht!]
\centering
\includegraphics[trim=0cm 0cm 0cm 0cm, width=0.9 \textwidth]{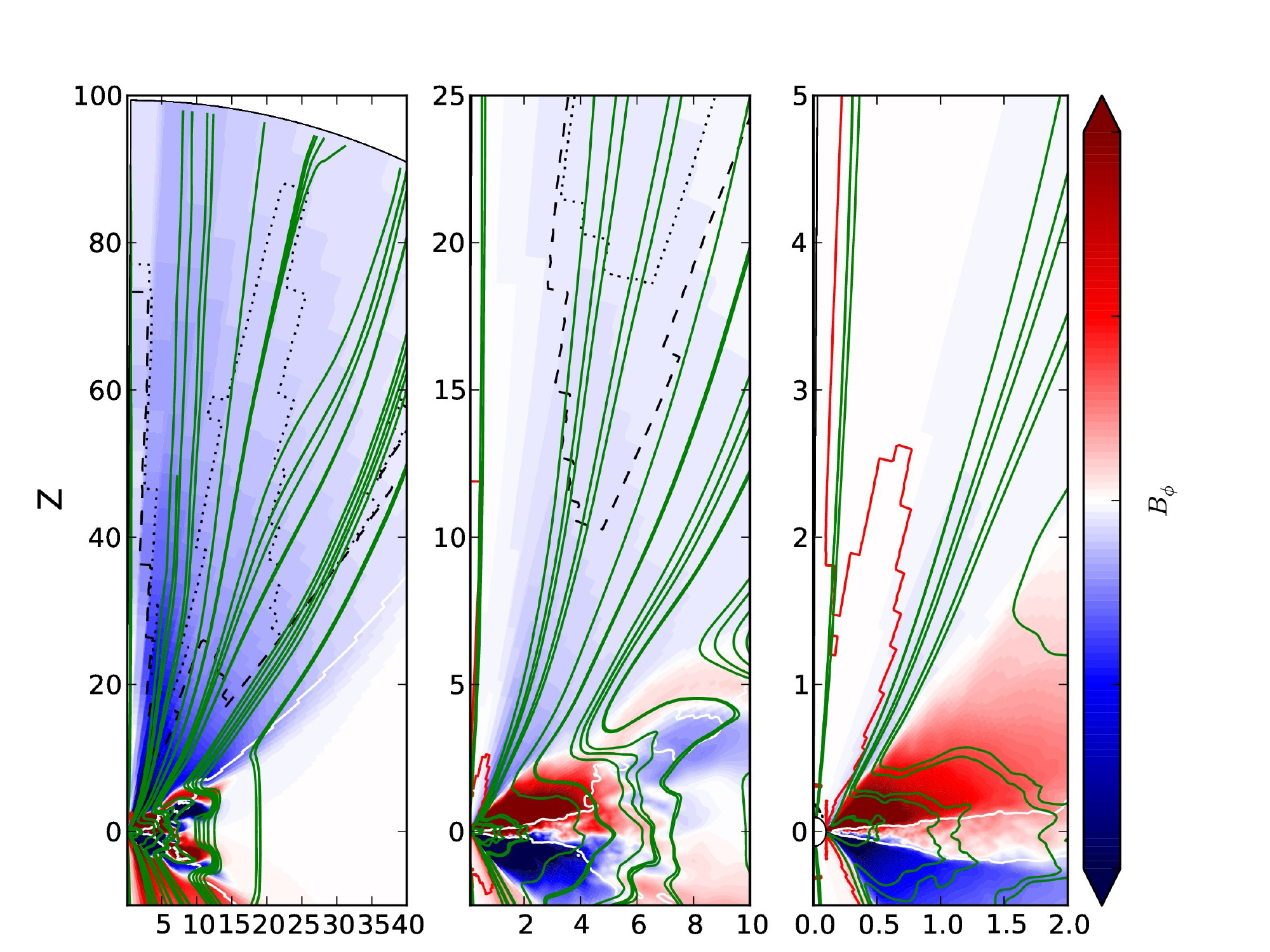} 
\vspace{-0.3 cm}
\caption{Similar to Figure \ref{fig:twodrhovel} but the color map is $B_{\phi}$ and 
the green curves show the magnetic field lines. The color bars in the left, middle, and right panel are from
-0.001 to 0.001, -0.01 to 0.01, and -0.1 to 0.1 respectively. } \label{fig:twodrhoalfven}
\end{figure*}

\begin{figure*}[ht!]
\centering
\includegraphics[width=0.9 \textwidth]{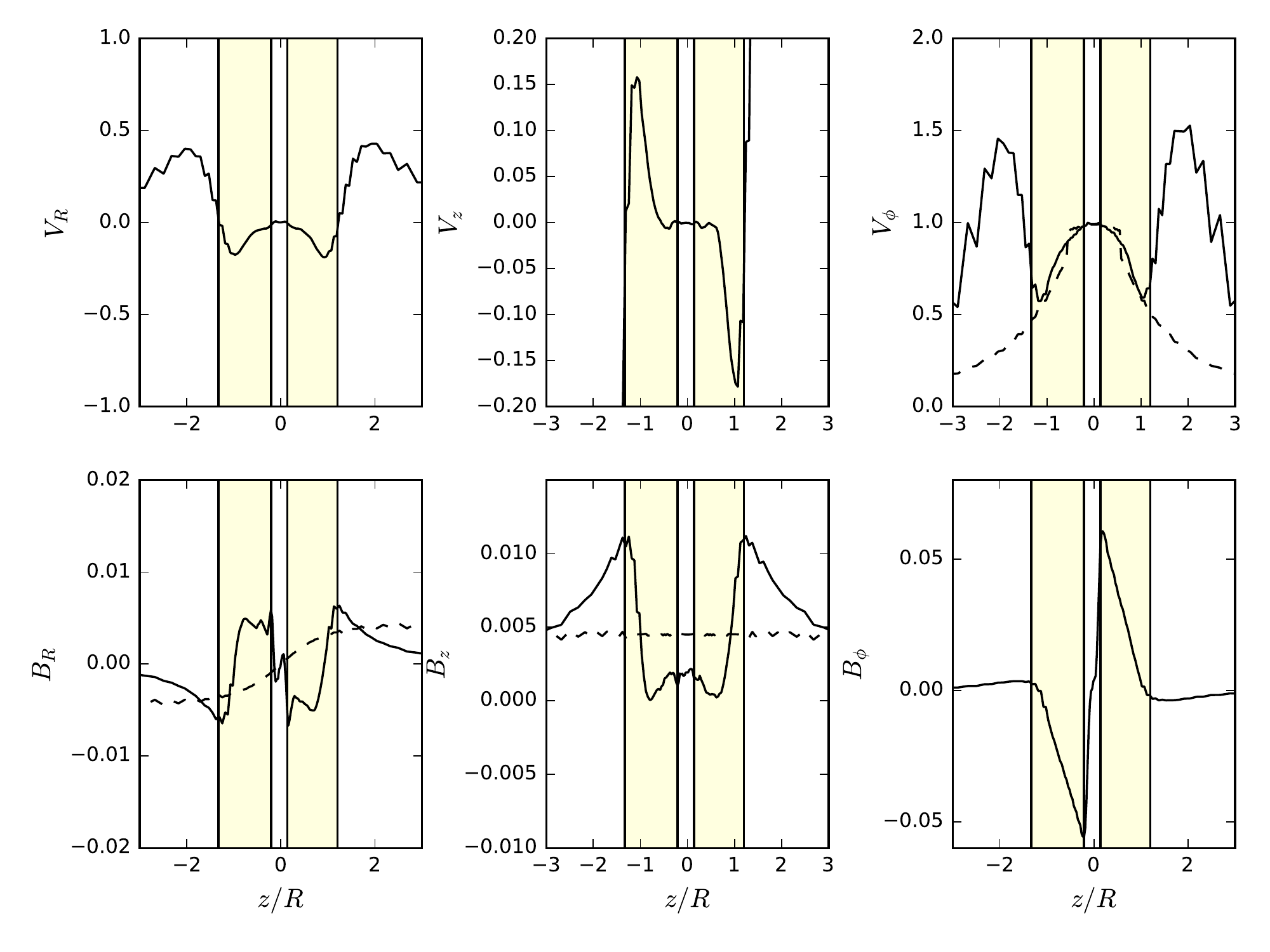} 
\vspace{-0.3 cm}
\caption{ Various velocity and magnetic components along the $z$ direction at R=1. The
quantities have been averaged both azimuthally and over time (t=35 to 42 $T_{0}$ with a $\Delta t$=0.1$T_{0}$ interval).
The dashed curves are from the initial condition. } \label{fig:coronal}
\end{figure*}

The fast coronal accretion at the disk surface carries the magnetic fields inwards, thus pinching magnetic fields 
at the disk surface. The magnetic field lines are shown in Figure \ref{fig:twodrhoalfven}.  The field lines change from
the vertical direction at the disk midplane to the horizontal direction at the disk surface, and then they reverse their directions 
higher above. The net fields along the height at $R=1$ are shown in Figure \ref{fig:coronal}. 
The yellow shaded region is the corona region which 
we define as the region that is magnetically dominated ($\beta<1$)
and has a negative radial velocity ($v_{r}<0$). The corona region separates the disk into three regions, with
the wind region above and the disk midplane region below.
Figure \ref{fig:coronal} shows that, away from the midplane, $B_{R}$ 
first becomes negative and then positive due to the radial drag from the surface accretion. Most importantly, the resulting
horizontal field lines are stretched azimuthally due to the Keplerian shear, similar to the growth of azimuthal fields in the linear
phase of MRI. At $z>0$, the negative $B_{R}$ close to the midplane is stretched to 
the positive $B_{\phi}$ and the positive $B_{R}$ at the upper corona region
is stretched to the negative $B_{\phi}$ (also shown in the color map of Figure \ref{fig:twodrhoalfven}). 
Net $B_{z}$ is always positive but it gets amplified at the upper boundary of the corona region
 since the field lines are pinched there. 
 
\begin{figure*}[ht!]
\centering
\includegraphics[trim=0cm 0.3cm 0cm 0cm, width=0.9 \textwidth]{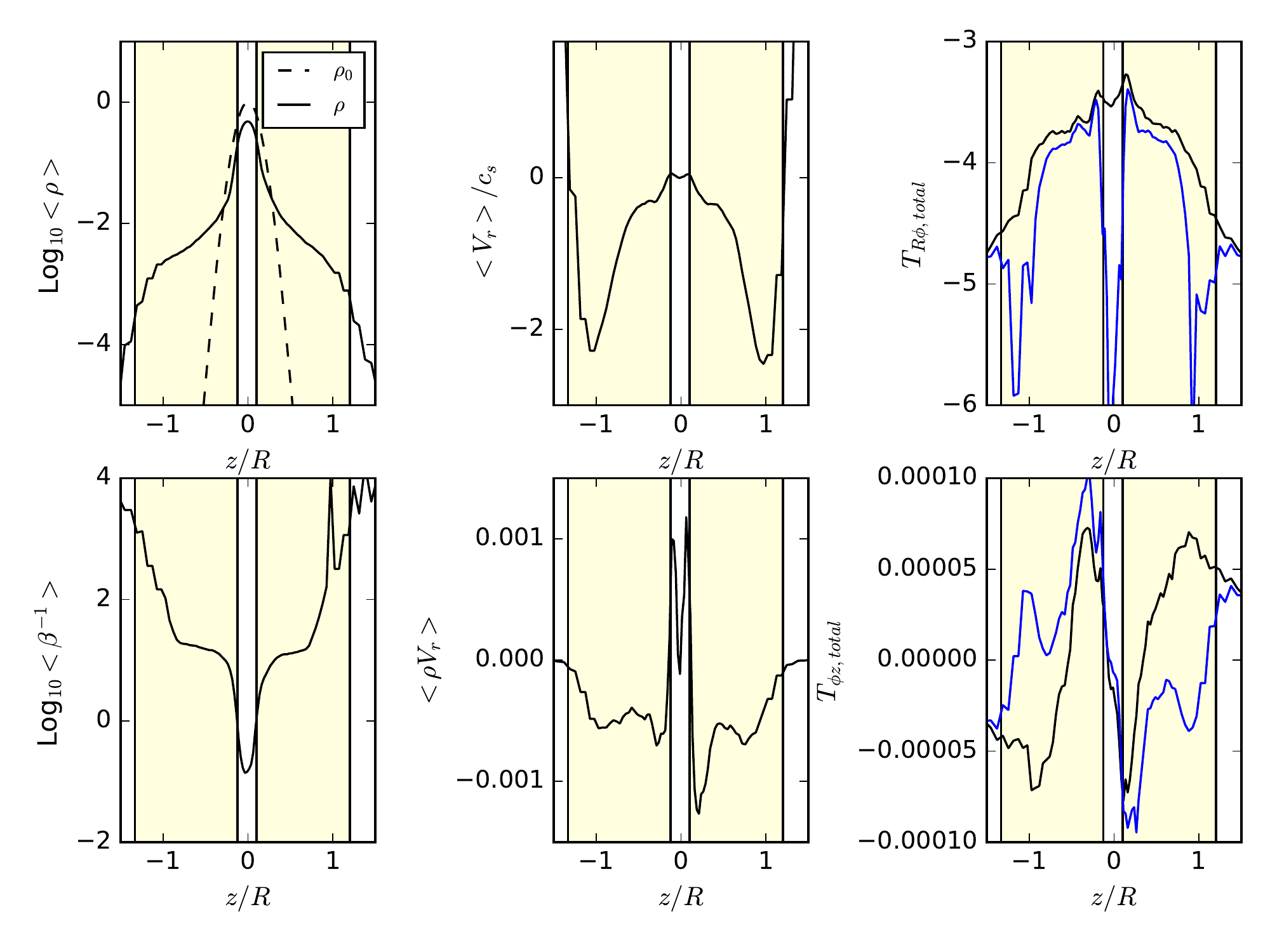} 
\vspace{-0.3 cm}
\caption{Various quantities along the vertical direction at R=1. The quantities are averaged over both the
azimuthal direction and time. The time average is taken from snapshots from t=35 to t=42 $T_{0}$ with $\Delta t=0.1 T_{0}$ interval. The blue curves
in the $T_{R\phi}$ and $T_{\phi z}$ panels are magnetic stresses that are calculated using the mean fields, -$\overline{B_{R}}\times\overline{B_{\phi}}$ and -$ \overline{B_{z}}\times\overline{B_{\phi}}$. The mean fields are both azimuthally and time averaged before being used to calculate the stress. The yellow shaded region labels the corona region. 
Note the fast inward flow at the disk corona. } \label{fig:onedr}
\end{figure*}

 Such net field geometry exerts strong stresses or torques to the disk, 
 crucial for maintaining coronal accretion and launching disk wind. For example,
since $B_{\phi}$ and $B_{R}$ have opposite signs as shown above, 
the $R\phi$ magnetic stress, -$B_{R}B_{\phi}$, is positive from the midplane
to the wind region, and the radial gradient of this stress leads to overall disk accretion. The $z-\phi$ stress -$B_{z}B_{\phi}$ has similar shapes as -$B_{\phi}$
since $B_{z}$ is always positive. At $z>0$, the stress changes from positive values in the lower corona region to negative values at the wind base, 
leading to vertically sheared motion which will be discussed below. 
These magnetic stresses are shown in  
Figure \ref{fig:onedr}.  We note that the net field dominates in the coronal region. 
Both $R\phi$ and $\phi z$ stresses are mostly from the mean field as shown
by the blue curves.

Figures \ref{fig:coronal} and \ref{fig:onedr} show disk quantities along the disk height ($z$) at $R=1$. 
The corona region is again marked as the yellow shaded area with the wind region above and the disk midplane below. We can see that the corona region has a 
supersonic inflow velocity, reaching $\sim$ 2 $c_{s}$ and transports
a significant amount of mass inwards. The density reaches a plateau in this magnetically supported corona region.
Figure \ref{fig:coronal} also suggests that the disk is sub-Keplerian in the corona region. Considering the corona region
is magnetically dominated, the sub-Keplerian motion could be due to that magnetic fields in the corona connect to the midplane at larger
radii. In other words, the midplane magnetically breaks the corona, as described in detail below. 
The $<\rho V_{r}>$ panel in figure \ref{fig:onedr} also reveals that, on average, the midplane region
transports mass outwards from t=35 to 42 $T_{0}$.
The wind region which is beyond the corona region carries little mass outwards since the density is low there (more
discussion on mass loss due to the wind is in \S 4.2).

\begin{figure*}[ht!]
\centering
\includegraphics[width=1.0 \textwidth]{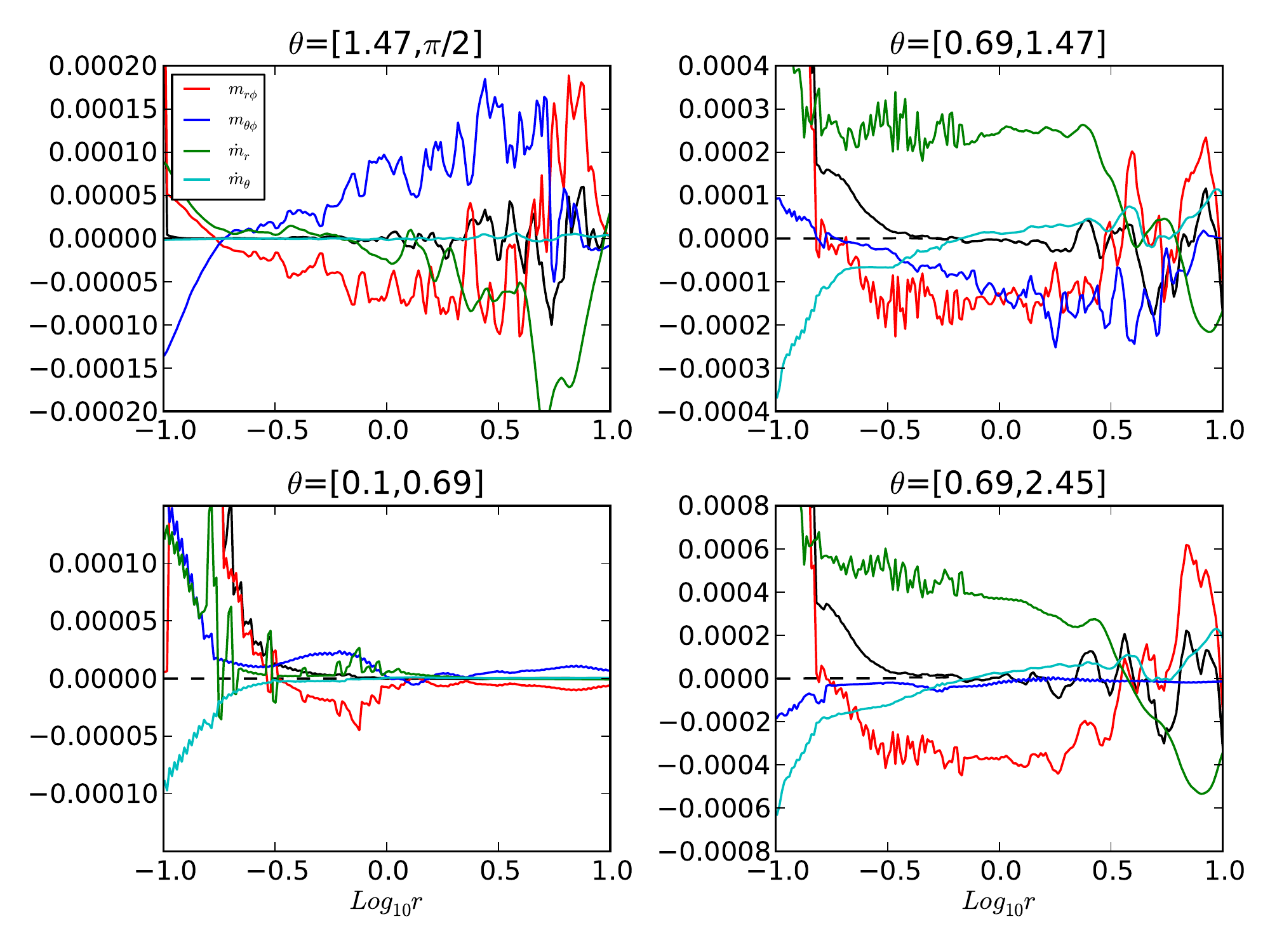} 
\vspace{-0.4 cm}
\caption{Angular momentum budgets within different $\theta$ wedges. Various components of the budgets
have been averaged over time (from t=35 to 42 $T_{0}$ over every timestep) and integrated over space 
($2\pi$ in $\phi$, and $\theta_{min}$ to $\theta_{max}$ with the weight of the $(sin\theta)^2$ geometry factor). 
The averaged quantities have also been multiplied by
($1/2\pi \times r^{3.5}$) so that these quantities are almost flat in radii and have similar scales.
} \label{fig:budget}
\end{figure*}

To understand how magnetic stresses can lead to the coronal accretion, we analyze 
the angular momentum budget due to these stresses in different regions.
Figure \ref{fig:twodrhovel} suggests that the boundary between the corona and wind region is around z$\sim$1.5 R,
and the boundary between the midplane and corona region is around z$\sim$0.1 R.
Thus, the midplane, corona, and wind regions are better represented by
$\theta$ wedges in the spherical-polar coordinate instead of $z$ slabs in cylindrical coordinates.
Thus, we rewrite and analyze the angular momentum equation
under the spherical-polar coordinate. After inserting the mass conservation equation and 
dividing the equation by $r$, we get
\begin{align}
\frac{\partial \langle\rho\delta v_{\phi}\rangle}{\partial t}&=-\frac{1}{r^3}\frac{\partial ( r^3  \langle T_{r\phi}\rangle)}{\partial r}-\frac{\langle\rho v_{r}\rangle}{r}\frac{\partial rv_{k}}{\partial r}\nonumber\\
&-\frac{1}{r {\rm sin}\theta^2}\frac{\partial ({\rm sin}^2\theta\langle T_{\theta\phi}\rangle)}{\partial \theta}-
\frac{\langle\rho v_{\theta}\rangle}{r {\rm sin}\theta}\frac{\partial ({\rm sin}\theta v_{k})}{\partial \theta}\label{eq:angsph}
\end{align}
where 
\begin{align}
T_{r\phi}\equiv \rho v_{r}\delta v_{\phi}-B_{r}B_{\phi}\nonumber\\
T_{\theta\phi}\equiv \rho v_{\theta}\delta v_{\phi}-B_{\theta}B_{\phi}\nonumber
\end{align}
We then write Equation \ref{eq:angsph} as
\begin{equation}
\frac{\partial \langle\rho\delta v_{\phi}\rangle}{\partial t}=m_{r\phi}+\dot{m_{r}}+m_{\theta\phi}+\dot{m_{\theta}}\,.\label{eq:budget}
\end{equation}
The left term is the change of the azimuthal momentum. 
 The first term on the right ($m_{r\phi}$) is the radial gradient of the $r-\phi$ stress. At the disk midplane, this term is associated with disk turbulence.
 However, within the wind region, this term accelerates the wind since  the magnetic fields are aligned with
 the radial direction. 
 The second term on the right ($\dot{m_{r}}$) is the angular momentum carried by the radial accretion flow. It is the radial accretion term. 
 When it is positive, the disk accretes inwards. The third term ($m_{\theta\phi}$) is the $\theta$ gradient of the $\theta-\phi$ stress. It can also be considered as 
 the torque between different layers in the disk.
 The forth term ($\dot{m_{\theta}}$) is the angular momentum loss due to the flow in 
the $\theta$ direction, such as disk wind. 

\begin{figure*}[ht!]
\centering
\includegraphics[trim=0cm 7cm 0cm 0cm, width=1.0 \textwidth]{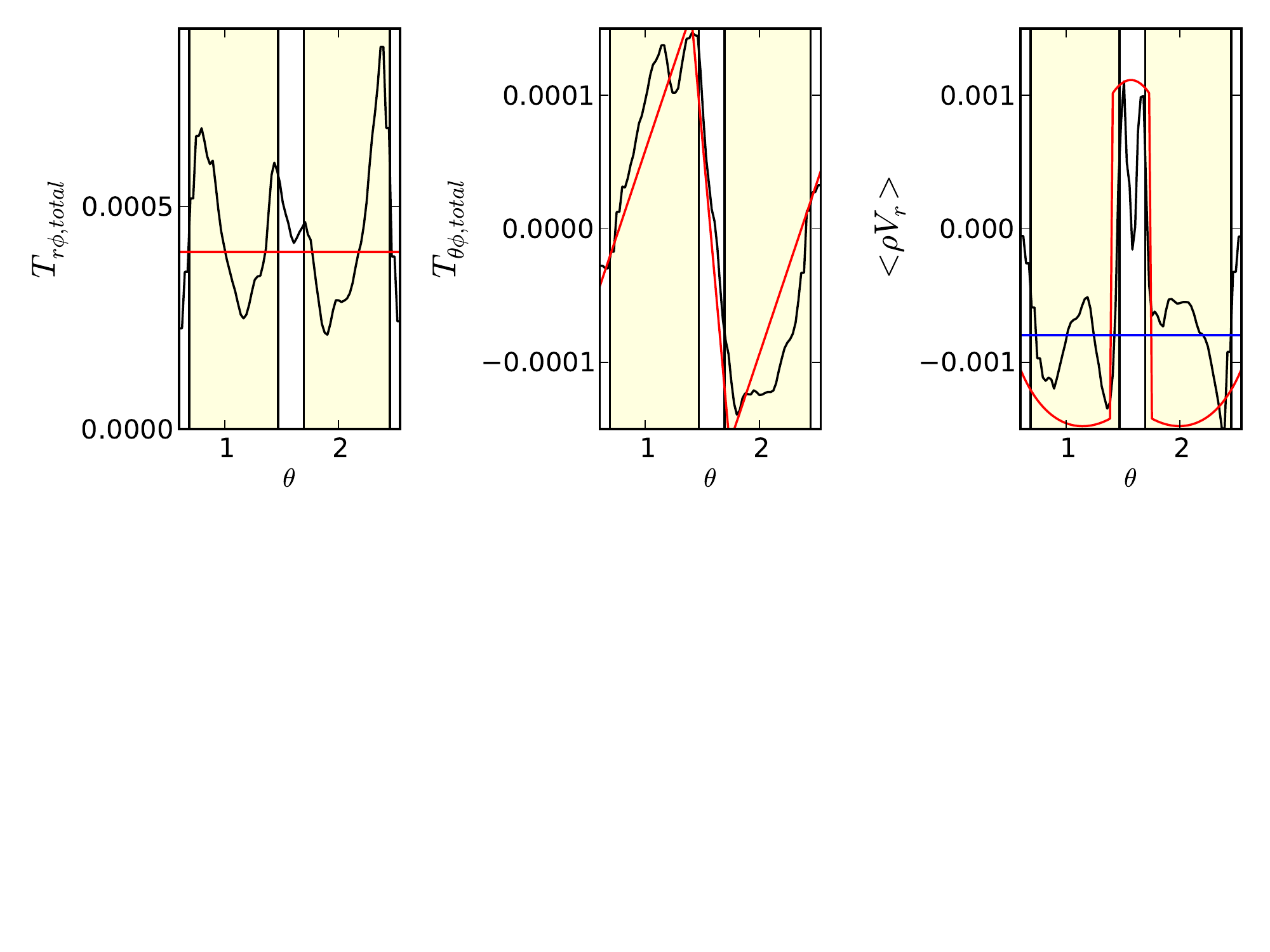} 
\vspace{-0.4 cm}
\caption{The stresses and radial mass flux along the $\theta$ direction at $r=1$. All quantities are averaged over every time step from t=35 to 42 $T_{0}$. 
The red curves in the left two panels are simple analytical fits to the stresses. Using these simple fits, the derived
mass flux is shown as the red curve in the right panel. The blue curve in the right panel is the mass flux calculated only
using the fit of the $T_{r\phi,total}$ stress.} \label{fig:onedrsphifov}
\end{figure*}

These different terms that are integrated over $\theta$ wedges are
 shown in Figure \ref{fig:budget}. The wedges are chosen in a way that they represent different disk regions at $R=1$.
Due to the symmetry of the problem,  we will only focus on regions at $z>0$ or $\theta<\pi/2$.
We can see that $m_{r\phi}$ ($\sim -1/r^3 \partial (r^3 T_{r\phi})/\partial r$) is negative in all these regions, thus trying to drive the disk inwards.
The radial profile of $T_{r\phi}$ is shown in 
Figure \ref{fig:dB2}, and it  follows $r^{-1.85}$ at the disk midplane and it is almost a constant
with height (Figure \ref{fig:onedrsphifov}). With the definition of $m_{r\phi}$ in Equation \ref{eq:budget}, $m_{r\phi}$ is thus $-1.15 T_{r\phi}/r$ 
which is negative.

At the disk midplane (upper left panel), the $\dot{m_{r}}$ term is negative at R=1, suggesting that 
the flow is outwards. With our setup, we can see that the $\dot{m_{r}}$ term balances the addition of two larger terms: $m_{r\phi}$
and $m_{\theta\phi}$. 
Although the $m_{r\phi}$ term is negative as shown above,
 the $\theta-\phi$ stress term ($m_{\theta\phi}\sim-\partial T_{\theta\phi}/\partial \theta$), which is mainly from net magnetic fields, is positive, trying to make
the midplane flow outwards.  The positive value of $m_{\theta\phi}$ can be understood from
 Figure \ref{fig:onedrsphifov}, where $T_{\theta\phi}$ becomes large moving away from the
midplane due to the azimuthal field stretch.  
We could also regard the $m_{\theta\phi}$ term as the magnetic breaking to the surface by the midplane. 
It tries to torque the surface to flow inwards and adds the momentum to the midplane driving midplane outwards.
Eventually, at the disk midplane, the $\theta-\phi$ stress term wins over the $r-\phi$ stress term and the disk flows outwards at $R\sim1$.
We note that 
the balance of two large terms not necessarily leads to the inward accretion. At some other radii shown in Figure \ref{fig:budget} or
possibly in disks with some other parameters, the midplane can flow inwards. 

In the corona region, both terms due to $r-\phi$ and $\theta-\phi$ stresses are negative, driving the disk to accrete inwards. 
We note that the  $m_{\theta\phi}$ term  that is vertically integrated with a weight of sin$\theta^2$ 
is basically the difference between the  $\theta-\phi$ stresses
at the upper and bottom surfaces of the corona region. Since $T_{\theta\phi}$ is negative and positive at the upper and bottom surfaces of
the corona region, the integrated $m_{\theta\phi}$ term is negative. Thus, the $\theta-\phi$ stress torques  the disk surface to flow inwards
and the disk midplane
to flow outwards, again like magnetic breaking. At R=1, the coronal accretion rate is $\sim$10 times larger than
the midplane outflow rate. 

Although it is interesting to observe the different flow patterns at different vertical heights, 
we are more interested in the disk's overall accretion rate as a whole.
For the whole disk, including both the midplane and the corona, the large $\theta\phi$ stress within the disk
cannot lead to the overall disk accretion. 
Only the $\theta\phi$ stress exerted at the upper and lower surface of the disk can torque the disk, leading to accretion.
The lower right panel in Figure \ref{fig:budget} 
shows that the total $m_{\theta\phi}$ term is negative.
Thus, the wind torque indeed contributes to the disk accretion.
However,  the  wind torque (the $\theta-\phi$ stress term) is smaller than the $r-\phi$ stress term by a factor of 20 by examining the lower right panel in more detail. 
Thus, 95\% of the disk accretion is due to the $r-\phi$ stress. Such stress is from MRI turbulence at the midplane and the global magnetic fields
in the corona (the T$_{R,\phi}$ panel in Figure \ref{fig:onedr}). 

To understand how the coronal accretion is determined by different components of stresses, we fit simple curves for both
$T_{r\phi}$ and $T_{\theta\phi}$ stresses and use Equation \ref{eq:angsph} to calculate the accretion rate
at different heights. We fit $T_{r\phi}$ with the constant of 0.00032 and $T_{\theta\phi}$ with
\begin{eqnarray}
T_{\theta\phi}=\left\{ \begin{array}{l} 
0.00025\times(\theta-0.77)\quad {\rm if}\quad \theta<1.4\nonumber\\
-0.00096\times(\theta-\pi/2)\quad {\rm if}\quad 1.4<\theta<1.74\nonumber\\
0.00025\times(\theta-2.37)\quad {\rm if}\quad 1.74<\theta\\
 \end{array}\right.
\end{eqnarray}
shown as the red curve in the middle panel of Figure \ref{fig:onedrsphifov}.
Then with the assumptions that the disk is in a steady state, $\dot{m_{\theta}}$ term is negligible,
and $T_{r\phi}\propto r^{-1.85}$,
we can derive the accretion rate as
\begin{eqnarray}
\rho v_{r}=\left\{ \begin{array}{l} 
-0.00073-0.0005-\\
\quad0.001\times{\rm ctg}\theta\times(\theta-0.77)\quad {\rm if}\quad \theta<1.4\nonumber\\
-0.00073+0.0019+\\
\quad0.0038\times{\rm ctg}\theta\times(\theta-\pi/2)\quad {\rm if}\quad 1.4<\theta<1.74\nonumber\\
-0.00073-0.0005-\\
\quad0.001\times{\rm ctg}\theta\times(\theta-2.37)\quad {\rm if}\quad 1.74<\theta\\
 \end{array}\right.
\end{eqnarray}
shown as the red curve in the right panel of Figure \ref{fig:onedrsphifov}. 
We can clearly see that the $T_{\theta\phi}$ stress torques the disk midplane to flow outwards
and the disk surface to flow inwards.
The integrated
mass flux (with the weight of sin$\theta^2$) at $r=1$ is thus $\sim-0.00073(\theta_{max}-\theta_{min})$-2$(T_{\theta\phi}(\theta_{max})-T_{\theta\phi}(\theta_{min}))$.
Since the disk region extends from $\theta_{min}$=0.69 to $\theta_{max}$=2.45, we can derive
the integrated flux to be -0.00135. 

If we only use the $T_{r\phi}$ stress, we can derive $\rho v_{r}$ as -0.00073 and the vertically integrated disk mass flux 
as $\sim$ -0.00073$(\theta_{max}-\theta_{min})$. With $\theta_{max}$ and $\theta_{min}$ plugged in, the mass flux
due to $T_{r\phi}$ stress is -0.00128. From this simple model, 95$\%$ of disk accretion is due to the $T_{r\phi}$ stress and $5\%$
from $T_{\theta\phi}$ stress exerted at the disk surface, which is consistent with the value by examining the components in the lower right panel of Figure \ref{fig:budget}.

\begin{figure*}[ht!]
\centering
\includegraphics[width=1.0 \textwidth]{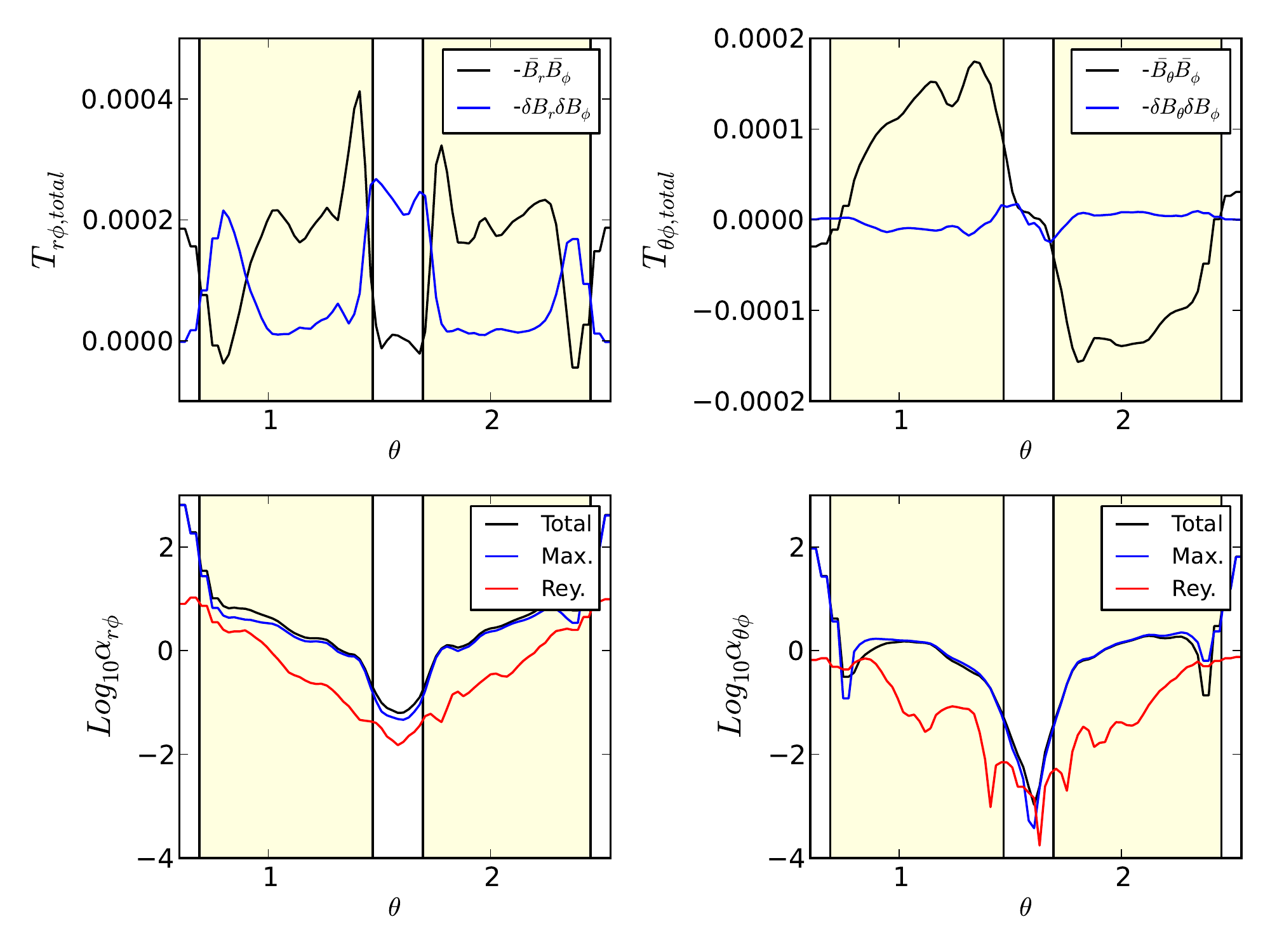} 
\vspace{-0.3 cm}
\caption{$r\phi$ and $\theta\phi$ components of the stress and $\alpha$ along the $\theta$ direction at r=1. Quantities
are averaged over every time step from t=35 to 42 $T_{0}$.
} \label{fig:onebr}
\end{figure*}

\begin{figure}[ht!]
\centering
\includegraphics[trim=0cm 6.cm 0cm 0cm, width=0.5\textwidth]{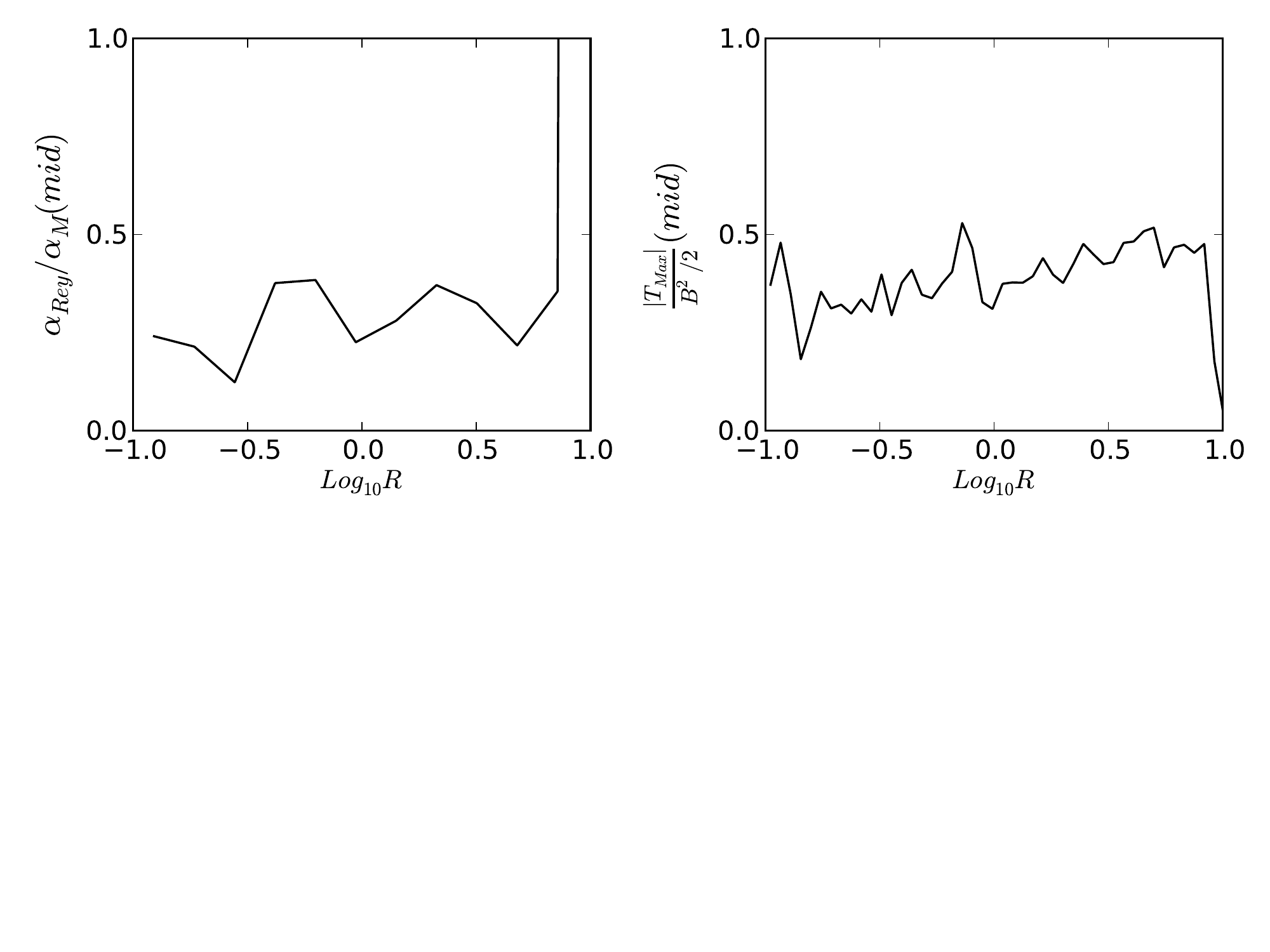} 
\vspace{-0.8 cm}
\caption{The ratio between the Reynolds and Maxwell stress (the left panel) and the ratio between the
Maxwell stress and magnetic pressure at the disk midplane at t=42 $T_{0}$. The stresses are also averaged along 
the radial direction over 4 disk scale heights in the left panel and over 1 disk scale height in the right panel.} \label{fig:onedradial}
\end{figure}

Since both turbulence and net magnetic fields can contribute to stresses, we would like to know their relative importance.
We separate the magnetic stress $B_{i}B_{j}$ into ($\langle B_{i}\rangle_{\phi}+\delta B_{i}$)($\langle B_{j}\rangle_{\phi}+\delta B_{j}$)
where $\langle B\rangle_{\phi}$ is the net field that has been averaged over the $\phi$ direction. Thus
the azimuthally averaged stress $\langle B_{i}B_{j}\rangle_{\phi}$ can be divided into the stress due to net fields 
$\langle B_{i}\rangle_{\phi}\langle B_{j}\rangle_{\phi}$ and the stress due to turbulence
 $\langle \delta B_{i} \delta B_{j}\rangle_{\phi}$. The stresses are shown in the upper panels of Figure \ref{fig:onebr}.
For the $r-\phi$ stress, we can see that the turbulent stress dominates at the midplane
and the transition between the corona and wind regions, while the net field stress dominates
in the corona and wind regions.
Previous work has 
 established that, in MRI turbulent models, $\alpha_{Rey}$ is $\sim$1/4 of $\alpha_{Max}$,
and the ratio between the Maxwell stress and the magnetic pressure $\alpha_{mag}=T_{Max}/P_{b}\sim 0.45$ \citep{Hawley1995}. 
In our simulation, the MRI turbulent midplane also satisfies these relationships (Figure \ref{fig:onedradial}).
Since the $r-\phi$ stress determines the radial inflow, the importance of both turbulence and net fields at different layers
leads us to conclude that both processes contribute to the radial accretion. 

On the other hand,  the $\theta-\phi$ stress is dominated by net fields in both
the wind and corona region and it is as important as the turbulent stress at the disk midplane. 
Since the $\theta-\phi$ stress determines the internal flow structure within the disk, we conclude that the net fields determine
the vertically-sheared motion.

\begin{figure}[ht!]
\centering
\includegraphics[trim=0cm 3cm 0cm 4cm, width=0.5 \textwidth]{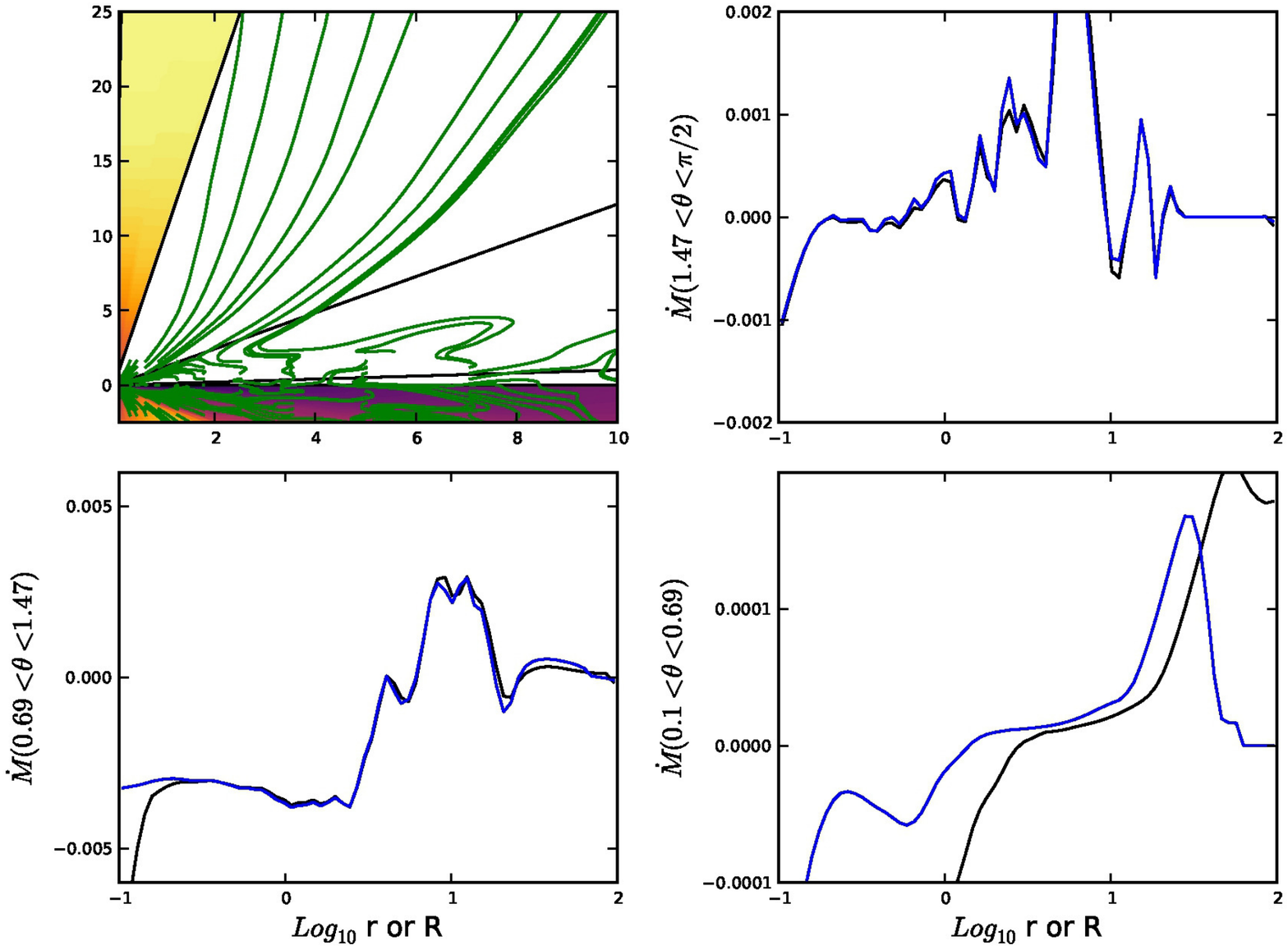} 
\vspace{-1.8 cm}
\caption{The radial mass flux within three different wedges ($\theta\in$[0.1,0.69], [0.69,1.47], [1.47, $\pi/2$]) that are labeled as the white regions in the upper  left panel.
The green curves in the upper left panel are velocity streamlines.
The mass flux is averaged over every time step from t=40 to 42 $T_{0}$.
The black curves are the mass flux in the $r$ direction with respect to $r$, while the blue curves are the mass flux in the $R$ direction with respect to $R$. } \label{fig:onedradialmdot}
\end{figure}

Finally, we would like to know the total mass flux in different regions.
Thus, we cut three wedges in the simulation domain and measure the radial mass fluxes, shown in Figure \ref{fig:onedradialmdot}.  
We can see that  the coronal accretion dominates. 
At $R=1$, the midplane outflow rate  is $10\%$ of the coronal accretion rate. 
However, we also note that, at some other radii, the disk flows inwards at the midplane. 
In any case, the net flow rate at the midplane is always much smaller 
than the inflow rate in the corona region.
The wind region has a very small mass flow rate. At larger distances, our cut
for the wind region includes part of the disk region so that the mass flow rate does not reflect the true wind loss rate. 
A more detailed analysis on the outflow properties is presented in the next subsection.

\subsection{Wind Region}
Although the torque by the wind at the disk surface only accounts for 5\% of the disk accretion rate, the disk wind is launched
at all the radii in the disk and it can be directly probed by observations \citep{Bjerkeli2016}. Thus, we will study its properties in more detail.

To properly study the disk outflow/wind, the simulation domain needs to be larger than the Alfven surface and the
fast magnetosonic surface \citep{BlandfordPayne1982}. This is satisfied in our simulation as shown in Figure \ref{fig:twodrhovel} and  \ref{fig:twodrhoalfven} 
where the dashed 
curves are the Alfven surface while the dotted curves are 
the fast magnetosonic surface.
Alfven surface is where the velocity in the poloidal plane $V_{p}\equiv\sqrt{\langle v_{r}\rangle^2+\langle v_{\theta}\rangle^2}$
equals the poloidal Alfven velocity $V_{Ap}\equiv\sqrt{(\langle B_{r}\rangle^2+\langle B_{\theta}\rangle^2)/\langle \rho\rangle}$. Fast magnetosonic surface is where $V_{p}$ equals the fast magnetosonic
speed $V_{F}\equiv \sqrt{1/2 ({{V_{A}}^2+c_{s}}^2)+1/2 |V_{A}^2-c_{s}^2|}$ where $V_{A}\equiv \sqrt{(\langle B_{r}\rangle^2+\langle B_{\theta}\rangle^2+\langle B_{\phi}\rangle^2)/\langle \rho\rangle}$. In the disk's atmosphere $V_{A}$ is much larger than $c_{s}$, so that $V_{F}\sim V_{A}$.

For a steady axisymmetric outflow, there are four conserved quantities along the magnetic and velocity field lines 
\citep{WeberDavis1967, BlandfordPayne1982}.
In a steady flow, the induction equation becomes $\nabla\times(\bf{v}\times\bf{B})$=0. If $\bf{v}$ and $\bf{B}$ are separated into
the poloidal  and the toroidal components ($\mathbf{v}=\mathbf{v_{p}}+\Omega R \hat{\phi}$ and $\bf{B}=\bf{B_{p}}+\bf{B_{\phi}}$), 
it has been shown that $\bf{v_{p}}$ and $\bf{B_{p}}$ are in the same direction, and using mass conservation equation we have the first
constant
\begin{equation}
k=\frac{\rho \bf{v_{p}}}{\bf{B_{p}}}\,,
\end{equation}
which is the mass load parameter,  affecting the dynamical properties of the wind \citep{OuyedPudritz1999}.
Using the same equations, the second constant can also be derived
\begin{equation}
\omega=\Omega-\frac{k B_{\phi}}{\rho R}\,.
\end{equation}
With the angular momentum equation, we have the third constant 
\begin{equation}
l=R(v_{\phi}-\frac{B_{\phi}}{k})\,
\end{equation}
which is the specific angular momentum of the wind.
In a barotropic fluid, we can use Bernoulli's equation to derive the fourth constant
\begin{equation}
e=\frac{1}{2}v^2+\Phi+h+\frac{B^2}{\rho}-\frac{{\bf B}\cdot{\bf v}}{k}
\end{equation}
where $h$ is $\int dp/\rho$. We can also write $e$ in some other ways using the first three constants above
\begin{align}
e&=\frac{1}{2}v^2+\Phi+h+\frac{B_{\phi}B_{\phi}}{\rho}-\frac{B_{\phi}v_{\phi}}{k}\\
&=\frac{1}{2}v^2+\Phi+h-\frac{RB_{\phi}\omega}{k}\,,
\end{align}
Although our disk is locally isothermal which is not barotropic, 
we will see that $h$ is much smaller than other terms (which means the wind is ``cold'' )
and we can still treat $e$ as a constant.

Using the azimuthally averaged velocity structure, we derive velocity streamlines and the four conserved quantities
along the streamlines. 
We find that the conserved quantities are not constants. For example, the values of $k$ and $\omega$  
have a factor of 2 peak at $r\sim 20$ at t=42 $T_{0}$. 
When we check these quantities over time, we notice that the peak in $k$ originates from the disk 
surface and propagates outwards. This is not surprising since the outflow is launched from the
turbulent disk. Previous simulations  have also observed the episodic disk wind \citep{OuyedPudritz1997b}.

\begin{figure*}[ht!]
\centering
\includegraphics[trim=0cm 0cm 0cm 0cm, width=0.8 \textwidth]{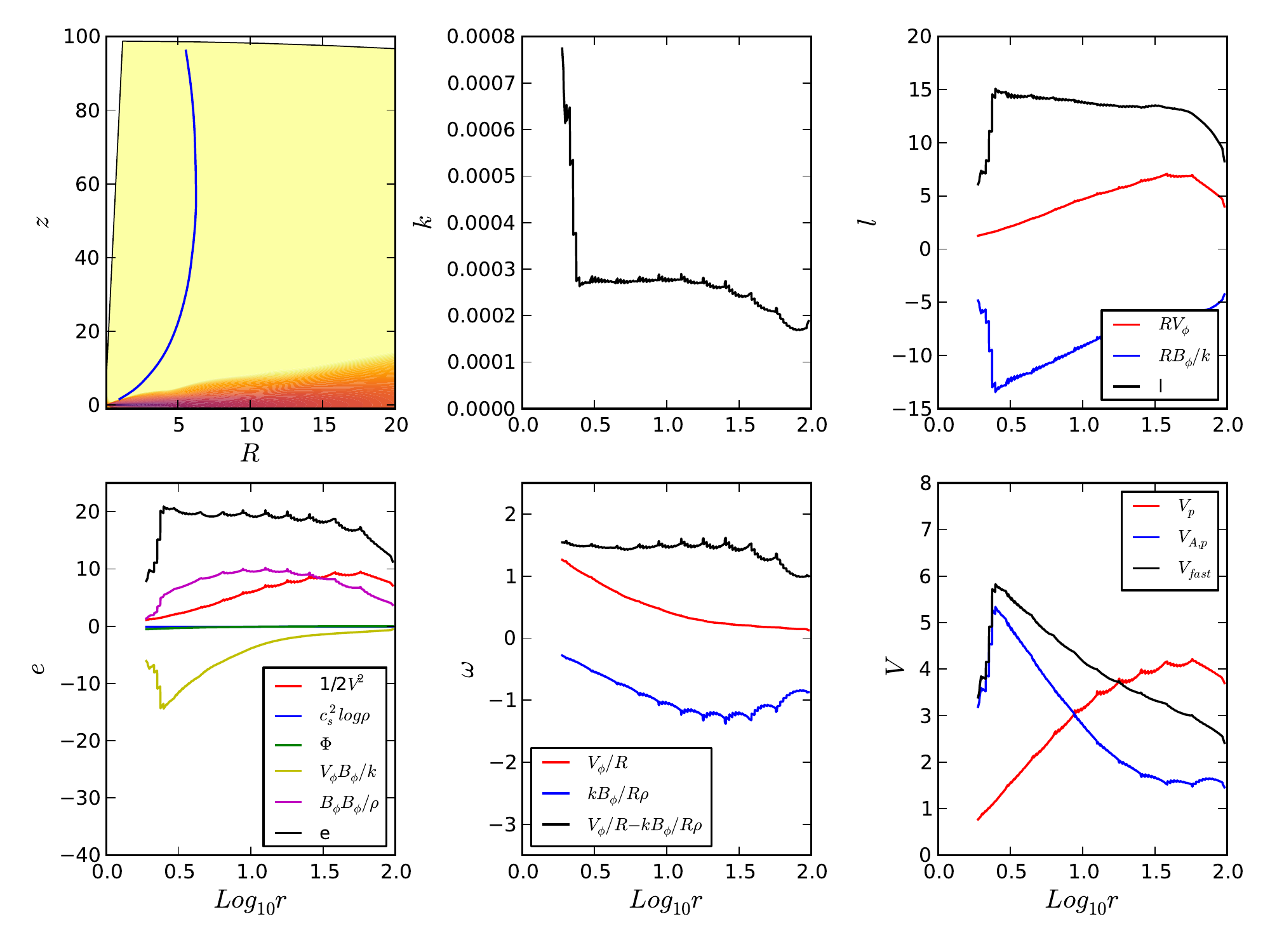} 
\vspace{-0.3 cm}
\caption{ Various conserved quantities ($k$, $l$, $e$, $\omega$) and characteristic speeds (poloidal speed, Alfven speed, and fast magnetosonic speed)
 along a fluid streamline shown as the blue curve in the upper left panel.
All the primitive variables have been averaged over both the azimuthal direction and time (using snapshots from t=40 to 45.6 $T_{0}$ with a $\Delta t=0.1T_{0}$ interval). 
 The streamline starts at $R$=1, $z$=1.6 and is 
derived using the azimuthally and time averaged
velocities. } \label{fig:twodstreamavg}
\end{figure*}

\begin{figure*}[ht!]
\centering
\includegraphics[trim=0cm 0cm 0cm 0cm, width=0.8 \textwidth]{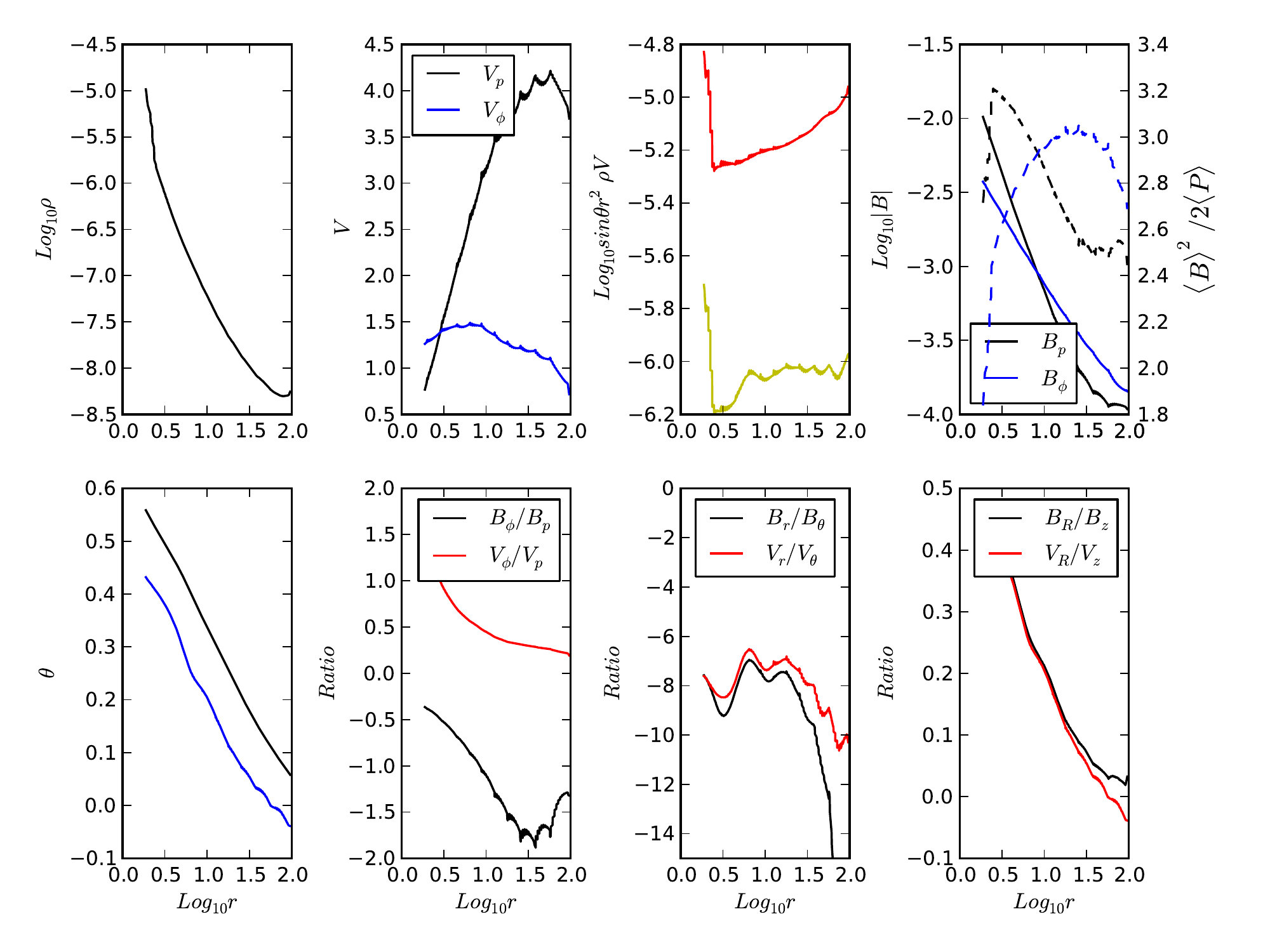} 
\vspace{-0.3 cm}
\caption{Time and azimuthally averaged quantities along the velocity streamline shown in Figure \ref{fig:twodstreamavg}.
In the upper right panel, $1/\beta$ that are calculated using $\langle B_{p}\rangle$ and  $\langle B_{\phi}\rangle$
are shown as the black and blue dashed curves respectively.  The black curve in the lower left panel shows
the $\theta$ coordinate of the fluid streamline and the blue curve shows the angle between the poloidal velocity direction and the vertical direction along the streamline.} \label{fig:twodstreamquanavg}
\end{figure*}

Thus, we average all primitive quantities over time, trying to smooth the unsteady wind and study the statistical properties
of the wind. The averaged disk structure and the conserved quantities are shown in Figure \ref{fig:twodstreamavg}. 
The conserved quantities are derived using the azimuthally and time averaged primitive quantities. 
The $k$ constant is calculated using the $r$ component of $\bf{v_{p}}$ and $\bf{B_{p}}$.
The conserved
quantities are almost constants. This is encouraging, suggesting that we may still be able to use
the traditional steady wind solution to study the statistical properties of the wind generated from the turbulent disk. Accordingly, all the relationships based on 
these four conserved quantities are also satisfied in our simulations. For example, $l=R_{A}^2\omega$ is satisfied. We can
verify this by noting that $l\sim 14$, $R_{A}\sim3$ (derived from $r_{A}=10$ in the lower right panel of
 Figure \ref{fig:twodstreamavg} and $\theta$=0.35 at r=10 in Figure \ref{fig:twodstreamquanavg}, or
 we can directly read the value of $R_{A}$ by following the streamlines in Figure \ref{fig:twodrhovel} ), and $\omega\sim 1.5$ in Figure \ref{fig:twodstreamavg}.
 Thus, the wind lever arm $\lambda\equiv(R_{A}/R_{0})^2$ is $\sim$ 10 for field lines launched at $R_{0}=1$. 
 
 Another useful relationship that should also be satisfied 
 in our simulation is $J=e-\omega l$ \citep{Anderson2003}, which reduces to
 \begin{equation}
 J=\frac{1}{2}v^2+\Phi+h-\omega Rv_{\phi}\,.
 \end{equation}
 At the wind base, $J\sim -3/2 v_{K}^2$, and at the distance far away from the base $v_{p,\infty}>>v_{\phi,\infty}$. Thus
 \begin{equation}
 -3/2 v_{K}^2\sim\frac{1}{2}v_{p,\infty}^2-\omega R_{\infty}v_{\phi,\infty}\,,
 \end{equation}
  Since $v_{p}$ is much larger than $v_{K}$ and both $\Omega$ and $\omega$ are $\sim\Omega_{K}$, this can be
 reduced to \citep{Anderson2003}
 \begin{equation}
 \frac{v_{p,\infty}^2}{2R_{\infty}v_{\phi,\infty}}=\sqrt{\frac{GM_{*}}{R_{0}^3}}\label{eq:rlaunch}
 \end{equation}
 which relates the wind properties at large distances to the launching point. 
 This relationship is particularly useful in observations to estimate the position of the launching point by using quantities
 far away from the disk \citep{Bjerkeli2016}.
We can test this relationship from our simulations.  
All primitive quantities ($\rho, v_{p}, v_{\phi}, B_{\phi}$) along the field are shown in
Figure \ref{fig:twodstreamquanavg}.
If we take $r=30$ as far enough away from the wind base, we have $v_{p}=4$, $v_{\phi}=1.2$, and $R=6$ (based on
Figure \ref{fig:twodstreamavg}). Thus, $R_{0}$ derived from Equation \ref{eq:rlaunch} is $\sim 1.1$, similar to the real
launching point $R_{0}=1$.

Some other wind properties can also be observed in Figure \ref{fig:twodstreamquanavg}. Clearly, the density and
magnetic fields drop off quickly along the streamline. The poloidal field is stronger than the toroidal field at the disk 
surface. But after the Alfven surface at $r\sim 10$,  the toroidal field is stronger than the poloidal field. At the wind launching point, 
1/$\beta$ 
calculated using the net field reaches 1000 and 240 for the $B_{p}$ component. 
\cite{Bai2016} pointed out that $\beta$ at the wind base is one important parameter, besides the global field geometry
and wind temperature, to determine the wind properties. In our particular case, the wind is highly magnetized when it
is launched. 
Beyond the launching point, both $B_{p}$ and $B_{\phi}$ decreases. But $B_{p}$ decreases faster
than $B_{\phi}$. At the Alfven point $B_{p}$ and $B_{\phi}$ are roughly equal and  the wind is dominated by $B_{\phi}$
beyond the Alfven point.
On the other hand, the poloidal velocity 
increases along the streamline while the azimuthal velocity is almost a constant. 
Eventually the poloidal velocity reaches the terminal velocity 4 ($\sim 2^{1/2}\Omega_{0}R_{A}$, \citealt{Pudritz2007} with
$R_{A}\sim$3 for wind launching from $R_{0}$=1 in our simulation).
Knowing $v$ and $B$, we can estimate the dimensionless $\mu$ parameter 
\begin{equation}
\mu\equiv\frac{\rho v_{p}v_{\phi}}{B_{p}^2}
\end{equation}
to see if the wind is ``light'' or ``heavy'',
where all the quantities are estimated at the wind base. Using our measured values, 
$\mu$ is $\sim 0.05$ suggesting that the wind is very light.
\cite{Anderson2005} has shown that the wind is still strongly collimated with such small $\mu$.

\begin{figure}[ht!]
\centering
\includegraphics[trim=0cm 0.cm 4.5cm 0cm, width=0.5\textwidth]{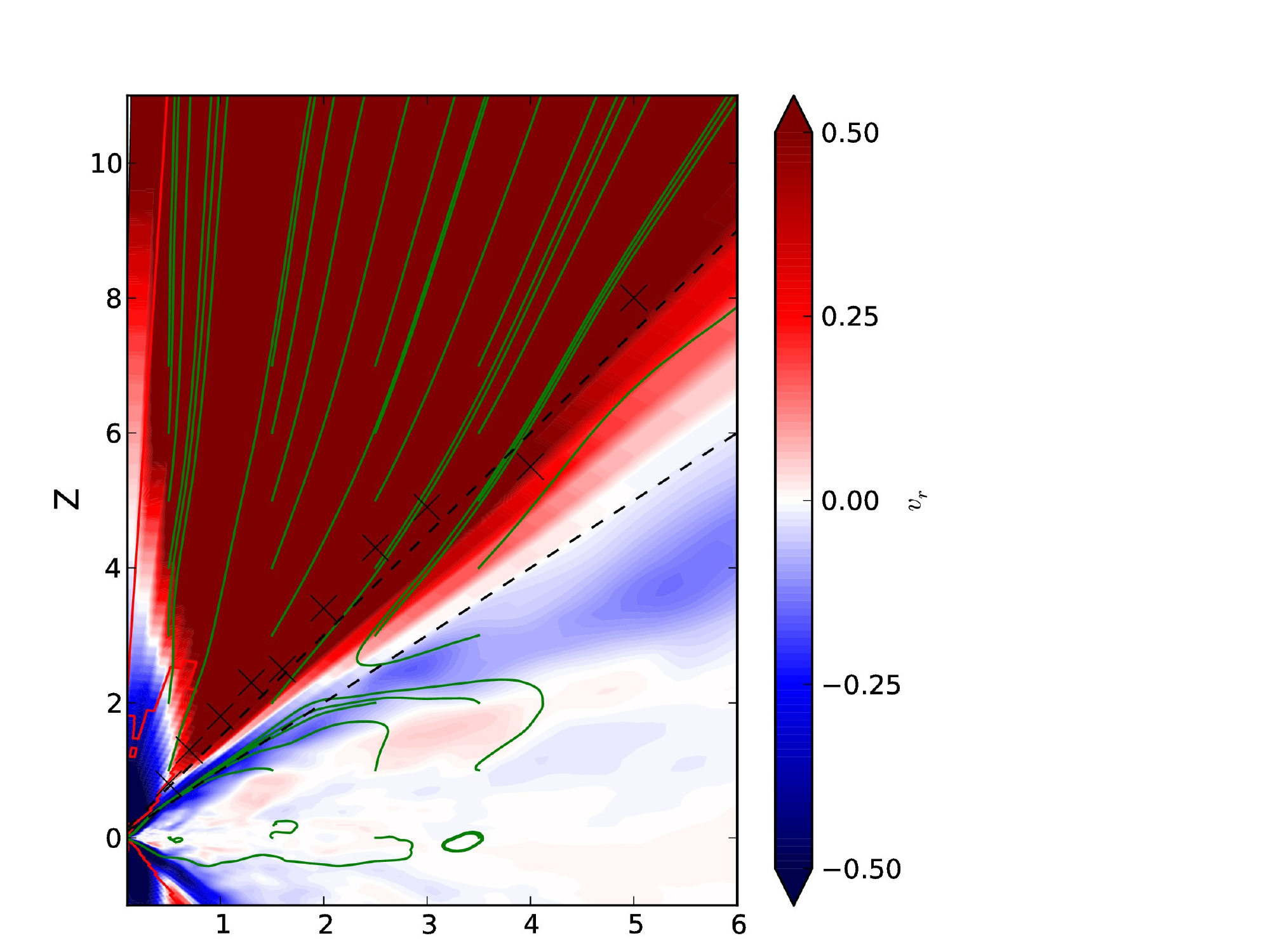} 
\vspace{-0.1 cm}
\caption{The $v_{r}$ averaged over both the azimuthal direction and time 
(using snapshots from t=40 to 45.6 $T_{0}$ with a $\Delta t=0.1T_{0}$ interval). The green curves are the magnetic field lines calculated
with azimuthally and time averaged velocities.  The two dashed lines show $z=R$ and $z=1.5 R$. The crosses
are the launching points at the wind base which are also shown in Figure \ref{fig:wind}.
} \label{fig:twodvz}
\end{figure}

\begin{figure*}[ht!]
\includegraphics[trim=0cm 0cm 0cm 0cm, width=1.0 \textwidth]{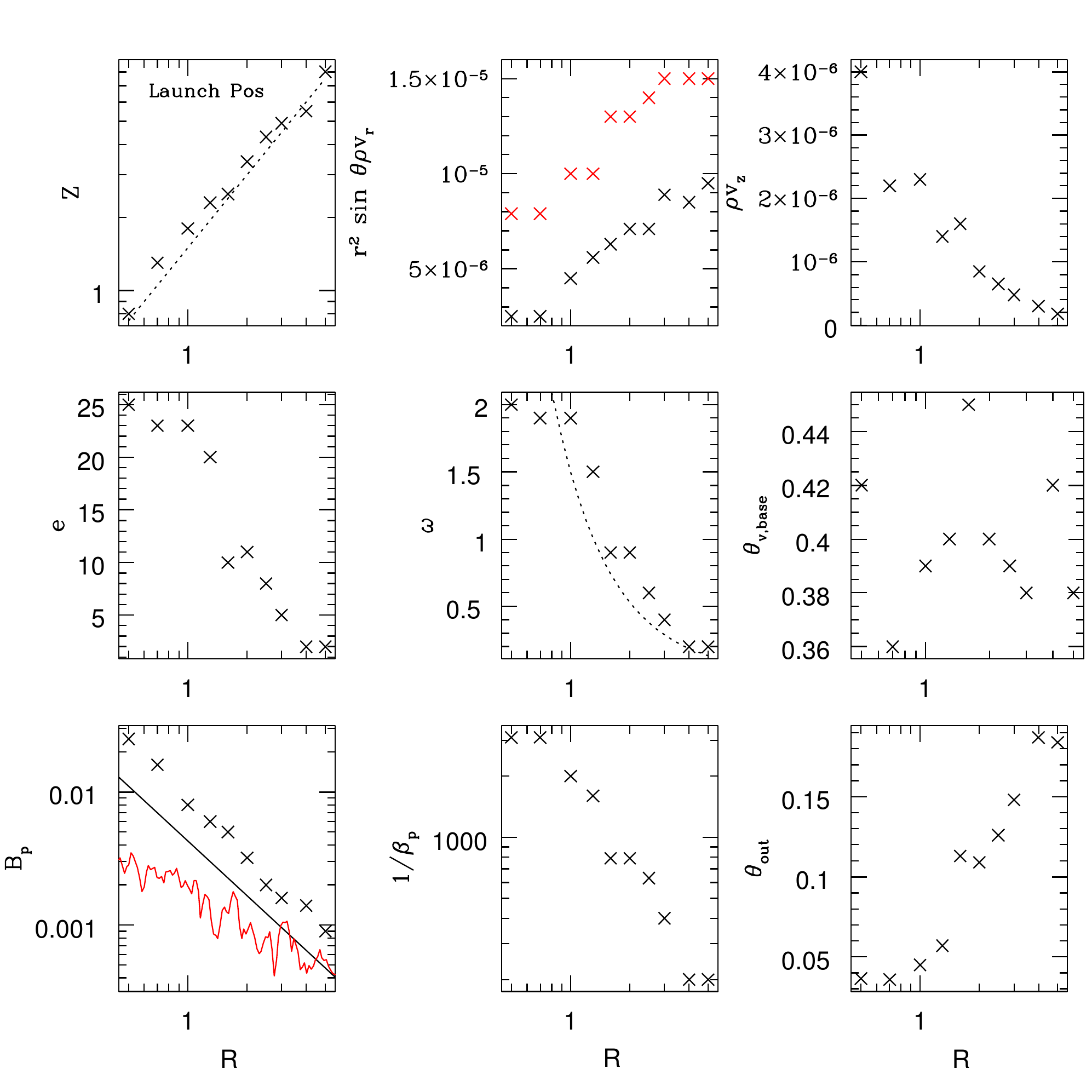} 
\vspace{-0.3 cm}
\caption{The properties of the wind that is launched from different positions in the disk. 
The wind launching points are shown in the upper left panel (the dotted curve is $z=1.5 R$) and also in Figure
\ref{fig:twodvz}. The upper middle panel shows
the mass flux ($\sin\theta r^{2}\rho v_{r}$)
along different wind streamlines with respect to the wind launching points 
(black crosses are measured at the wind base
while the red crosses are measured when the wind is leaving the simulation domain). $\rho v_{z}$ 
at the wind base is shown in the upper right panel. The constants of $e$ and $\omega$ along different
streamlines are shown in the middle left and center panels. In the $\omega$ panel, the dotted curve is 1.5 $\Omega_{k}$.
  The  angle between 
the poloidal velocity at the lunching point and the vertical direction is shown in the middle right panel.
At the wind base, the poloidal magnetic fields $B_{p}$  and the plasma $\beta$ calculated with $B_{p}$
are shown in the lower left and middle panels.  The black line is $B_{p}$ at the disk midplane
in the initial condition, and the red curve is the $B_{p}$ at the disk midplane which is averaged 
from t=40 to 45.6 $T_{0}$ with
$\Delta$t=0.1 $T_{0}$  interval. The $\theta$ position when the wind streamlines
leave the simulation domain are shown in the lower right panel.
} \label{fig:wind}
\end{figure*}

Since the wind that is launched from different disk positions has different properties, 
we apply the same approximation to streamlines that are launched from other disk positions as shown in Figure \ref{fig:wind}. 
We can see that the wind is launched high in the disk around $z\sim 1.5 R$, beyond the corona region.
The launching points are also overplotted as crosses in Figure \ref{fig:twodvz}.  We can see that the material starts to flow
outwards beyond $z\sim R$. But the region between $z\sim R$ and $z\sim 1.5 R$ sometimes flow inwards
as the corona and sometimes flows outwards as the wind. Only the region beyond $z\sim 1.5 R$ always flows outwards
as the wind so that the conserved quantities are constant beyond $z\sim 1.5 R$. 
We find that the two conserved quantities $l$ and $k$ for other streamlines are almost the same as
the streamlines originated from $R=1$, while $\omega$ scales roughly $\sim 1.5 \Omega_{K}$ and $e$ is roughly $\sim 10\omega$
as shown in Figure \ref{fig:wind}. 
$B_{p}$ at the wind base roughly follows $R^{-1.5}$. 
The angle between the velocity vector and the vertical direction is 0.36-0.45 ($20^{o}-25^{o}$). Although
this angle is smaller than 30$^{o}$ required for launching disk wind from the disk midplane, the wind in our simulation
is actually launched from high above the disk atmosphere ($z\sim 1.5 R$). If we use the marginally stable equipotential surface
\begin{equation}
\phi(R,z)=-\frac{GM}{R_{0}}\left[\frac{1}{2}\left(\frac{R}{R_{0}}\right)^2+\frac{R_{0}}{(R^2+z^2)^{1/2}}\right]\,,
\end{equation}
we can derive that the critical angle to launch the wind at $z=1.5R_{0}$ is only 17$^{o}$. Thus, field lines in our simulations are tilted 
enough to launch disk winds from the disk surface.

We can also use quantities along different streamlines in Figure \ref{fig:wind} to estimate the total mass loss rate from the wind.
If the wind is in a steady state, 
the mass loss rate between two streamlines will be a constant at different $r$. Thus 
2$\pi r^2\int {\rm sin}\theta \rho v_{r}{\rm d}\theta $ between two streamlines will be a constant along $r$. 
Since the $\theta$ separation between two streamlines does not change dramatically along r,
we expect that $r^2 {\rm sin}\theta \rho v_{r}$ should be roughly a constant. 
As shown in the upper middle panel of Figure \ref{fig:wind},   
$r^2 {\rm sin}\theta \rho v_{r}$ only changes by a factor of 2 from the launching point (black crosses) to the domain
boundary (red crosses). We also know that wind that is launched at larger disk radii has a larger opening angle, 
as shown in the lower right panel. Thus, we can integrate 
  $r^2 {\rm sin}\theta \rho v_{r}$ over the opening angle at the domain boundary to derive
the total mass loss rate from two sides of the the disk region [$R_{in}$, $R_{out}$],
\begin{equation}
\dot{M}_{loss}=4\pi r^2\int_{\theta_{ri}}^{\theta_{ro}} {\rm sin}\theta \rho v_{r}{\rm d}\theta\,.
\end{equation} 
where $\theta_{ri}$ and $\theta_{ro}$ are the $\theta$ positions of the streamlines that are originated from $R_{in}$
and $R_{out}$. If we take $r^2 {\rm sin}\theta \rho v_{r}\sim 10^{-5}$, $\theta_{ri}=0.05$ at $R_{in}=0.5$, and $\theta_{ro}=0.2$
at $R_{out}=5$,
we can estimate the $\dot{M}_{loss}\sim 2\times10^{-5}$ from two sides of the disk region at [0.5,5]. Considering the total
disk accretion rate is $5\times 10^{-3}$, the wind loss rate from this disk region is only 0.4\% of the disk accretion rate. 

We can also use the mass flux at the disk surface to estimate the two sided mass loss rate using
\begin{equation}
\dot{M}_{loss}=\int 4\pi R \rho v_{z} dR\,.
\end{equation}
The $\rho v_{z}$  panel in Figure \ref{fig:wind} suggests that
$ \rho v_{z}\sim 2\times 10^{-6} r ^{-1.5}$. Thus,
 $\dot{M_{out}}\sim 2\times 10^{-5}$, which is consistent with the estimation above.

We would also like to know how much angular momentum is carried away by the wind 
if we derive this wind torque using the measured wind properties, and if this calculated torque
 is consistent with our direct measurement in \S 4.1.
At the wind base, the $\phi z$ stress term in Equation
\ref{eq:angcyl} can be written as
\begin{equation}
\langle \rho v_{z}v_{\phi}\rangle-\langle B_{z}B_{\phi}\rangle=\frac{\rho v_{z} l}{R}=\frac{1}{4\pi}\frac{\partial \dot{M}_{loss}}{\partial R}\frac{R_{A}^2}{R^2} \omega\label{eq:windstress}
\end{equation}
 using the conserved constants
 \footnote{Note that the Reynolds stress in the wind uses $v_{\phi}$ instead of $\delta v_{\phi}$. But the stress at the disk surface 
won't change much if we calculate
the Reynolds stress 
using $\delta v_{\phi}$ since that the magnetic stress dominates. }.  Figure \ref{fig:wind} suggests that 
$\rho v_{z}$ is around $2\times 10^{-6}$
at the wind base around $R=1$. Figure \ref{fig:twodstreamavg} suggests that $l\sim 10$. Thus
 the total stress at the wind base is around $2\times 10^{-5}$, which is roughly consistent with the direct measurement in Figure \ref{fig:onedr}.
This confirms that the wind torque is very small compared with the radial stress and the wind torque contributes little ($\sim 5\%$) to the disk accretion.

\subsection{How to Maintain Global Magnetic Fields?}
\begin{figure}[ht!]
\centering
\includegraphics[width=0.45 \textwidth]{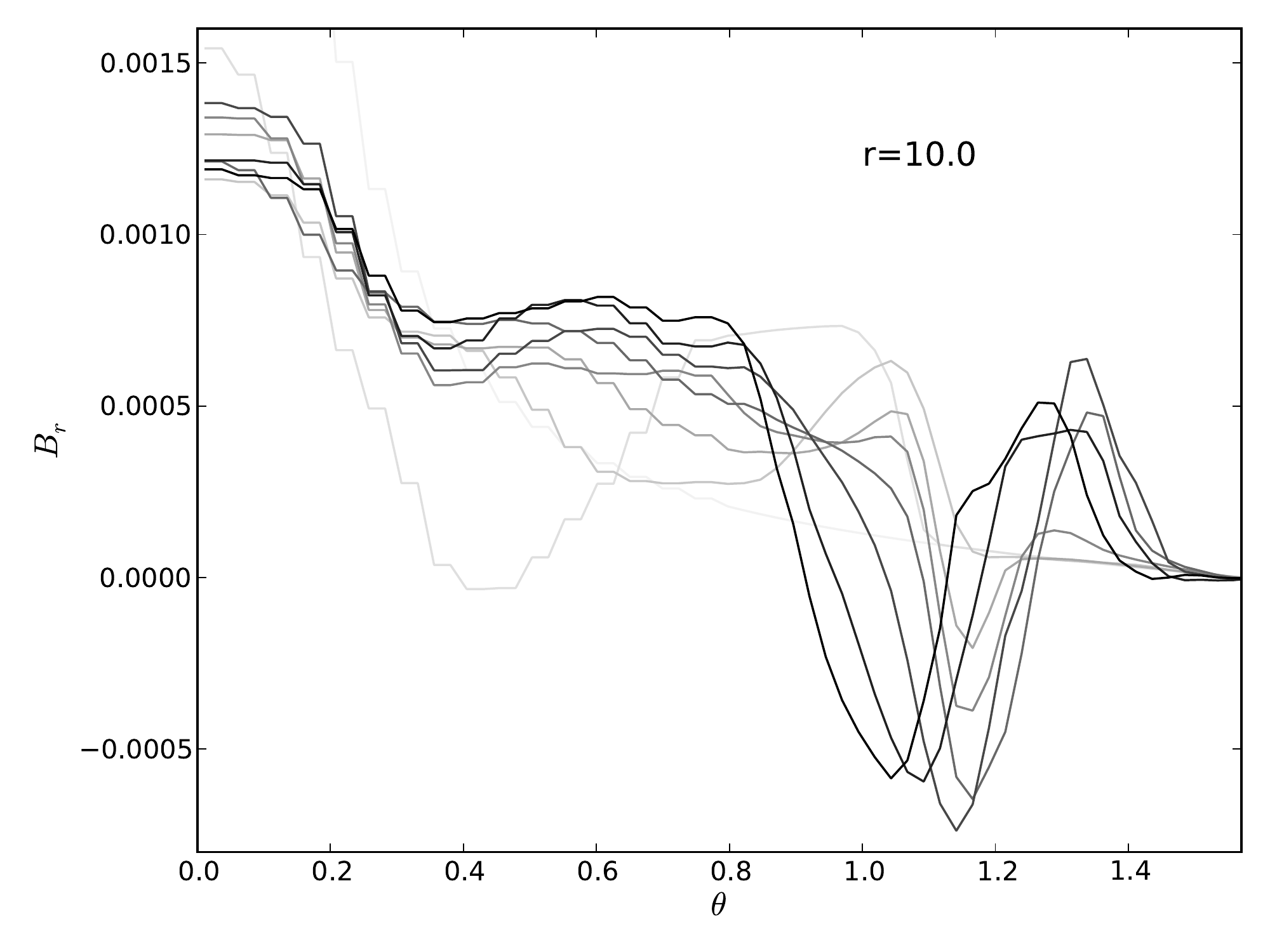} \\
\includegraphics[width=0.45 \textwidth]{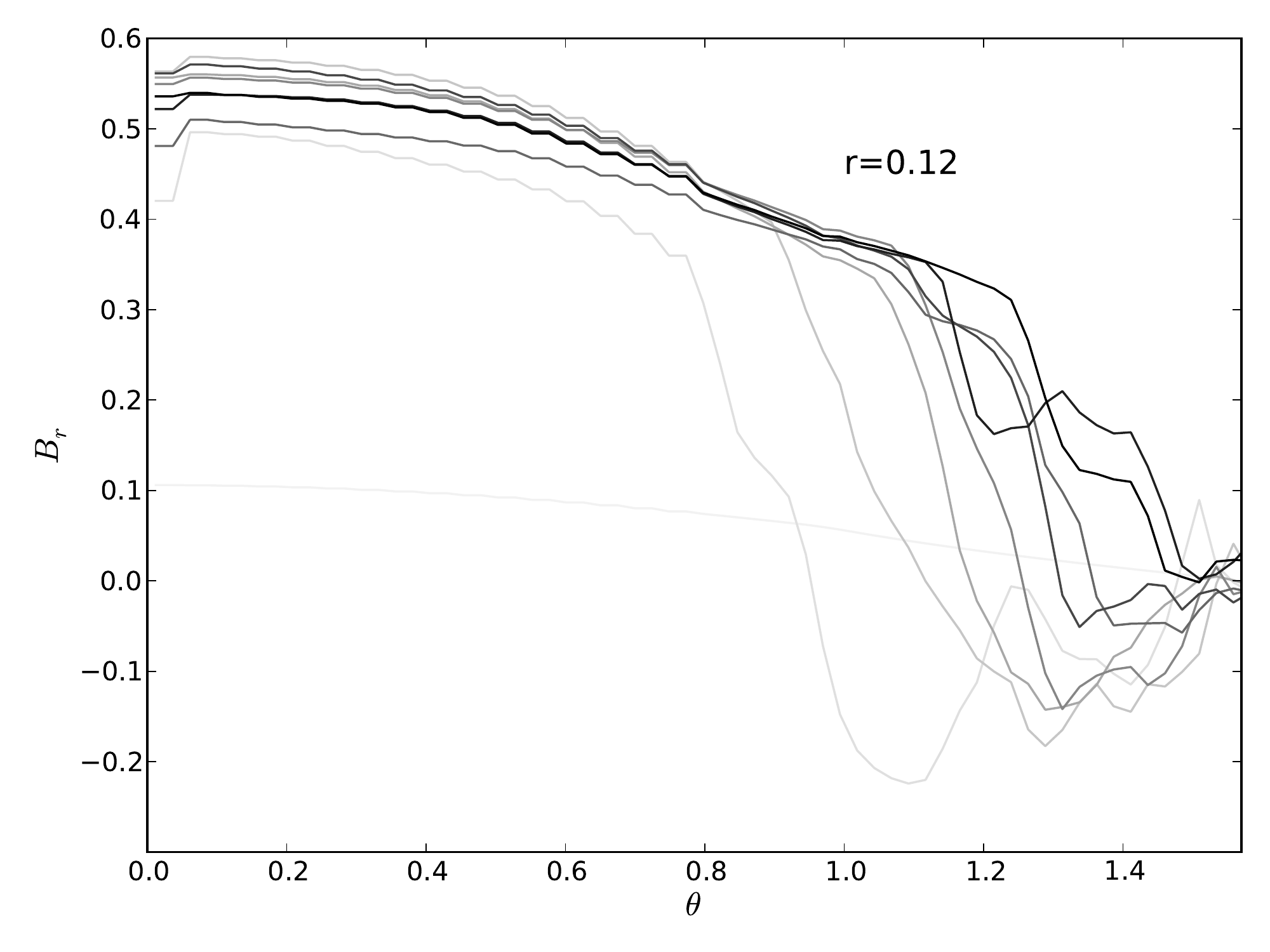} 
\vspace{-0.3 cm}
\caption{Azimuthally averaged $B_{r}$ with respect to $\theta$ at
r=10 (upper panel) and r=0.12 (lower panel). From light to dark,
the curves are from t=0, 2, 10, 20, 25, 30, 35, 40, 45 $T_{0}$. 
} \label{fig:br10}
\end{figure}

Since both MRI and disk wind depend on the strength of net magnetic fields,
an important question to understand disk accetion is what is the rate of inward transport
of net vertical magnetic flux.
To answer this question, we plot $B_{r}$ both near and far away from the inner boundary at $r=0.12$ and $r=10$ in Figure \ref{fig:br10}.
The total magnetic flux through the sphere at $r$ is $\int 2\pi r^2 {\rm sin}\theta \langle B_{r}\rangle {\rm d}\theta$.
In previous 3-D MHD simulations which have not covered the polar region, magnetic fields can be lost
at the $\theta$ boundary close to the pole. However, our simulations cover the full 4$\pi$ sphere, 
magnetic fields cannot be lost at the poles. If no field is being accreted, the total flux should remain
a constant. In Figure \ref{fig:br10}, we can see that, after the initial relaxation, $B_{r}$ at the inner boundary increases with time,
which suggests that magnetic fields are accreted to the central star while mass is being accreted.
Simulations with longer timescale
are needed to see if the accumulation of flux will saturate at some point. We also observe that
magnetic field is strongest at the equator, and not the pole.  However, we caution that a density floor is applied in the polar regions which may affect migration of the field from the equator.

\begin{figure*}[ht!]
\centering
\includegraphics[width=1.0 \textwidth]{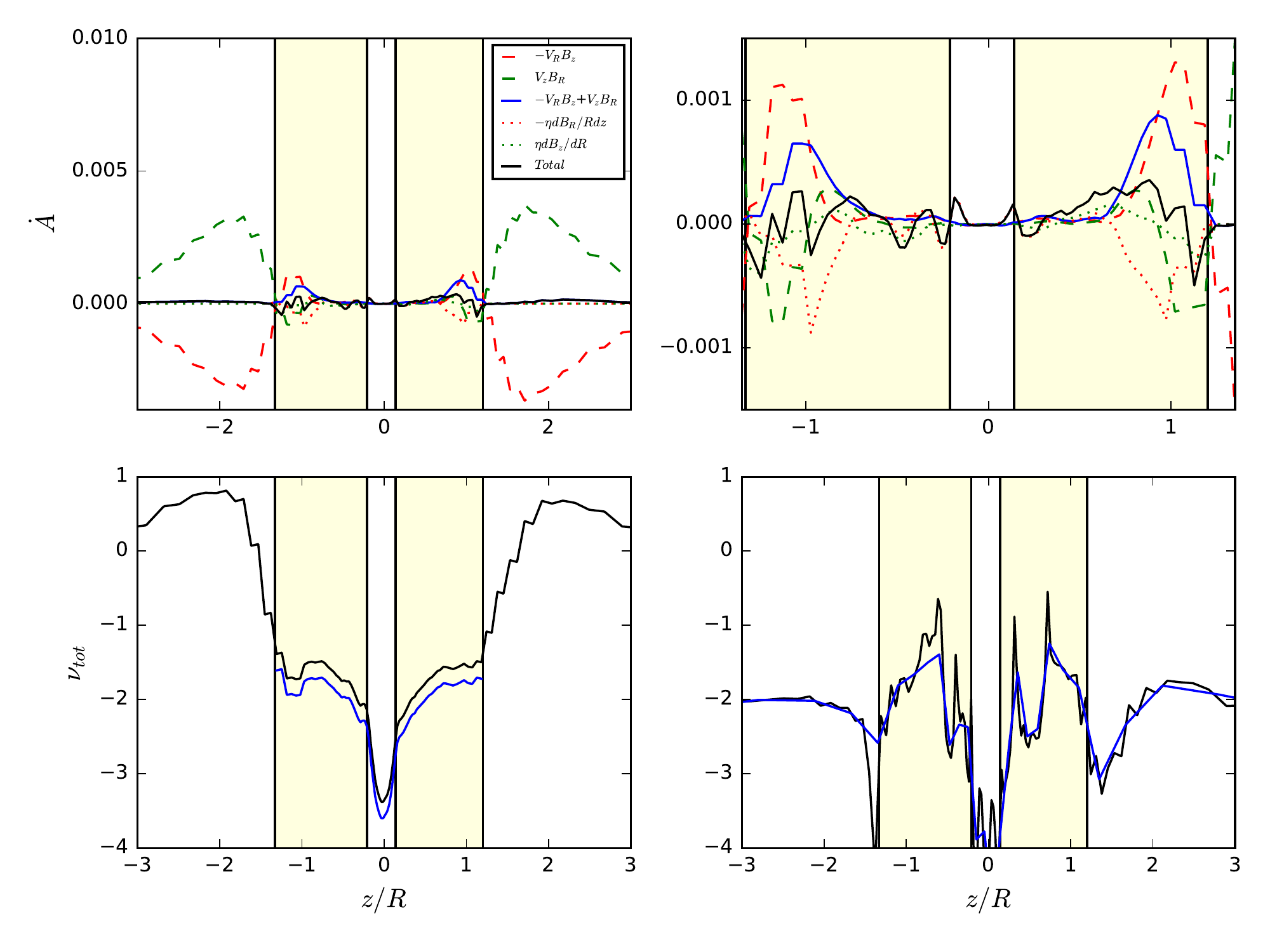} 
\vspace{-0.3 cm}
\caption{ Various contributions to the change of $A_{\phi}$ at R=1 from z=-3 to z=3 (the upper left panel),
from z=-1.35 to z=1.35 (the upper right panel). The equivalent turbulent viscosity ($\nu=\langle\rho v_{R} (v_{\phi}-\langle v_{\phi}\rangle)-B_{R}B_{\phi}\rangle/(1.5\rho\Omega)$) has been shown as the black curve in the lower left panel. The adopted $\eta$ is shown
as the blue curve in the lower left panel (The disk region below the wind region has 0.6 $\nu$ while $\eta$ is set to be 0 in the wind region.).  The derived resistivity is shown in the lower right panel. The blue curve in the lower right panel 
shows $\eta$ averaged over 1 disk scale height.  All primitive variables
have been averaged over both the azimuthal direction and time (t=35 to 42 $T_{0}$ with a $\Delta t$=0.1$T_{0}$ interval).  } \label{fig:vectorA}
\end{figure*}

Another important question regarding net magnetic fields is how to maintain
such a large scale field in an accreting disk. The balance between the field advection and diffusion
will determine the global field strength \citep{Okuzumi2014,TakeuchiOkuzumi2014}.
As discussed in the introduction, if Pr$\sim$1,  
large-scale fields will diffuse outwards faster than the inward advection and the disk quickly loses magnetic flux 
\citep{Lubow1994}. 

To understand how magnetic fields are maintained in our simulations,
we can write down the
induction equation using the magnetic vector potential,
\begin{equation}
\frac{\partial \bf{A}}{\partial t}=\bf{v}\times\bf{B}-\eta_{turb}\times\nabla\times B
\end{equation}
where $\eta_{turb}$ is the resistivity due to turbulence.

The poloidal field is determined by $A_{\phi}$, which is
\begin{equation}
\frac{\partial A_{\phi}}{\partial t}=-v_{R} B_{z}+v_{z} B_{R}-\eta_{turb}\frac{\partial B_{R}}{R\partial z}+\eta_{turb}\frac{\partial B_{z}}{\partial R}\,.\label{eq:vectorp}
\end{equation}
When the field is steady, the advection of fields (the first two terms on the right side) is balanced by
the field diffusion due to turbulence. These four terms on the right side of the equation are plotted
in Figure \ref{fig:vectorA}. In the wind region, the poloidal components of the velocity and magnetic vectors are parallel
to each other so that ${\bf v_{p}}\times{\bf B_{p}}=0$. The two advection terms ($v_{R} B_{z}$ and $v_{z} B_{R}$) are balanced by each other. 
To calculate the turbulent diffusion terms, we assume the turbulent resistivity is $\eta_{turb}=$0.6 $\nu$ 
(where $\nu=T_{total, R\phi}/(1.5 \rho\Omega)$) below the wind 
region and zero in the wind region,  shown as the blue curve in the lower left panel. 
In the upper corona region ($|z/R|>1$),  $v_{p}$ and $B_{p}$ are still parallel to each other so that the two advection terms are balanced by each other. 
But at the lower corona region ($|z/R|<1$), the radial inflow carries fields inwards ($-v_{R}B_{z}<0$). 
With our choice of turbulent resistivity, this inward motion
is balanced by the turbulence diffusion, especially the $dB_{R}/Rdz$  term there (the black curves in the top panels are almost zero.). 
Thus, the turbulent diffusion seems to balance the advection very well with the assumption that $\eta_{turb}/\nu$ is on the order of unity. 

On the other hand, we can derive $\eta_{turb}$ directly using equation \ref{eq:vectorp} if we assume 
that the magnetic field structure has reached a quasi-steady state, which is evident in our simulations. 
In other words, fields advection
and diffusion occur at a much shorter timescale than the secular evolution of the fields. 
Thus, by setting $\partial_{t}A_{\phi}$=0, we have
\begin{equation}
\eta_{turb}=\frac{-v_{R} B_{z}+v_{z} B_{R}}{\frac{\partial B_{R}}{R\partial z}-\frac{\partial B_{z}}{\partial R}}\,.
\end{equation}
Using this equation, the derived $\eta_{turb}$ is shown in the lower right panel of Figure \ref{fig:vectorA}. 
This profile is quite similar to the $\nu$ profile, consistent with local simulations that $\eta_{turb}\sim\nu$.
The quasi-steady state with $\eta_{turb}\sim\nu$ seems to be in contradiction to previous studies that the disk loses
magnetic fields due to the turbulent diffusion if $\eta_{turb}\sim\nu$. However, two factors affect this.
First, the inflow velocity in our simulation (e.g. 0.2 at R=1)
is faster than $v_{R}\sim\nu/R$ (e.g. 0.03 from Figure \ref{fig:vectorA}) based on the viscous theory. This is due
to the internal $\phi z$ stress which creates the vertical shear. Second, since the disk extends to $z\sim R$, the vertical diffusion of the field is on the order of $\eta B_{R}/R$
instead of $\eta B_{R}/H$. Thus, the faster inflow and the slower diffusion allow the disk to maintain a global
field even if $\eta\sim\nu$. We want to caution that the curves in Figure \ref{fig:vectorA} have large uncertainties (e.g. a factor of 2) since
we need to average over 70 snapshots and also over the azimuthal direction to derive the mean fields out of the turbulent fields, and then we need to calculate the derivatives  
for these non-smooth curves to get the vector potential.

To confirm that the disk has evolved to a quasi-steady state when the secular evolution of the global field is much slower than the diffusion and advection, 
we checked the field strength in the simulation.
From t=200 to t=420, $B_{\theta}$ at $R=1$ changes by at most 0.002. Then, $\partial A_{\phi}/\partial t \sim B_{\theta}R/\Delta t$ is
$\sim 10^{-5}$, which is two orders of magnitude smaller than other terms.

Besides studying the field structure within the disk, we would like to know 
the secular evolution of the global fields. As \cite{OgilvieLivio2001} and \cite{Okuzumi2014}
point out, the poloidal field evolution is determined by the balance between the conductivity-weighted radial velocity and 
conductivity-weighted resistivity. 
If we define the effective viscosity as the product of the conductivity-weighted radial velocity and R,
we can calculate the effective Prandtl number as the ratio between the effective viscosity and the
conductivity-weighted resistivity \footnote{This Pr$_{eff}$ is different from the definition in \cite{TakeuchiOkuzumi2014}
by a factor of $C_{u}$.}, and it can be reduced to
\begin{equation}
Pr_{eff}=\frac{1}{2}\int_{-R}^{R}\frac{v_{r}(z)}{\eta(z)}dz\,.
\end{equation}
We assume that the disk extends from z=-R to R. 
Using time and azimuthally averaged $v_{r}$ and assuming $\eta(z)=T_{R\phi}/(1.5 \rho\Omega)$, we calculate
$Pr_{eff}\sim 2.7$. If we use the averaged $\eta(z)$ in the lower right panel of Figure \ref{fig:vectorA},
we calculate $Pr_{eff}\sim 10$, although this value has a large uncertainty since the derived $\eta(z)$
is very uncertain. Nevertheless, the effective Prandtl number is on the order of unity. 

\subsection{Different Net Fields and $H/R$}

\begin{figure*}[ht!]
\centering
\includegraphics[trim=0cm 0cm 0cm 6cm, width=0.9\textwidth]{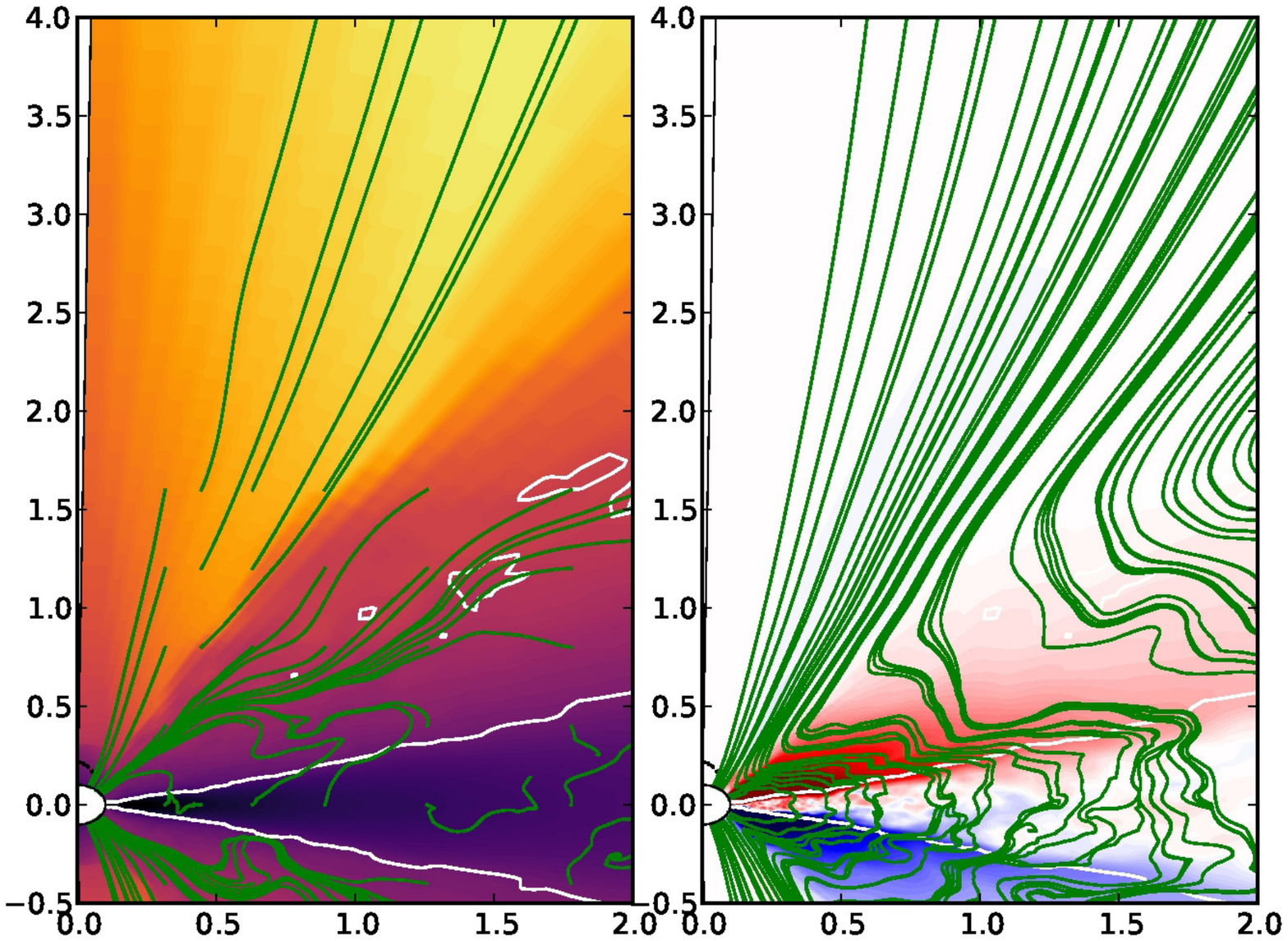} 
\vspace{-4.7 cm}
\caption{Similar to the right panels of Figure \ref{fig:twodrhovel} and \ref{fig:twodrhoalfven} but for the $\beta_{0}=10^4$ case. 
The snapshot is at t=42 $T_{0}$.  } \label{fig:twodrhovelBweak}
\end{figure*}

To explore how our results depend on the imposed magnetic fields strength, we have carried out simulations with 
initial $\beta_{0}=10^4$ but keeping $(H/R)_{R=R_{0}}=0.1$ \footnote{ We have also carried out a case with $\beta_{0}=100$. But that case behaves
very differently from our fiducial case. The accretion rate is so high that the disk quickly loses
mass and becomes magnetically dominated. We leave the discussion for the strongly magnetized
disk  to a later paper.}.
The velocity and magnetic fields structure are shown in Figure \ref{fig:twodrhovelBweak}. 
Velocity streamlines are shown in the left panel and the magnetic streamlines are shown in the right panel.
The color map in the right panel shows $B_{\phi}$. We can clearly see that coronal accretion is still present
in this case. The magnetically dominated 
corona still extends to $z\sim 1.5 R$. Similar to our fiducial case, field lines are pinched at the coronal 
and weak winds are launched. 

\begin{figure*}[ht!]
\centering
\includegraphics[trim=0cm 0cm 0cm 0cm, width=0.9\textwidth]{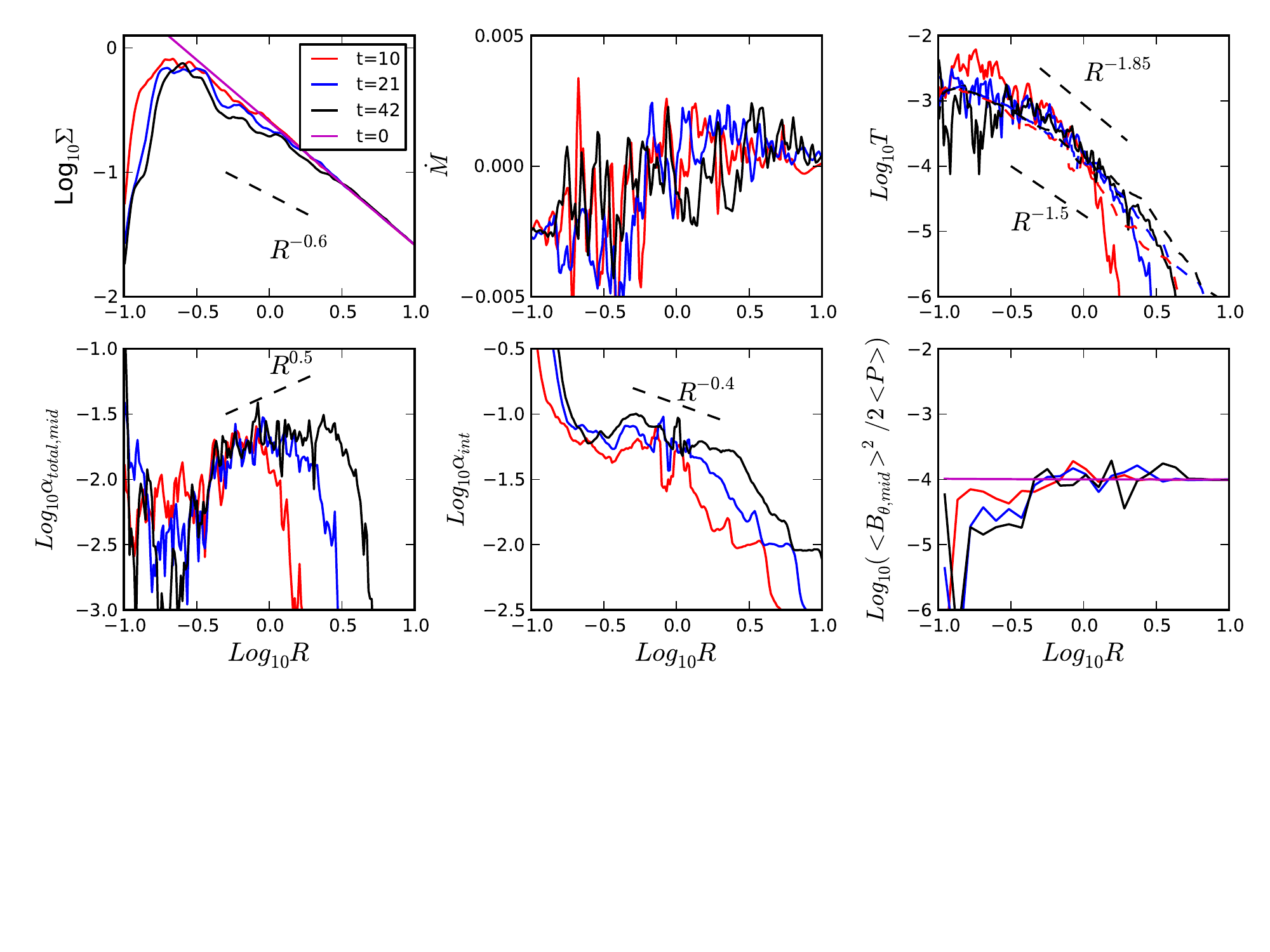} 
\vspace{-3.7 cm}
\caption{Similar to  Figure \ref{fig:dB2} but for the $\beta_{0}=10^4$ case. 
} \label{fig:onedradialweak}
\end{figure*}

The radial disk structure is presented in Figure \ref{fig:onedradialweak}. 
The disk evolves much slower than our fiducial case since the accretion is less efficient
with $\dot{M}\sim -0.002$ instead of -0.005 in our fiducial case. 
Similar to our fiducial case, the vertically integrated
$\alpha$ ($\alpha_{int}$) is significantly larger than the midplane $\alpha$, suggesting that most accretion
occurs at the disk surface. On the other hand, $\alpha_{int}$ is
one order of magnitude smaller than our fiducial case, implying that $\alpha_{int}$
is proportional to the net field strength.  

\begin{figure*}[ht!]
\centering
\includegraphics[trim=0cm 0cm 0cm 0cm, width=1.\textwidth]{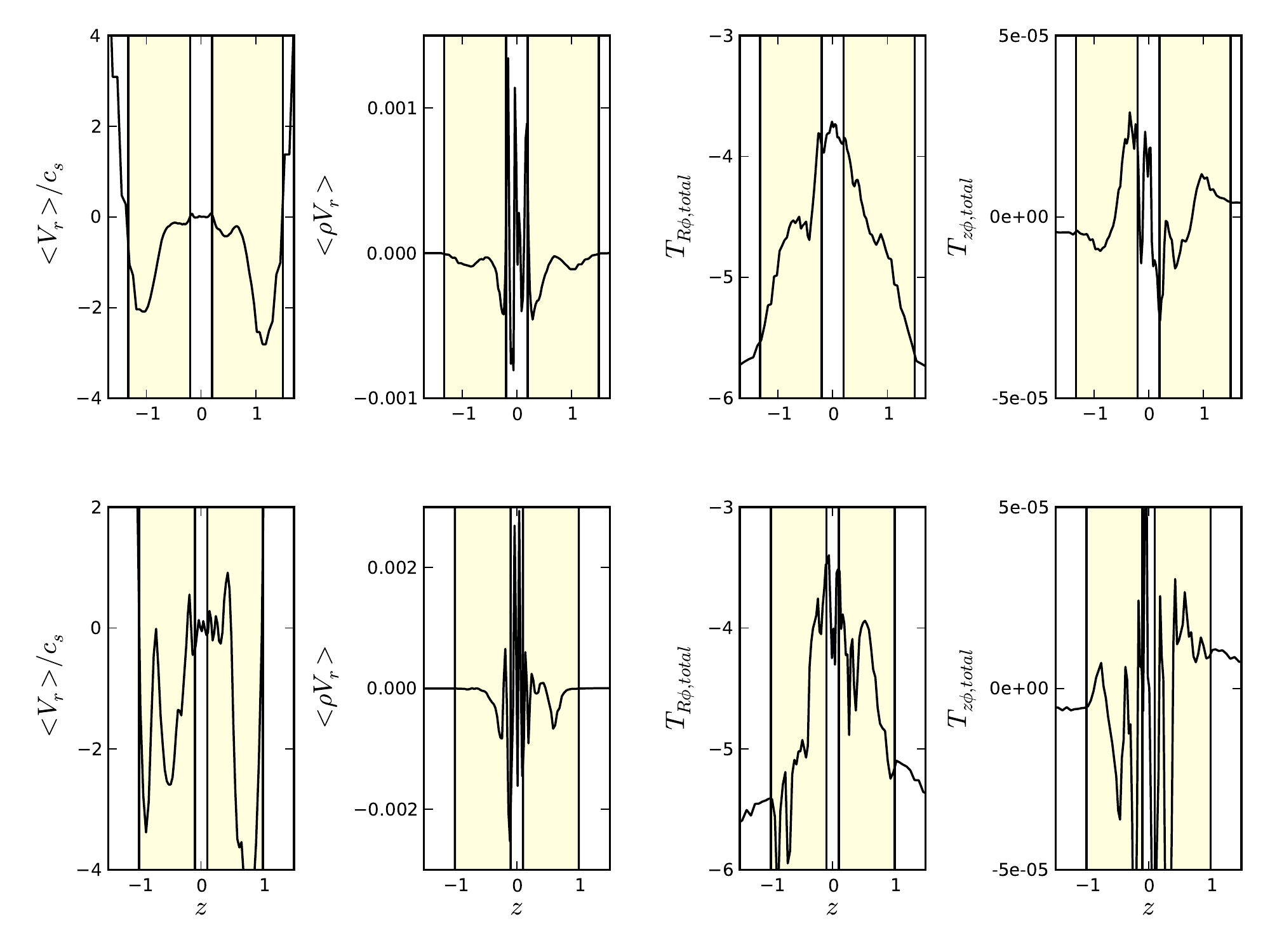} 
\vspace{-0.7 cm}
\caption{The radial velocity, radial mass flux,  $T_{R\phi}$ and $T_{\phi z}$ at $R$=1 along the disk height for both the $\beta_{0}=10^4$ case (upper panels) and the thin disk case (lower panels). The  quantities have been averaged both azimuthally and over time (t=40 to 42 $T_{0}$ with a $\Delta t$=0.1$T_{0}$ interval for the upper panels, and t=20 to 20.9 $T_{0}$ with the same interval for the lower panels ).
} \label{fig:twoplot}
\end{figure*}

We also did the angular momentum budget analysis as the fiducial case. 
Similar to our fiducial case, most of the accretion occurs at the surface. 
Less than 2\% of disk accretion is due to the wind torque. Figure \ref{fig:twoplot}
shows that the inflow is still supersonic in the corona region, and some material
is transported outwards at the disk midplane. The $T_{R\phi}$ stress at the disk midplane
is smaller than the value in the fiducial case by a factor of $\sim$3, consistent 
with local shearing box simulations \citep{Hawley1995} that $\alpha$
is proportional to the initial $v_{A}$. On the other hand, the $T_{\phi z}$ stress
at the wind base is smaller than the value in our fiducial case by a factor of $\sim$10.
Thus, the disk wind seems to play a less important role in the disk threaded by a weaker field.

\begin{figure*}[ht!]
\centering
\includegraphics[trim=0cm 0cm 0cm 6cm, width=0.9\textwidth]{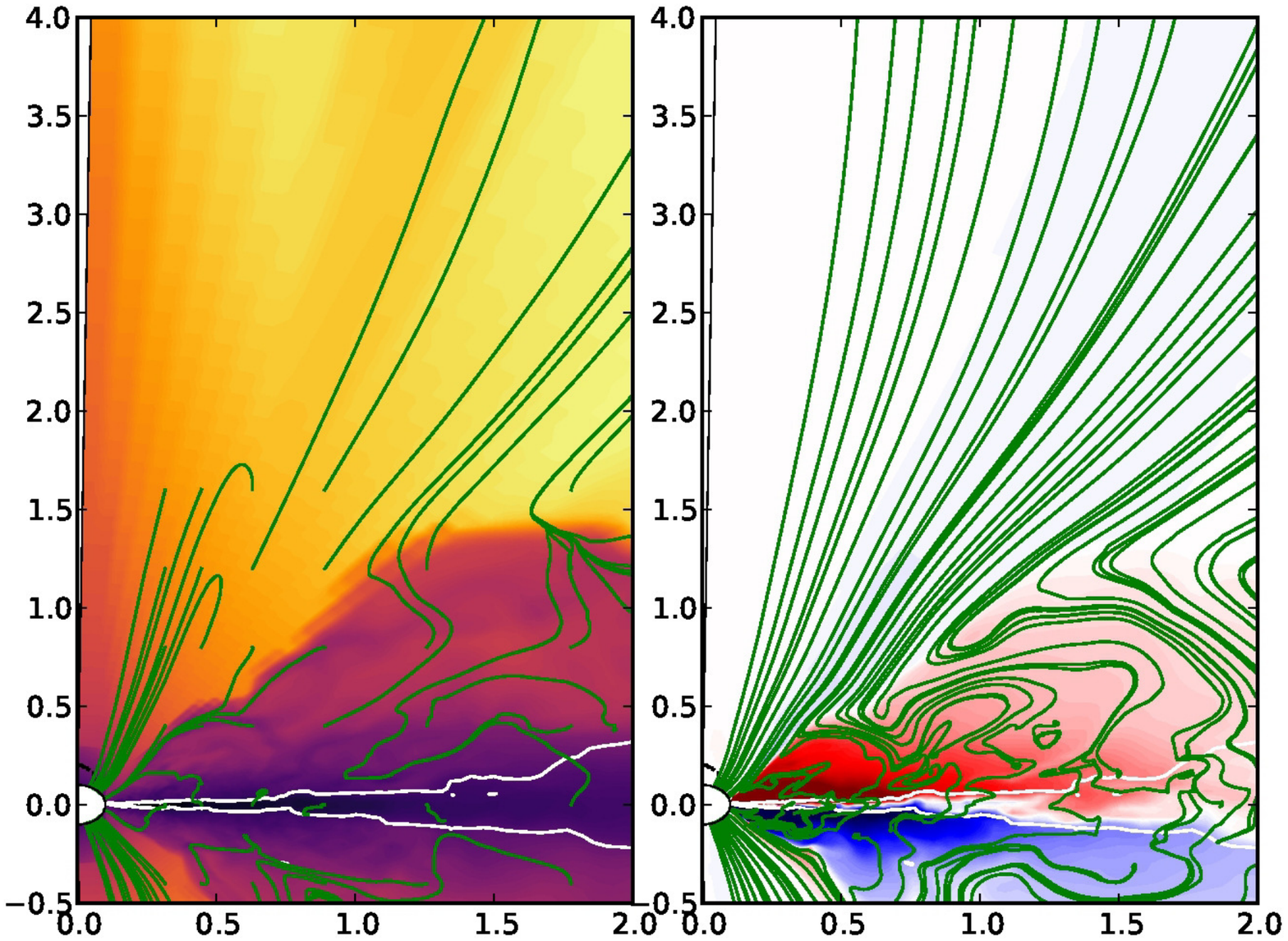} 
\vspace{-4.7 cm}
\caption{Similar to the right panels of Figure \ref{fig:twodrhovel} and \ref{fig:twodrhoalfven} but for the $(H/R)_{R=R_{0}}$=0.05 case. 
The snapshot is at t=20.9 $T_{0}$.  } \label{fig:twodrhovelBthin}
\end{figure*}

\begin{figure*}[ht!]
\centering
\includegraphics[trim=0cm 0cm 0cm 0cm, width=0.9\textwidth]{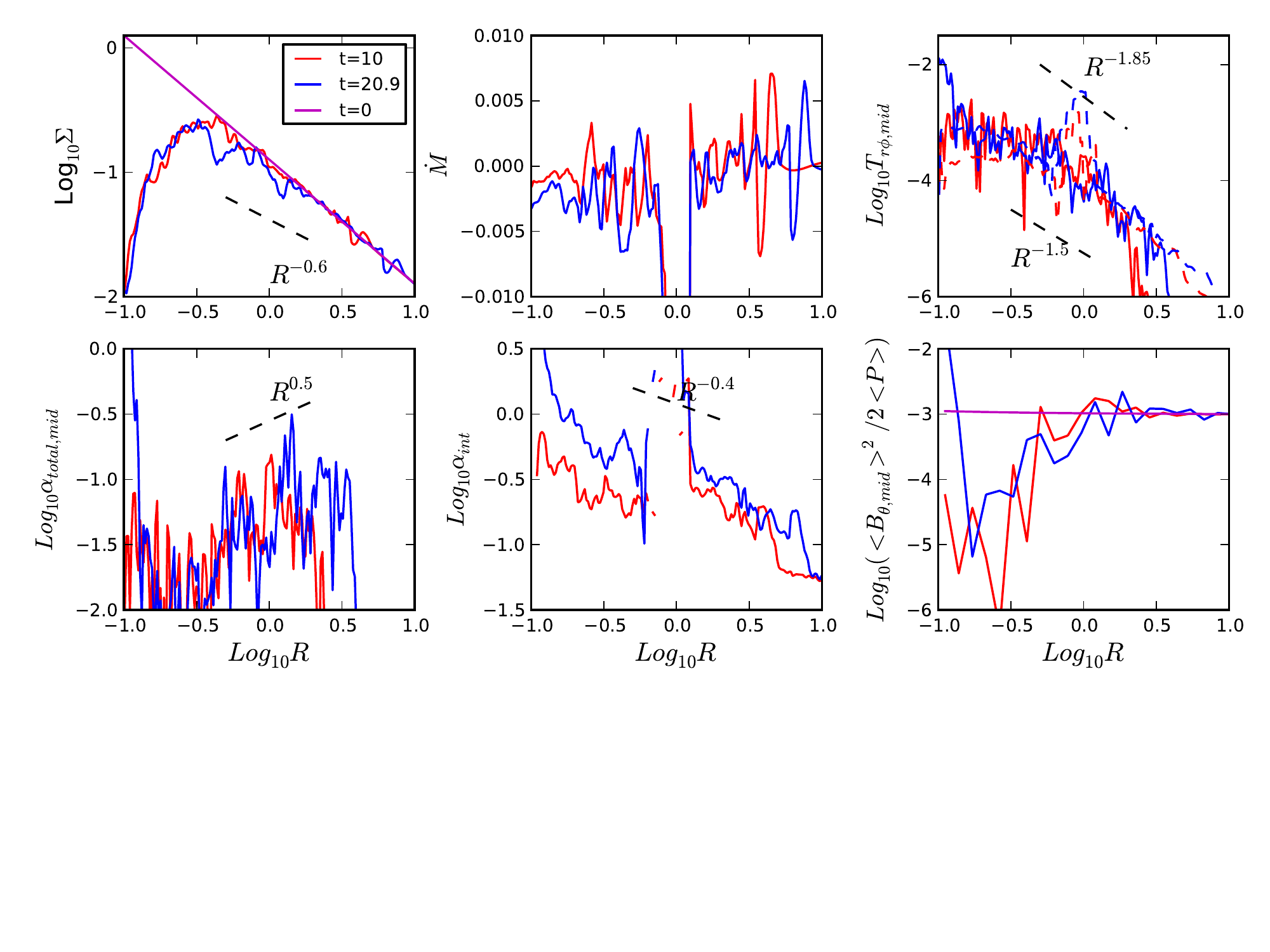} 
\vspace{-3.7 cm}
\caption{Similar to  Figure \ref{fig:dB2} but for the $(H/R)_{R=R_{0}}$=0.05 case. 
} \label{fig:onedradialthin}
\end{figure*}

To see if the coronal accretion picture will hold for thin disks, we have also tried one case with $(H/R)_{R=R_{0}}=0.05$ and
$\beta=1000$. As shown in Figure \ref{fig:twodrhovelBthin}, the coronal accretion still dominates the disk accretion. The corona
still extends to $z\sim R$. On the other hand, the disk accretion rate is $\sim$-0.002,
which is similar to the weak field case but
smaller than the fiducial case
(Figure \ref{fig:onedradialthin}). This lower accretion rate is mainly due to a smaller stress associated with a weaker field. 
Even though the midplane $\beta_{0}$ is the same as the fiducial case, the 4 times smaller gas pressure means that the initial
magnetic field is weaker by a factor 2, which leads to a weaker stress. 

smaller $c_{s}$ in 
the disk considering $\alpha_{int}$ is similar to the value in the fiducial case. 
Figure \ref{fig:twoplot} also suggests that the $\phi z$ stress at the wind base is $\sim$6 times smaller than the fiducial case. Considering the total accretion rate is only 2-3 times smaller than the fiducial case, the wind seems to play a less important role in thinner disks too. 

\section{Discussion}

\subsection{Meridian Circulation}
\begin{figure*}[ht!]
\centering
\includegraphics[trim=0cm 0cm 0cm 0cm, width=0.8\textwidth]{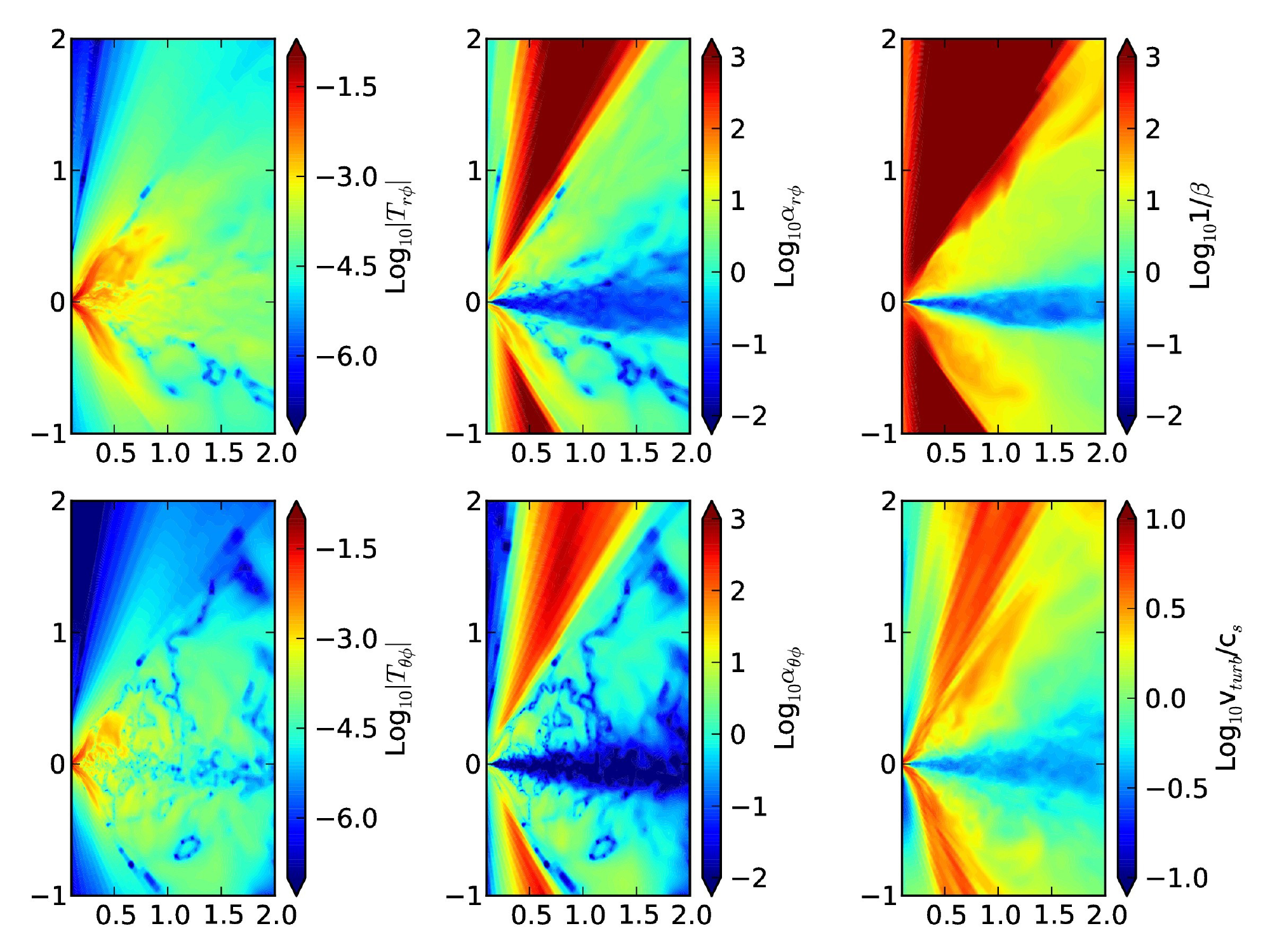} 
\vspace{-0.1 cm}
\caption{Various azimuthally averaged quantities at t=42 $T_{0}$ for the fiducial case.  } \label{fig:twodb}
\end{figure*}

How mass is transported in an accretion disk is important not only for understanding star and planet formation, but
also for explaining
components of primitive meteorites, or chondrites, in our solar system \citep{Cassen1996}.
 The refractory inclusions in chondrites are formed at $\sim$1400-1800 K \citep{Grossman2010}.  Such high temperature
 environment exists at the inner disk within 1 AU. In order to explain their presence in chondrites and even in comets (e.g.
 \citealt{Simon2008}), outward mass transfer is needed. In viscous disks, mass can flow outwards at the disk midplane,
 so-called ``meridian circulation''  ( \citealt{Urpin1984, TakeuchiLin2002, Jacquet2013, PhilippovRafikov2017}). Such outward mass transfer could explain the refractory
 inclusions in meteorites \citep{Ciesla2007,Hughes2010}. However, this ``meridian circulation'' pattern with the midplane going out and surface going in
 is not supported in MHD simulations with net toroidal magnetic fields \citep{Flock2011, Fromang2011}. \cite{Fromang2011}
 found that disk material is going out at all disk heights. 
 
In this paper, a  ``meridian circulation'' pattern is found in our MHD simulations with net vertical fields ( a similar finding is reported in
\citealt{SuzukiInutsuka2014} despite that their surface inflow is very close to their $\theta$ boundary). 
However, the driving mechanism in our simulations is entirely different from the traditional ``meridian circulation''
in viscous disks.

\cite{TakeuchiLin2002} have shown that, in viscous disks with the stress $T_{R\phi}=\rho\nu R\partial \Omega/\partial R$
and $T_{\phi z}=\rho\nu R\partial \Omega/\partial z$,
 the radial velocity at the disk midplane is positive whenever 3p+2q+6$<$0. With the normal disk
 parameters of q=-1/2 and p=-2.25,  3p+2q+6 is -1.75 and the disk flows outwards at the midplane due to viscous stresses.  
 At larger $z$,
$v_{R}$ becomes negative and the disk accretes inwards. \cite{Fromang2011} have shown that such meridian circulation in the viscous disk
is due to the $R-\phi$ stress,  while the $z-\phi$ stress actually tries to drive the midplane inwards. When $T_{\phi z}$ stress is completely ignored, the radial velocity is 
\begin{align}
\frac{v_{R}}{c_{0}}=&-\alpha_{visc}\left(\frac{H_{0}}{R_{0}}\right)\left(\frac{R}{R_{0}}\right)^{q+1/2}\nonumber\\
&\left[3p+3q+6+
\frac{3q+9}{2}\left(\frac{z}{H}\right)^2\right]\,.
\end{align}
where $\nu=\alpha_{visc}c_{s}h$ \footnote{$\alpha_{R\phi, visc}$ defined in this way is smaller than $\alpha_{R\phi}$ 
defined in our paper ($\alpha_{R\phi}=T_{R\phi}/\rho c_{s}^2$)
 by a factor of 1.5 }. With q=-1/2 and p=-2.25, the disk still flows outwards at the midplane and flows inwards at the surface. 

However, for the coronal accretion presented here, the vertically sheared motion is mostly due to the anomalous $z-\phi$ stress instead of the $T_{R,\phi}$ stress as in viscous disks.
In most MHD simulations, the turbulent $T_{R,\phi}$ stress is almost uniform along the vertical direction instead of
being proportional to the density \citep{Fromang2011}. In Figure \ref{fig:twodb}, we can see that the stress is even higher at the atmosphere at our inner disk, and the stress is mostly
due to the mean fields at the atmosphere (Figure \ref{fig:onedr}). For an order-of-magnitude estimate, we assume that the stress is uniform vertically, and the $\alpha$ parameter at the disk midplane vary radially as 
$\alpha=\alpha_{0}(R/R_{0})^{\gamma}$. Then the disk accretion rate is
\begin{align}
\frac{v_{R}}{c_{s,0}}=&-2\alpha(\gamma+p+q+2)\nonumber\\
&\left(\frac{H_{0}}{R_{0}}\right)\left(\frac{R}{R_{0}}\right)^{\gamma+q+1/2}{\rm exp}\left(\frac{z^2}{2 H^2}\right)\,,\label{eq:vrmhd}
\end{align}
based on Equation 25 in \cite{Fromang2011}. 
In our fiducial case with $\gamma$=0.5, $p$=-1.85, $q$=-0.5, the whole disk should flow inwards. 
Our angular momentum analysis in Figure \ref{fig:budget} confirms 
that the $r-\phi$ stress term is negative, trying to drive the disk to accrete inwards. However, Figure \ref{fig:budget}  also reveals that the $z-\phi$ (or $\theta-\phi$) stress 
transports the angular momentum from the disk surface to the midplane, or in other words the midplane magnetically breaks the surface,  leading to the surface coronal accretion.

\begin{figure*}[ht!]
\centering
\includegraphics[trim=0cm 0cm 0cm 0cm, width=1.\textwidth]{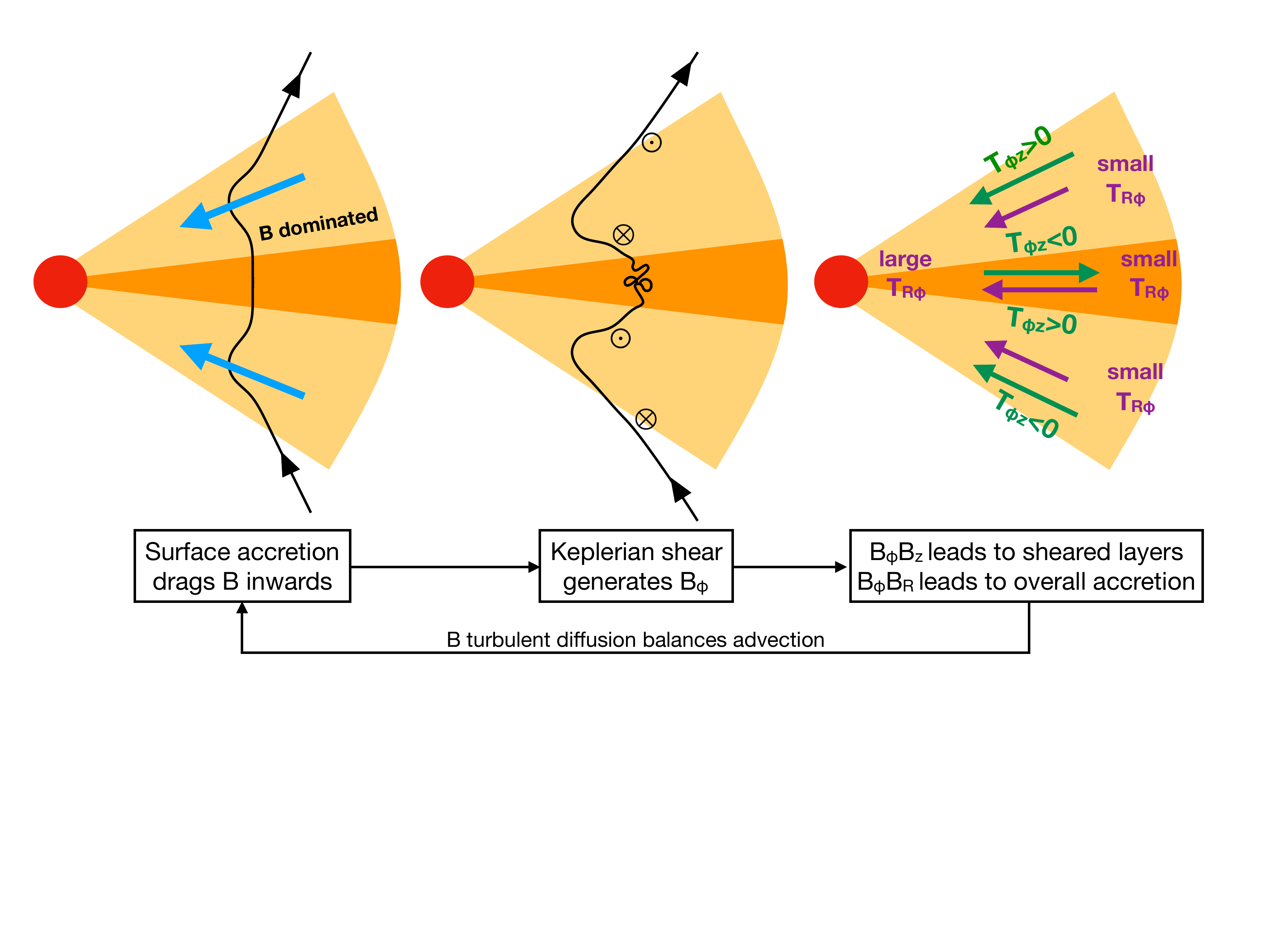} 
\vspace{-4. cm}
\caption{The schematic diagram shows the proposed mechanism for the coronal accretion at $z\sim R$ in the disk atmosphere.  In the rightmost panel, the T$_{\phi z}$ stresses at the transition between the outflow, corona, and disk midplane regions are shown.
The arrows represent the flow driven by various stresses. 
The green arrows show the flow direction at the corona and disk midplane driven by the T$_{\phi z}$ stresses, while the purple arrows show the flow in these regions driven by the T$_{R\phi}$ stresses. Overall, both T$_{\phi z}$ and T$_{R\phi}$ drive surface accretion, while their effects at the midplane work against each other and the flow is slowed down. } \label{fig:scheme}
\end{figure*}

 We have summarized how such $z-\phi$ stress term is produced and how it feeds back to the disk accretion flow in Figure \ref{fig:scheme}. 
When the disk accretes at the surface,  the magnetic fields at the surface are dragged inwards with the flow. This produces the negative $B_{R}$ component close to the midplane 
at $z>0$ (the lower left
panel in Figure \ref{fig:coronal}). Due to the Keplerian shear, the inward magnetic fields is twisted in the azimuthal direction, producing positive $B_{\phi}$ close to the midplane 
(the lower right panel in 
Figure \ref{fig:coronal}). With the positive net $B_{z}$, the magnetic stress $-B_{\phi}B_{z}$ is negative at $z>0$ and $\partial T_{\phi z}/\partial z$  (or $\partial T_{\theta\phi}/\partial \theta$) is negative, driving the midplane to flow outwards (Equation \ref{eq:angsph}). At the top of the corona region, $B_{R}$ is positive and it is twisted to generate
 negative $B_{\phi}$ due to the Keplerian shear. Thus $B_{\phi}$ decreases from positive to negative within the corona region, thus producing  
positive $\partial T_{\phi z}/\partial z$, which leads to the inward
surface accretion. Such vertically sheared motion then feeds back to the field geometry. 
This picture is quite similar to the linear growth phase of MRI, but operating at the global scale.

To make the connection between the $\theta-\phi$ stress (Figure \ref{fig:onebr}) and the classical viscous disk model, where 
\begin{equation}
T_{\phi z}=-\alpha P\frac{R}{\Omega}\frac{\partial \Omega}{\partial z}\sim \alpha P\frac{z}{R}\,
\end{equation}
using the relationship $\partial \Omega/\partial z\sim -\Omega_{K}z/R^2$ at the midplane from Equation \ref{eq:vphi}, we can calculate $\alpha_{vis,\theta\phi}$ as 
$T_{\theta\phi}/(P(\theta-\pi/2))$ which is similar to $T_{\phi z}R/(Pz)$.
Figure \ref{fig:onebr} shows that $T_{\theta\phi}$ is negative at z$>$0 and positive at z$<$0. Thus, $\alpha_{vis,\theta\phi}$ is negative, which is qualitatively
different from the viscous model. In the viscous model, the positive $\alpha_{vis,\theta\phi}$ tries to drive the midplane flow inwards while $\alpha_{vis,r\phi}$ tries
to drive the midplane flow outwards. In our MHD models,
the negative $\alpha_{vis,\theta\phi}$, due to the mechanism illustrated in Figure \ref{fig:scheme}, drives the midplane flow outwards. 
The magnitude of the $\alpha_{vis,\theta\phi}$ in our model is shown as the dashed curve in the lower right panel of Figure \ref{fig:onebr}, which is comparable
to $\alpha_{r\phi}$. 

Overall, the vertically sheared flow is mostly determined by the internal $z-\phi$ (or $\theta-\phi$) stress from net magnetic fields.
However, the  $z-\phi$ stress contributes little to the disk overall accretion since this stress at the corona upper surface due to  the wind torque is weak.
As described in \S 4, the inward accretion of the disk is mostly due to the $r-\phi$ stress (illustrated in Figure \ref{fig:scheme}). In the corona region, the  $r-\phi$ stress is provided by the net fields,
while at the disk midplane it is from MRI turbulence. 

\subsection{Accretion Mechanism and Wind vs Turbulence}
If we are only interested in the net disk accretion rate, we can vertically integrate the angular momentum equation and it is
 the $z-\phi$ stress at the disk surface that matters (Equation \ref{eq:mdot}). If we plug 
Equation \ref{eq:mdot} into the mass conservation equation
\begin{equation}
2\pi R\frac{\partial \Sigma}{\partial t}=-\frac{\partial \dot{M}_{acc}}{\partial R}-\frac{\partial \dot{M}_{loss}}{\partial R}\,,
\end{equation}
and assume $v_{k}\propto R^{1/2}$, we have
\begin{align}
2\pi R\frac{\partial \Sigma}{\partial t}=&\frac{\partial}{\partial R}\left[ \frac{4\pi}{ v_{k}}\frac{\partial}{\partial R}\left( R^2 \alpha_{R\phi,int}\Sigma c_{s}^2\right)\right] \nonumber\\
&-\frac{\partial}{\partial R}\left(\frac{4\pi}{v_{k}}R^2\langle B_{z}B_{\phi}\rangle\bigg |_{z_{min}}^{z_{max}}\right)-\frac{\partial \dot{M}_{loss}}{\partial R}\,,
\end{align}
Inserting Equation \ref{eq:windstress}, we get
\begin{align}
2\pi R\frac{\partial \Sigma}{\partial t}=&\frac{\partial}{\partial R}\left[ \frac{4\pi}{ v_{k}}\frac{\partial}{\partial R}\left( R^2 \alpha_{R\phi,int}\Sigma c_{s}^2\right)\right] \nonumber\\
&+\frac{\partial}{\partial R}\left(\frac{2R}{\Omega_{k}}\frac{\partial \dot{M}_{loss}}{\partial R}(\frac{R_{A}^2}{R^2} \omega-\Omega_{k})\right)-\frac{\partial \dot{M}_{loss}}{\partial R}\,.
\end{align}
where $(R_{A}/R)^2\equiv\lambda$ is the wind lever arm. 
The terms on the right are accretion due to the radial stress gradient, the accretion due to the wind torque, and the mass loss rate due to the wind. 
Since $\omega\sim\Omega_{K}$, the accretion rate led by the wind torque is 
\begin{equation}
 \dot{M}_{acc,wind}\sim -2(\lambda-1)R \frac{\partial \dot{M}_{loss}}{\partial R}\label{eq:maccwind}
\end{equation}

In our fiducial run, $\alpha_{R\phi,int}$ is 0.5 at $R=1$ and can be larger at the inner disk (Figure \ref{fig:dB2}). The accretion 
rate due to the radial stress gradient is on the order of $-0.005$, the accretion due to the wind torque is $-2.5\times 10^{-4}$,
and the mass loss from a large disk region [0.5,5] is $2\times 10^{-5}$. Thus, we can see that the importance of the terms
decreases on their orders at the right side. Most disk accretion is led by the radial stress.  Such stress is from the MRI turbulence at the disk midplane
and the net global field at the disk corona. 
The wind carries the angular momentum away but it only leads to 5\% of the accretion. The direct mass loss 
by the wind is another order of magnitude smaller. 

We can also relate the accretion due to the
wind and the wind mass loss rate by using Equation \ref{eq:maccwind} and $\lambda\sim 10$ for the wind that launches at $R=1$. Considering
$\partial \dot{M}_{loss}/\partial R=4\pi R\rho v_{z}$ and $\rho v_{z}\sim 2\times 10^{-6}$ (Figure \ref{fig:wind}), we can derive the mass
accretion rate due to the wind is $\sim 3\times 10^{-4}$, consistent with our direct measurement.  

Overall, the accretion mechanism in our simulations is very different from most previous works. The accretion is mostly driven by neither the turbulence nor the wind. 
Instead, it is mainly driven by the large scale net magnetic fields in the atmosphere, with some contribution from MRI turbulence at the midplane. 
The net field is sustained by the interplay between the  accretion and diffusion.
Large scale net magnetic fields have been found in some previous local MHD simulations. \cite{TurnerSano2008} has suggested that such fields can lead to accretion in the ``undead zone''. 
We can also make an analogy between the field structure in our simulations with channel flows in magnetized disks \citep{GoodmanXu1994,Latter2010} where the stretch between different channel layers creates large scale magnetic fields which transport angular momentum.

\subsection{Time and Spatial Variability}
\begin{figure*}[ht!]
\centering
\includegraphics[trim=0cm 0cm 0cm 0cm, width=0.95\textwidth]{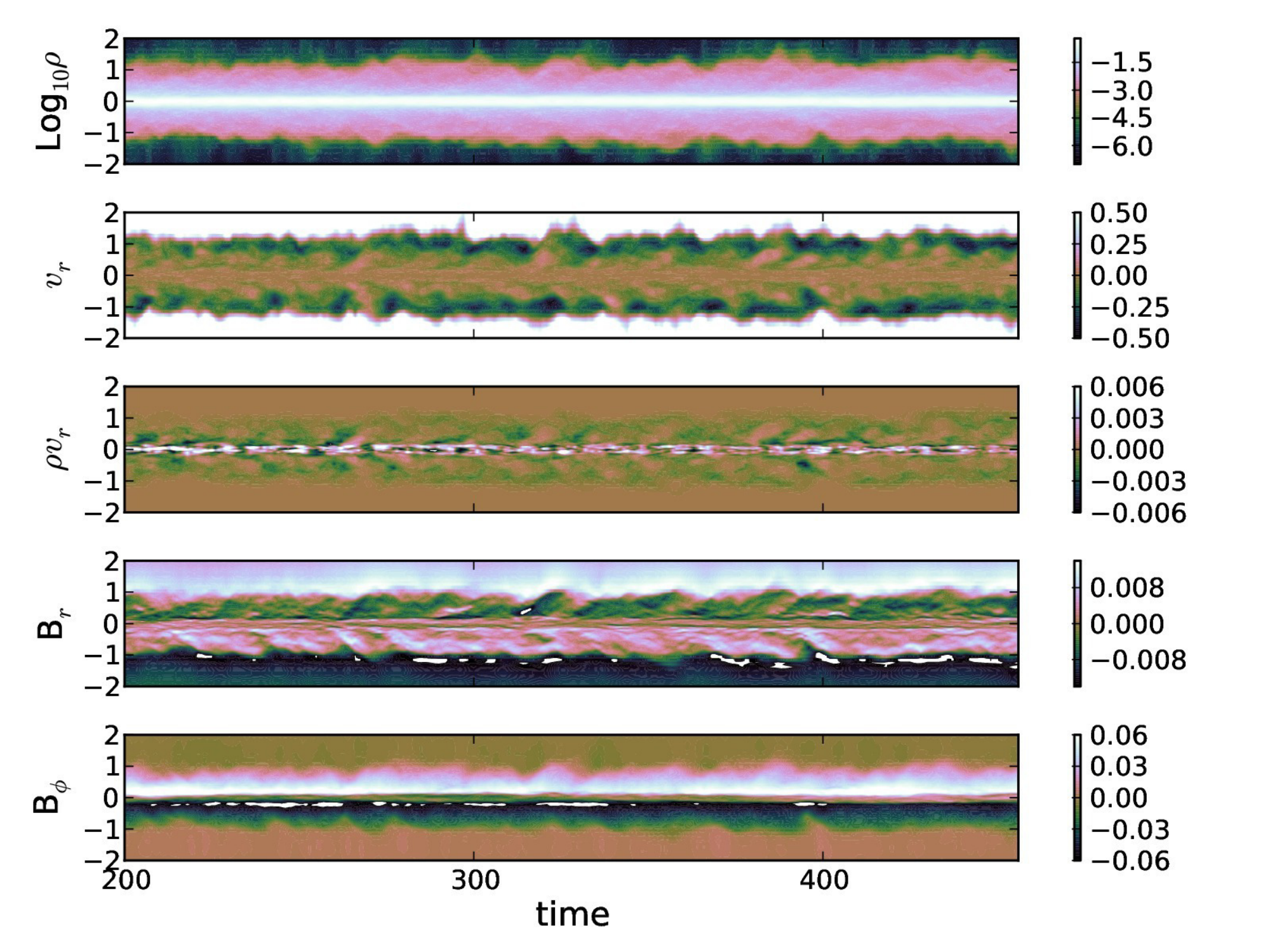} 
\vspace{0 cm}
\caption{Space-time diagram for various quantities along the $z$ direction. The time unit is 0.1 T$_{0}$.} \label{fig:space-time}
\end{figure*}

\begin{figure}[ht!]
\centering
\includegraphics[trim=0cm 0cm 0cm 0cm, width=0.5\textwidth]{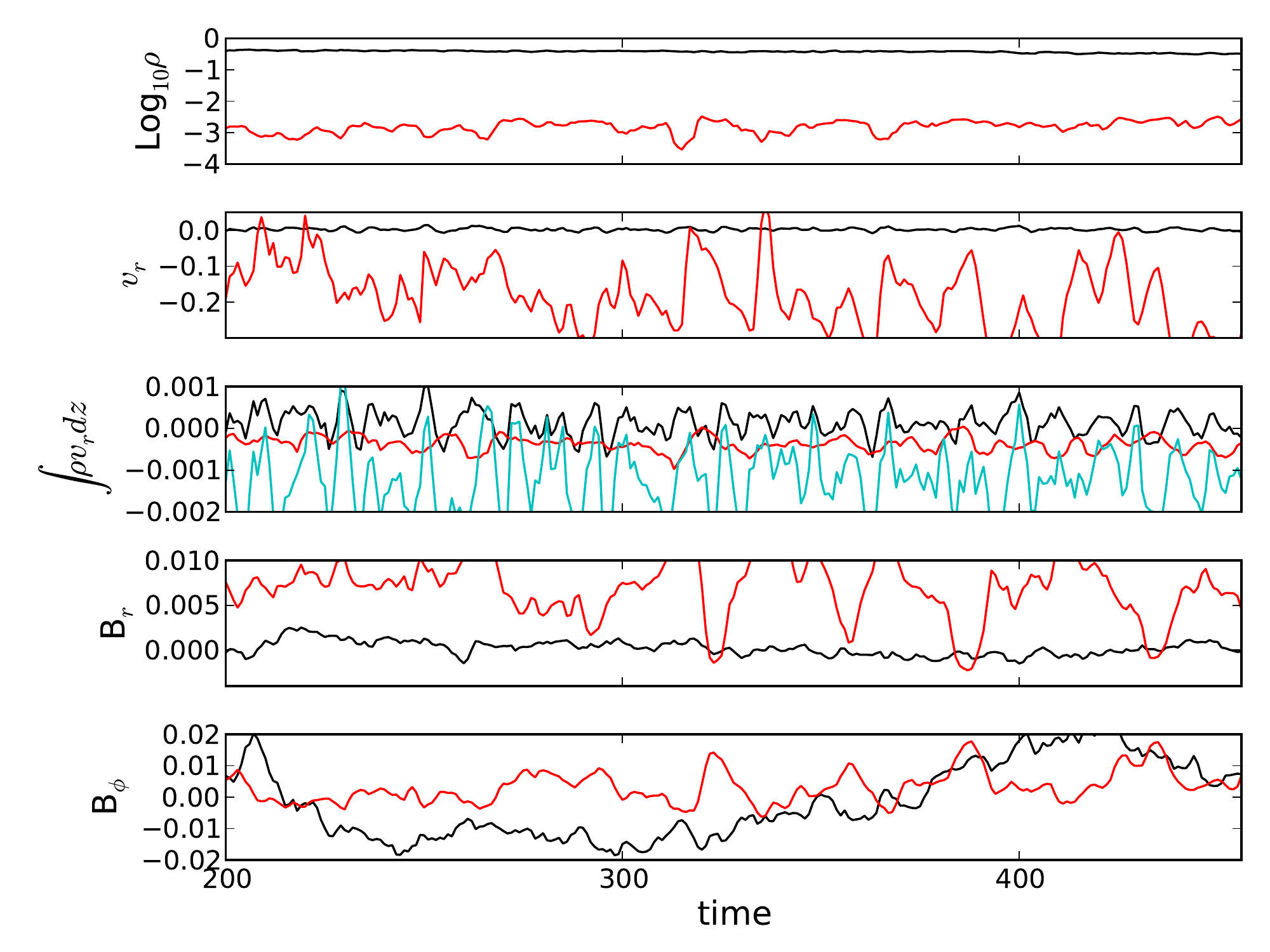} 
\vspace{0. cm}
\caption{The change of various quantities with time. The time unit is 0.1 T$_{0}$. Except the $\int \rho v_{r} dz$ panel, the black curves
are azimuthally and vertically averaged quantities from $z$=-0.1 to 0.1, while the red curves
are azimuthally and vertically averaged quantities from $z$=0.9 to 1.1. In the $\int \rho v_{r} dz$ panel,
the black curves are the azimuthally averaged but vertically integrated $\rho v_{r}$ from $z$=-0.1 to 0.1,
the red curves are the same qualities integrated from $z$=0.5 to 1.5, and the cyan curve is
the quantities integrated from $z$=-2 to 2.  } \label{fig:spacetimeoned}
\end{figure}

Due to MRI turbulence, both disk accretion and wind launching are stochastic. To asses the time variability of
various quantities in our fiducial run, the space time diagram and the time evolution of various quantities are shown in Figures
 \ref{fig:space-time} and \ref{fig:spacetimeoned}. We can see that the flow at the disk midplane is not always outwards.
There seems to be a cyclic behavior on a timescale of 1.5 T$_{0}$. Both $B_{r}$ and $B_{\phi}$ have similar cyclic patterns. 
However, different from zero field or net toroidal field simulations \citep{Stone1996}, 
the net $B_{\phi}$ field does not exhibit negative and positive
alternation during one cycle. Instead $B_{\phi}$ is dominated by the stress from the coronal accretion and always positive
at $z>0$ and negative at $z<0$. Previous shearing box simulations 
 suggests that the cyclic dynamo behavior disappears when $\beta_{0}<1000$ \citep{BaiStone2013} or $\beta_{0}<10$ \citep{Salvesen2016}. The $\beta$ calculated by the
net $B_{z}$ in our simulation is slightly larger than 1000 during the quasi-steady state. However, our disk is also threaded
by a strong net $B_{\phi}$ due to the azimuthal stretch of $B_{R}$ and it is unclear how the net $B_{\phi}$ will affect the cyclic 
dynamo. 

\subsection{Compared with Previous work}
Our simulation indicates that the corona region extends to z$\sim$1.5 R and beyond which the magnetocentrifugal
wind is launched. This suggests that many previous simulations do not have enough vertical range to capture
the disk wind. Being unable to capture the wind region could explain why  the mass loss rate in \cite{Fromang2013} is 
reduced by one order of magnitude when the shearing
box extends from 5 to 10 H above the midplane (even $10$H is within our corona region).
The tallest box simulation in \cite{Fromang2013} suggests that the disk depletion time is 700 orbits. Our disk patch
at $R\sim$1 has a depletion timescale of $\Sigma/(\rho v_{z})\sim 0.7/2\times10^{-6}\sim 5\times 10^4$ orbits, much longer
than the depletion timescale in shearing box simulations. This implies that shearing box simulation is inadequate for
studying disk wind. 

The wind exerts a torque to the upper surface of the corona region. Since the corona region is magnetically dominated, it can
 accrete inwards as a whole. The corona region shares similarities with the sub-Keplerian disk region proposed by \cite{WardleKoenigl1993}
 where the disk is confined by magnetic stresses.  
 
\begin{figure*}[ht!]
\centering
\includegraphics[trim=0cm 0cm 0cm 0cm, width=1.\textwidth]{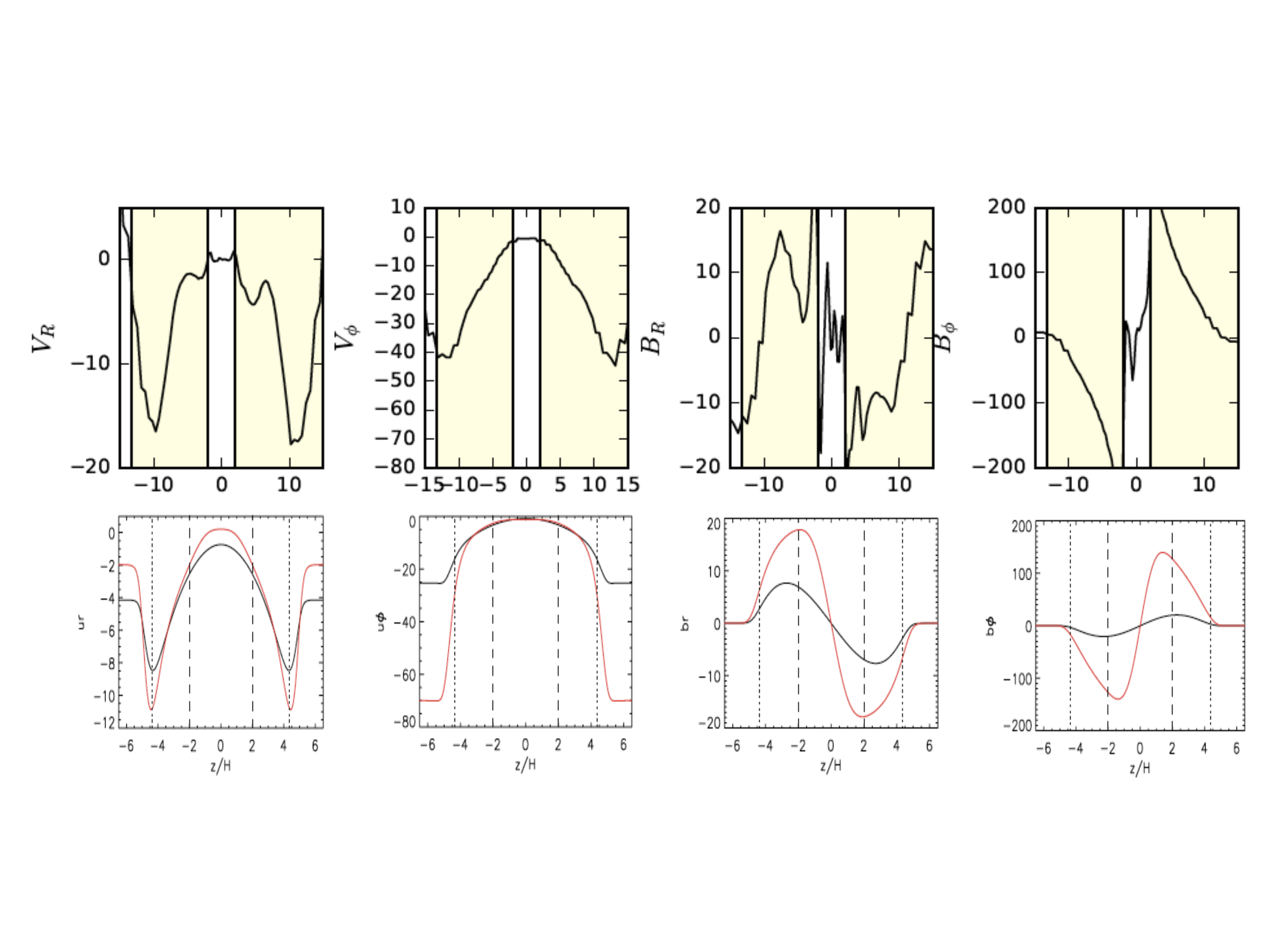} 
\vspace{-2.5 cm}
\caption{Upper panels: similar to Figure \ref{fig:coronal} but for our weak field simulation ($\beta_{0}=1000$). The
quantities have been averaged both azimuthally and over time (t=40 to 42 $T_{0}$ with a $\Delta t$=0.1$T_{0}$ interval).
Lower panels are from Figure 4 in \cite{GuiletOgilvie2013}. The red curves in the lower panel use the $\alpha$ profile which is more similar to our numerical simulations.
The unit system in \cite{GuiletOgilvie2013} is adopted. } \label{fig:gordon}
\end{figure*} 
 
The vertically sheared disk motion in our simulation has been observed in 3-D global MHD simulations by \cite{SuzukiInutsuka2014}. However,
their simulations only extend to $z\sim0.5 R$, well within our corona region. Thus, it is unclear if the surface inflow they observed 
is from part of the corona, due to the wind as they suggested, or due to the imposed boundary. The coronal accretion in our simulation
is very similar to the results in GRMHD simulation by \cite{Beckwith2009} and earlier simulations by \cite{StoneNorman1994}, 
despite the very different numerical setups.  \cite{StoneNorman1994}
make an analogy with the channel mode of MRI. However, the processes on how the flow motion 
determines the global fields and how the global
fields sustain the flow motion have not been studied in detail. 
In this paper, we have presented a careful analysis of the angular momentum budget and have studied the interplay between flow motion and global fields. 
We also point out that although the $z-\phi$ stress drives coronal accretion, it is the $R-\phi$ stress that determines the total accretion rate. 
Furthermore, although \cite{Beckwith2009} 
suggests that the parameterization of local turbulent viscosity and turbulent resistivity may not be sufficient to describe the field transport, 
our simulations agree quite well with R-z 2-D disk models (espeically \citealt{GuiletOgilvie2012, GuiletOgilvie2013}) 
that still use local turbulent viscosity and resistivity but consider
R-Z 2-D mean field and flow dynamics.

Although our simulations are qualitatively similar to the surface inflow picture suggested by these analytical models, the simulations do not support some
assumptions made in some of these models.  \cite{RothsteinLovelace2008} and \cite{Lovelace2009} assume conducting non-turbulent surface layers, while the simulations
imply a turbulent transition region between the coronal and the wind region. On the other hand, our simulations agree with \cite{GuiletOgilvie2012} and \cite{GuiletOgilvie2013}
surprisingly well. Our $\alpha$ profile (Figure \ref{fig:onebr}) is similar to that assumed by \cite{GuiletOgilvie2013} (GO, the red curve in their Figure 1) and the resulting
accretion structures are similar (the comparison is in Figure \ref{fig:gordon}). 
For example, our maximum surface inflow velocity is 2 $c_{s}$ while 
it is $c_{s}$ in GO with H/R=0.1. The $v_{\phi}$ in our coronal region decreases to 5 $c_{s}$, while it decreases to 3 $c_{s}$ in GO. The maximum
$B_{R}/B_{z,0 }$ is 1 in our model and 2 in GO. The maximum $B_{\phi}/B_{z}$ is 10 in both models. Despite the surprising similarity, there are some
notable differences.  First, the maximum
inflow occurs at 10 H in our models while it is 4 H in GO. We think such difference is due to that the coronal region is magnetically supported and
the density there is different from the gaussian profile assumed by GO. Second, we have the wind region beyond the coronal region, which is not the case in GO
due to the specific boundary condition assumed. Future analytical calculations with these considerations are desired.

\subsection{Observables}
Our simulations have observational implications for cataclysmic variables (CVs) and protoplanetary disks. 

By fitting the light curves of CVs, previous studies have
constrained that the $\alpha$ parameter is $\sim$0.1-0.3 during the outburst state (e.g. \citealt{MineshigeOsaki1983,Smak1984,Smak1999,MeyerMeyer-Hofmeister1984,Cannizzo1993,KotkoLasota2012}).
Since the disk should be fully ionized during the outburst state, MRI is thought to be the main angular momentum transport mechanism in such disks \citep{GammieMenou1998}.
However, MRI in zero net field simulations can only generate   $\alpha\sim0.01$ which is much weaker than the $\alpha$
constrained from observations. There are two main solutions to this problem:1) MRI is enhanced by thermal convection 
\citep{Hirose2014}; 2) the disk is threaded by net vertical magnetic fields. The second solution has met several challenges \citep{King2007}: 
a relatively strong net vertical field (e.g. $\beta\sim10$) is required  to generate
$\alpha\sim$0.1 based on local shearing box simulations, a strong disk wind will be generated if the disk is threaded by large fields, 
and the global field
can be lost in the disk.  
On the other hand, our simulations suggest that we can generate vertically integrated $\alpha\sim0.5$ even with
a weak net field ($\beta\sim10^3-10^4$). Part of this large stress is due to the geometry of the global magnetic fields. 
A very weak wind is launched in the disk and the net magnetic fields are maintained in our simulations. 
Thus, a weak net vertical field which could be carried by the inflow remains one solution to the large $\alpha$ in outbursting CV disks. 

\begin{figure}[ht!]
\centering
\includegraphics[trim=0cm 0cm 0cm 0cm, width=0.5\textwidth]{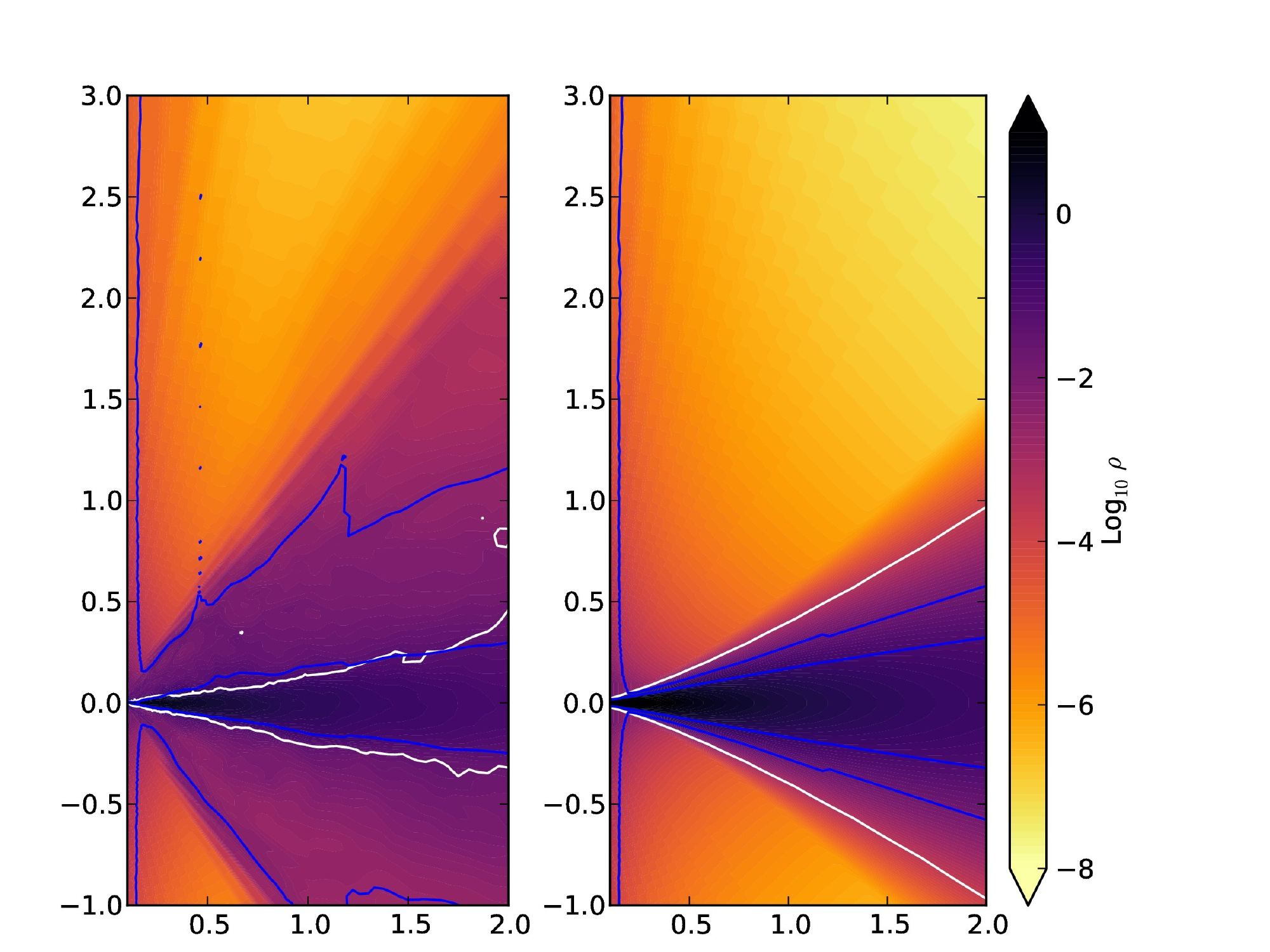} 
\vspace{0 cm}
\caption{Blue curves label where the disk's vertically integrated surface density is 0.01 and 0.001. 
The left panel is at t=42 $T_{0}$ and the right panel is the initial condition.
The white curve is where $\langle\beta\rangle$=1. } \label{fig:tau0}
\end{figure}

One puzzle for disks in Herbig stars is that their near-IR fluxes are several times larger than those predicted from
radiation hydrodynamic models \citep{MuldersDominik2012,Flock2016}. Since most near-IR flux comes
from the inner dust rim where dust sublimates,  this discrepancy could be due to a high disk atmosphere that is 
magnetically supported \citep{Turner2014b}. Our simulations support this picture where the magnetic dominated 
corona extends to z$\sim$R.  Figure \ref{fig:tau0} shows the position where the disk's vertically 
integrated density is 0.001 and 0.01. For comparison, the two-sided surface density at $R=1$ is 0.1.
As long as the $\tau=1$ surface is above the 0.01 curve, the corona will play an important role on the 
spectrum energy distribution. If we scale our fiducial simulation by choosing the length unit as 0.1 AU, the central star mass as 1 solar mass,
and the disk accretion rate (-0.005 code unit) as $-10^{-8}\msunyr$, we can calculate the unit of the surface density as 
8.9 g cm$^{-2}$. Thus, the disk surface density at $R=1$ is 0.1*8.9$\sim$ 1 g cm$^{-2}$. If we assume that the Rosseland mean
opacity is 10 $cm^2 g^{-1}$, the disk's optical depth is 10. Thus, the blue curves have optical depths of 0.1 and 1. 
 If the disk has an accretion rate of $10^{-7}\msunyr$, the highest blue curve will correspond to 
  the $\tau=1$ surface, which is
 twice higher than the $\tau=1$ surface in the hydrodynamical model.   
We note that our disk's aspect ratio at R=1 is larger than the aspect ratio of a real disk at 0.1 AU which is normally $\sim$ 0.03.
Although MHD simulations with realistic thermal dynamics is needed for addressing this problem properly in future, 
our simulations suggest that magnetically supported corona may play an important role  in explaining the  strong near-IR flux in Herbig Ae/Be stars.

Another puzzle our simulations may shed light on is the fast inflow in transitional disks. Transitional disks have moderate
accretion rates but low surface density \citep{Espaillat2014,vandermarel2016}. This implies a large
$\alpha$ in the disk or even supersonic inflow. Fast inflow could also explain the twist of channel maps observed 
in some transitional disks \citep{Rosenfeld2014,Pineda2014,Casassus2015,VanderPlas2016}. 
\cite{Rosenfeld2014} derive that the inflow in HD 142527  approaches the infall velocity or the disk is warped. 
In our simulation, the inflow velocity in the corona region approaches 0.2 v$_{K}$. On the other hand, 
ambipolar diffusion, which should operate in low density regions, may also lead to fast inflow \citep{WangGoodman2016}. 
MHD simulations with ambipolar diffusion included are desired in future to solve this puzzle.

\begin{figure}[ht!]
\centering
\includegraphics[trim=0cm 0cm 0cm 0cm, width=0.5\textwidth]{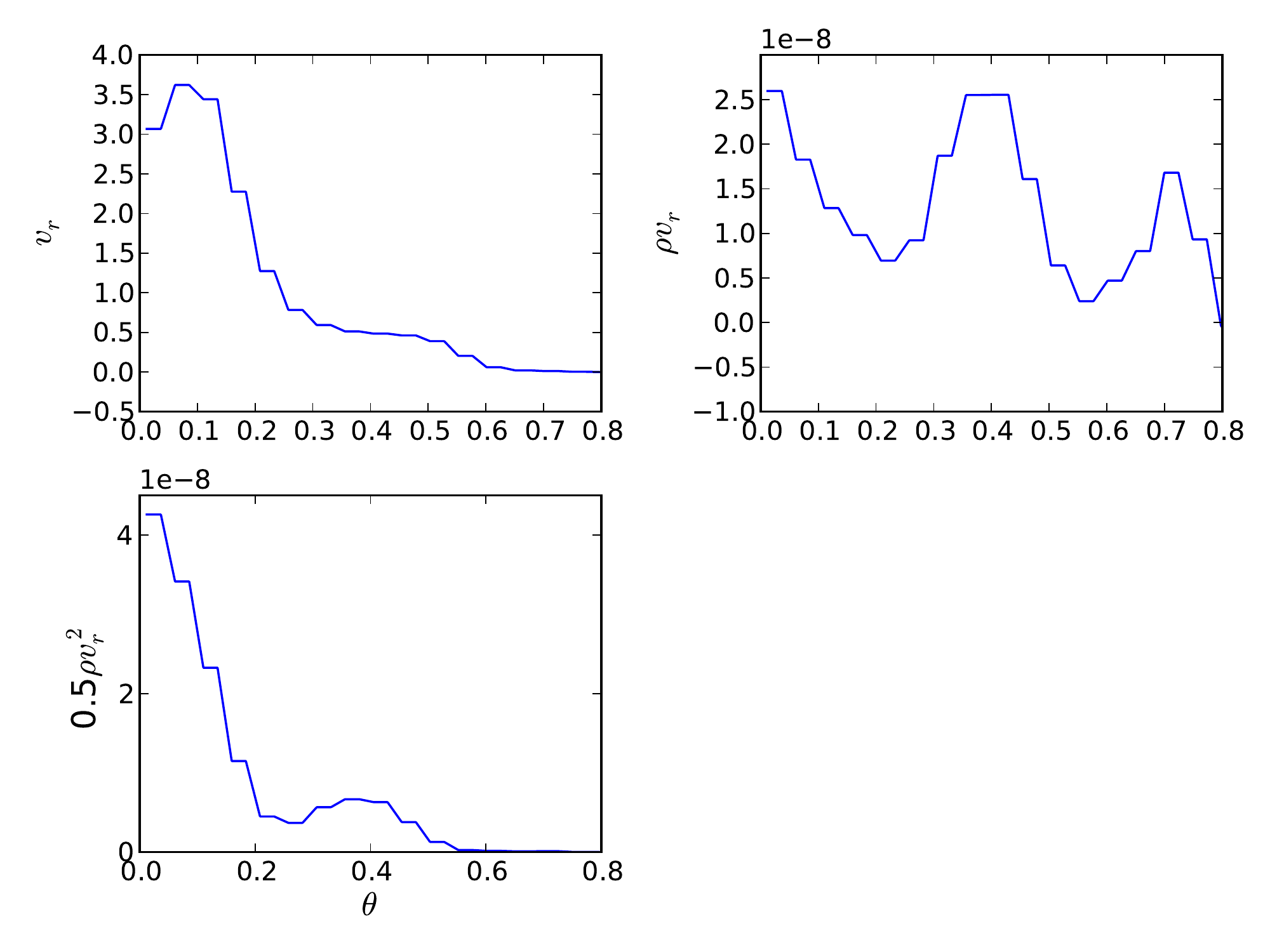} 
\vspace{0 cm}
\caption{The azimuthally and time averaged (from t=40 to 45.6 $T_{0}$ with $\Delta$t=0.1$T_{0}$) 
quantities at r=100 with respect to $\theta$.  } \label{fig:outerboundary}
\end{figure}

Strong outflows have been observed in FU Orionis system. Since these disks are fully ionized within $\sim$AU \citep{Zhu2007},
we should be able to directly compare our simulations with observations. Recent ALMA high angular resolution
observations \citep{Zurlo2017} reveal a wide hourglass shape outflow with an outflow velocity of several
km/s. Our outflow velocity is at a comparable rate.  As shown in Figure \ref{fig:outerboundary},
the terminal velocity of the outflow ranges from 0.5 to 3. If we assume that $R=1$ in our simulation
corresponds to the ionized disk size ($\sim$1 AU) and the central star mass is 0.3 $M_{\odot}$  \citep{Zhu2007}, 
the terminal velocity in our simulations ranges from 8 km/s
to 48 km/s.
On the other hand, our simulations cannot explain some
observables. For example, 
\citep{Calvet1993} have estimated an outflow rate
of 10$^{-5}\msunyr$ for FU Ori which has an accretion rate of $2\times 10^{-4}\msunyr$ . 
In our simulations, the outflow rate from R=0.5 to 5 is only 0.4\% of the accretion rate,
while observations suggest that the outflow rate is 5\% of the accretion rate. One solution is that the disk
is threaded by a stronger net vertical magnetic fields. Strong magnetic fields have been
observed in FU Orionis system \citep{Donati2005}. Another solution is that the wind strength will be much stronger 
in a thicker disk whose gravitational potential is smaller at the wind launching points so that more mass can escape the disk.

\section{Conclusions}
We have carried out global ideal MHD simulations to study accretion disks threaded by net vertical
magnetic fields.  Static mesh refinement has 
been adopted at the disk midplane to capture the growth of magnetorotational
instability (MRI), and special boundary conditions have been used to prevent the loss of magnetic
fields at the polar region. 

For our fiducial case which has an initial field of $\beta=1000$ at the midplane,
after running for 1442 orbits at the inner edge, the accretion flow reaches a steady state from R=0.1 to 3. The vertically integrated
$\alpha$ follows $R^{-0.4}$, reaching  almost 1 at the inner disk. Due to this $\alpha$ profile,
the disk surface density
follows $R^{-0.6}$ which is shallower than $R^{-1}$ in a viscous disk having a constant $\alpha$. 

The disk exhibits a complicated accretion pattern with supersonic inflow at the corona region and little inflow or even outflow at the midplane.
The corona region is magnetically dominated and the inflow velocity can reach $\sim 2 c_{s}$.  Although the inward surface flow is filamentary, 
it drags magnetic fields inwards, pinching fields at the disk surface. The Keplerian shear stretches the radial fields into azimuthal
fields and creates large $z-\phi$ stress between the midplane and the corona. Such $z-\phi$ stress torques the corona inwards and the midplane outwards, thus
sustaining the vertically sheared flow motion. If we treat the disk as a whole and neglect the internal sheared motion,
only 5\% of the disk accretion is due to the wind torque. 
95\% of the disk accretion is driven by the radial $r-\phi$ stress. Such stress is from MRI turbulence at the disk midplane
and large scale net magnetic fields at the disk atmosphere. However, disk wind can play a more important role when non-ideal
MHD effects have weakened MRI at the disk midplane \citep{Bai2017, Bethune2017}.

Even with such strong net vertical fields, a very weak disk wind is launched beyond the corona region
at $z\sim 1.5 R$. 
Although the wind is episodic, the time averaged wind properties can be fully described 
by the four conserved quantities from the steady wind theory. 
 The launching angle  (20$^{o}$-25$^{o}$) is smaller than 30$^{o}$ required to launch
wind from the disk midplane, but it is enough to launch wind from $z\sim 1.5 R$.
The wind is highly magnetized and very weakly loaded. 
The mass loss rate from R=0.5 to 5 is only 0.4\% of the disk accretion rate. 

When a weaker net field has been applied ($\beta_{0}=10^4$) or a thinner disk has been considered,
the wind torque accounts for less of the total accretion. 
Supersonic inflow at the disk surface also occurs and the picture of coronal accretion remains the same. 
The corona still extends to $z\sim 1.5 R$. $\alpha_{int}$ is significantly smaller when
a weaker field is applied. However,
for a thinner disk which is also threaded with $\beta_{0}=10^3$ as in our fiducial case, $\alpha_{int}$ is similar to the value in the fiducial case.

Magnetic fields are accreted to the central star along with mass. 
The global magnetic field geometry in the disk is sustained by the inward accretion and turbulence.
$Pr\sim$1 still seems to hold throughout the disk. 
Such steady field configuration is possible since 1) the surface inflow
is faster than the viscous flow and 2) the magnetic structure in the disk, which extends to $z\sim R$, is much thicker than the thermal scale height 
so that magnetic fields diffuse much slower.  Our simulations show excellent agreement
with some previous analytical studies which consider magnetic flux transport in 2-D turbulent disks (e.g. \citealt{GuiletOgilvie2012} and \citealt{GuiletOgilvie2013}). 

Our simulations may shed light on some astrophysical problems.
The large $\alpha$ in our simulations may be applicable to  cataclysmic variables.
The puffed corona may help to explain the high near-infrared flux in Herbig stars.
Fast surface inflow may be the cause for the fast inflow in transitional disks. 
The self-consistent accretion and outflow geometry may be applied to FU Orionis systems. 
Finally, the vertically sheared mass transport may play an important role on transporting
chondrite components in protoplanetary disks.

\acknowledgments
All  simulations are carried out using computer supported by the 
Princeton Institute of Computational Science and Engineering, and the 
Texas Advanced Computing Center (TACC) 
at The University of Texas at Austin through XSEDE grant TG-
AST130002. 
Z. Z. acknowledges support from the National Aeronautics and Space Administration 
through the Astrophysics Theory Program with Grant No. NNX17AK40G and support
from the Sloan Research Fellowship.
We thank Kengo Tomida and Christopher J. White for their contributions to the Athena++ code.
We thank Jerome Guilet  and Gordon Ogilvie for sharing their figures with us.
We also thank Satoshi Okuzumi, Roman Rafikov, and Sebastien Fromang for very helpful discussions and comments. 
Finally, we thank the referee for very helpful suggestions.

\appendix
To prevent the loss of magnetic fields at the polar region, 
we implement a special polar boundary in the $\theta$ direction. The implementation of this boundary depends on 
the domain size in the $\phi$ direction. We will first describe this boundary condition when the domain extends over the full $2\pi$ in the $\phi$ direction. 
In this case,
at the pole, the ghost zones in the $\theta$ direction at (r$_{i}$, $\theta_{j=-1}$, $\phi_{k}$) overlap with the active zones at 
(r$_{i}$, $\theta_{j=0}$, $2\pi-\phi_{k}$) where $-1$ and 0 refer to the ghost and active zone. 
Thus, we first assign all cell-centered and face-centered quantities in the ghost zones 
using the quantities from the corresponding active zones at the opposite $\phi$ position. Due to the positive direction 
adopted in the spherical-polar system, cell-centered $v_{\theta}$,  $v_{\phi}$ and face-centered $B_{\theta}$, $B_{\phi}$ 
need to flip their signs when they
are copied to the ghost zones. Since the first active zone in the $\theta$ direction basically has a
zero area at the pole, we do not
use the CT scheme to update $B_{\theta}$ at the pole. Instead $B_{\theta}$
there is the average of $B_{\theta}$ from the second active zone and the first ghost zone. Finally, $E_{r}$
at the pole is taken as the average of $E_{r}$ from all the grids touching the pole and the averaged value is shared among all these grids.  

If the domain only extends over a wedge in the $\phi$ direction, all above steps are the same except the first step where
quantities at the ghost zones are directly copied from the active zones at the same $\phi$ slice 
instead of copying from the $2\pi-\phi$ slice. 

For mesh-refinement , the prolongation and restriction methods in \cite{TothRoe2002}  can 
preserve both $\nabla \cdot \bf{B}$ and $\nabla \times \bf{B}$ in Cartesian coordinates. 
We adopt the same methods for mesh-refinement with spherical-polar coordinates. 
Theoretically, we could follow the steps in  \cite{TothRoe2002}  to design algorithms preserving both $\nabla \cdot \bf{B}$ and $\nabla \times \bf{B}$ under
the spherical-polar coordinate system. However, due to the additional geometrical factor in $\nabla \cdot \bf{B}$ and $\nabla \times \bf{B}$,
the final restriction function is very complicated. Thus, we only require the restriction to satisfy $\nabla \cdot \bf{B}$=0, which is crucial
for CT.
It can be shown that Equation (8)-(12) in \cite{TothRoe2002} can guarantee $\nabla \cdot \bf{B}$=0 even in cylindrical and spherical-polar coordinates 
as long as $\nabla \cdot \bf{B}$=0
initially. Note that it cannot conserve  $\nabla \cdot \bf{B}$ if $\nabla \cdot \bf{B}$ is not zero initially, which is different when this method
is used in Cartesian coordinates.

\end{document}